\documentclass[a4paper,11pt]{article}
\usepackage{jheppub}
 \usepackage[utf8]{inputenc}
\usepackage{graphics}
\usepackage{epsfig}
\usepackage{amsfonts,amsmath,amssymb,amsthm,mathtools}
\usepackage{calc}
\usepackage{caption}
\usepackage{subcaption}
\newcommand{\beq}{\begin{equation}}
\newcommand{\eeq}{\end{equation}}
\newcommand{\bea}{\begin{eqnarray}}
\newcommand{\eea}{\end{eqnarray}}
\usepackage{booktabs}

\begin{document}

\author[a,b]{
Bercel Boldis
}
\author[c]{P\'eter L\'evay}

\affiliation[a]{
Department of Theoretical Physics
\\Budapest University of Technology and Economics
\\M\H uegyetem rkp. 3., H-1111 Budapest, Hungary
}
\affiliation[b]{
HUN-REN Wigner Research Centre for Physics\\Konkoly-Thege Miklós u. 29-33, 1121 Budapest, Hungary
}
\affiliation[c]{
MTA-BME Quantum Dynamics and Correlations Research Group\\
Eötvös Loránd Research Network (ELKH)\\
Budapest University of Technology and Economics
\\M\H uegyetem rkp. 3., H-1111 Budapest, Hungary
}

\emailAdd{boldis.bercel@wigner.hu}
\emailAdd{levay.peter@ttk.bme.hu}

\title{A holographic connection between strings and causal diamonds}

\abstract{
In this paper we explore ideas of holography and strings living in the $d+1$ dimensional Anti-de Sitter space $AdS_{d+1}$ in a unified framework borrowed from twistor theory. In our treatise of correspondences between geometric structures of the bulk $AdS_{d+1}$, its boundary and the moduli space of boundary causal diamonds aka the kinematic space ${\mathbb K}$, we adopt a perspective offered by projective geometry. From this viewpoint certain lines in the $d+1$ dimensional real projective space, defined by two light-like vectors in ${\mathbb R}^{d,2}$ play an important role. In these projective geometric elaborations objects like Ryu-Takayanagi surfaces, spacelike geodesics with horospheres providing regularizators for them and the metric on ${\mathbb K}$ all find a natural place. Then we establish a correspondence between classical strings in $AdS_{d+1}$ and causal diamonds of its asymptotic boundary. At each point on the worldsheet, the tangent vectors $\partial_\pm X$ are projected onto boundary coordinates that identify the past and future tips of a causal diamond. Under this projection, the string equations of motion translate into a dynamics of boundary causal diamonds. A procedure for lifting up a causal diamond to get a proper string world sheet is also developed. In this context we identify an emerging $SO(1,1)\times SO(1,d-1)$ gauge structure incorporated into a Grassmannian $\sigma$-model targeted in ${\mathbb K}$. The $d=2$ case is worked out in detail. Surprisingly in this case $AdS_3$ with its strings seems to be a natural object which is living inside projective twistor space. On the other hand ${\mathbb K}$ (comprising two copies of two dimensional de Sitter spaces)  is a one which is living inside the Klein quadric, as a real section of a complexified space time.
}


\maketitle

\section{Introduction}

Since the discovery of holography and the AdS/CFT correspondence\cite{Maldacena2} physicists have realized that in order to understand the properties of a physical system sometimes one can find it useful to rephrase such properties within the realm of another system of a wildly different character.
In mathematics, an idea of similar kind have already been followed by classical geometers of the 19th century.
Namely there it has been found that it is sometimes insightful to reiterate problems connected to geometrical structures of a space  in terms of geometrical structures of another space of a very different kind.
For example the so called Klein correspondence\cite{Ward,Hurd} revealed that the lines of the three dimensional (projective) space can be parametrized by the points of a four dimensional one.
To physics this idea has made its debut via twistor theory of Penrose\cite{Penrose} as a possible approach for studying four dimensional quantum gravity  in terms of complex geometry living in the dual three dimensional projective twistor space.
Since then twistor techniques have also catalysed many interesting developments such as twistor strings\cite{Wittenstring} a topic inherently  connected to the study of scattering amplitudes. Such developments have culminated in the introduction of revolutionary ideas such as the amplituhedron\cite{Nima}.
Results with connections to the idea of holography has also been established\cite{Adamo}.
Although the power and elegance of such techniques is inherently connected to complex geometry, the spirit of this theory also motivated studies not only within the complex but also
in the real domain \cite{Krasnov}.

Recently in the AdS/CFT correspondence it has been realized that apart from studying  geometrical correspondences between the usual spaces (namely the bulk and its boundary) it is also useful to explore the physical implications of the geometric data encoded into yet another space: the moduli space of boundary causal diamonds. For this space the term kinematic space ${\mathbb K}$ has been coined\cite{Czech,Myers}.
In a previous paper\cite{Levay4} we have observed that there is a correspondence between the causal diamonds of the boundary of $AdS_{d+1}$, represented by points of ${\mathbb K}$, and segmented strings\cite{Gubser,DV1,DV2,Dv3} in $AdS_{d+1}$.
Now
$AdS_{d+1}$
can be regarded as a space embedded into the projective space ${\mathbb R}{\mathbb P}^{d+1}$ with its boundary being a quadric there\cite{Seppi}. In this respect one can regard AdS geometry as a specialization of projective geometry.
Then in this context string segments trace out two dimensional timelike planes, and projectively such planes correspond to special lines in ${\mathbb R}{\mathbb P}^{d+1}$.  Then 
 one is given a correspondence between certain lines of a space and points of another type of a space namely ${\mathbb K}$. Moreover, this correspondence for $d=2$ is
reminiscent of a truncation of the Klein correspondence.  

In this paper we would like to present a detailed elaboration on this observation by connecting ideas of holography, strings and kinematic space in a unified framework  motivated by some of the methods of twistor theory. However, our considerations in this paper will be purely classical, the very interesting quantum aspects are to be explored in a later work.
Moreover, in order to present our ideas within the simplest context in this paper we consider strings propagating in pure AdS geometry.
Except for some hints presented in the conclusions, we refrain from
elaborations concerning the applicability of our results to asymptotically AdS geometries.
Such explorations, interesting for readers looking for applicability of our ideas to such more general spacetimes, will be pursued in future work.

In our treatise of correspondences between the geometric structures of different spaces we emphasize the pivotal role of certain planes spanned by two light-like vectors $U$ and $V$ in ${\mathbb R}^{d,2}$ satisfying the constraint $U\cdot V<0$. These planes are time-like ones i.e. ones of (1,1) signature. Alternatively in the projective picture such planes define special lines in ${\mathbb R}{\mathbb P}^{d+1}$.  Employing certain constraints we realize that in the Poincaré patch these planes are associated with geometric objects of two characteristic types. Such objects has already been familiar to physicists as the main actors on the holographic scene.
In the first type
we find {\it disconnected space-like geodesics} and {\it minimal surfaces}. 
The latter type of objects are well known from the covariant generalization of Ryu-Takayanagi (RT) surfaces\cite{RT1,RT2,RT3}.
On the other hand in the other type we find  {\it connected spacelike geodesics}  that are half ellipses. In this case the pair $(U,V)$ also determines the {\it horospheres}\cite{Seppi} that can be used as geometric regularizators\cite{Levay1} for the diverging length of such geodesics. 
In the $d=2$ case via the Ryu-Takayanagi correspondence\cite{RT1,RT2,RT3} such regularizators can also be related to regularizators of diverging entanglement entropies\cite{Levay1}.
Hence in this class, the pair $(U,V)$ encodes {\it connected geodesics} taken together with their {\it geometric regularizators}. 

Interestingly such pairs of null vectors $(U,V)$ and their associated projective lines are also natural objects for presenting a new characterization of the metric structure of the moduli space of boundary causal diamonds.
The novelty of our reformulation of this metric structure is that it is manifestly invariant under the  gauge transformations associated with the ambiguity of defining the pair $(U,V)$
spanning the time-like plane.
Indeed a natural gauge invariant object to consider is the separable bivector ${\mathcal P}=U\wedge V/(U\cdot V)$. In terms of this bivector the metric on kinematic space takes the nice (\ref{endmetrik}) form.

Motivated by these ideas next we upgrade the $(U,V)$ pair of vectors to {\it fields} defined on a $2D$ space.  Then all our geometric objects can naturally be related to  strings that propagate in $AdS_{d+1}$.
In order to achieve this goal we will regard the pair $(U,V)$ not merely as a pair of null vectors, but rather as null vector fields parametrized by some set of coordinates. Explicitly,  we will use the parametrization $U^a(\sigma, \tau)$ or $U^a(\sigma_+,\sigma_-)$ where $\sigma_{\pm}=\tau\pm\sigma$ are familiar from string theory. 
Then by relating the string action to an action incorporating the moduli space metric, we show that strings in the AdS bulk can alternatively be described by a Grassmannian $\sigma$ model targeted on ${\mathbb K}$.

The fact that string motion in the bulk is projected to a dynamics of causal diamonds can easily be understood.
This result simply follows from the observation that to each point of the string world sheet one can associate two tangent null vectors subject to the Virasoro constraints characterizing the conformal gauge.
However, one can even succeed in the other way round, namely in the construction of the bulk world-sheet from the boundary data provided by the future and past tips of a causal diamond.
This is the problem of how to lift up a diamond to get a proper string world sheet.
The solution of this problem amounts to a clever choice of the gauge degree of freedom associated with our timelike planes.
As an extra bonus one gets a nice interpretation of the so called $\alpha$ field. As is well-known, in the AdS$_3$ case this field is featuring the null-polygonal Wilson loop calculations of Ref.\cite{Maldacena}. In our fully general AdS$_{d+1}$ context this object shows up as the sum of the two gauge fixing
functions associated to the future and past tips of the diamonds.
This unique lift is reminiscent of a unique horizontal curve which is a lift to the boundle space of a base space curve familiar from parallel transport.

We recall that the applicability of the idea of lifting up mathematical objects from a space to another one is rooted in a gauge structure. In our case this structure is coming from the possibility of considering in ${\mathbb R}^{d,2}$ the bundle of pseudo-orthogonal frames which has its base space precisely our ${\mathbb K}$.
The basis vectors of this pseudo-orthogonal frame are the ones spanning the world sheet, and the vectors spanning the normal directions. These $d+1$ vectors taken together with the vector running on the world-sheet itself gives the correct count: $d+2$.
Then the emerging gauge structure is the one that is modeled by a Stiefel bundle considered over the Grassmannian ${\mathbb K}=SO(2,d)/SO(1,d-1)\times SO(1,1)$. This structure is then responsible for the possibility of representing string dynamics 
by $SO(1,d-1)\times SO(1,1)$ gauge invariant equations.
As far as physics is concerned this is basically the approach followed by Ref.\cite{Maldacena} for the $d=2$ case having its origin in the works of Refs.\cite{Pohlmeyer,Vega,Jevicki}.
In this paper we present a reformulation of this gauge structure for arbitrary $d$ convenient for our purposes.

The possible physical implications of the  gauge theory connected to a $\sigma$ model targeted on ${\mathbb K}$ has not fully been explored yet.
Apart from clarifying mathematical aspects of holography, the results of this paper clearly show that this model incorporating gauge degrees-of-freedom is inherently connected to bulk strings.
It is common wisdom that point particles can be used to test the curvature of space-time 
by looking at deviation equations for congruences of geodesics. On the other hand it is also well-known that being extended objects strings can offer a test for bulk geometry in a profoundly different manner.
Our contribution to this knowledge is that even at the level of classical geometry bulk strings are holographically connected to boundary causal diamonds of the boundary in an explicit geometric manner.
At first sight since strings have provided the first successful implementation of the idea of holography\cite{Maldacena2} this should not come as a surprise.  In any case we hope that we can convince the reader that our elaboration of the simple correspondence  between bulk strings and kinematic space is worth exploring further, especially within the quantum realm.

The organization of this paper is as follows.
In Section 2. we begin by presenting a study of planes, geodesics and minimal surfaces of AdS$_{d+1}$ incorporated into a projective geometric formalism familiar from twistor theory.  
We introduce Plücker coordinates for projective lines, their principal null directions, and connect these notions to objects well-known from holography.
{Kinematic space, our main object of scrutiny, shows up in this context as a special Grassmannian of two planes of signature $(1,1)$ or alternatively as the space of special lines in the projective space ${\mathbb R}{\mathbb P}^{d+1}$. A gauge degree of freedom is identified here as the ambiguity in defining the null directions $U$ and $V$ for such planes.
Armed with these ideas in Section 3. we present a twistor geometric rederivation of the metric\cite{Myers} in kinematic space.
The local gauge group of the previous section, which is ${\mathbb R}\times SO(1,1)$ plays an important role here.
The new formula for the metric displays a gauge invariant separable bivector ${\mathcal P}=U\wedge V/U\cdot V$.
The distance between two infinitesimally separated planes is measured by the cross-ratio.

Using this formalism in Section 4. we turn to strings in AdS$_{d+1}$.
We elevate the null vectors $U$ and $V$ to vector fields by identifying them with the fields arising from the Virasoro constraints of string theory in the conformal gauge.
Then working in the Poincaré patch, starting from the string action we derive a $\sigma$-model action featuring fields targeted in $\mathbb K$. 
This model encapsulates the correspondence between boundary causal diamonds
and bulk strings. We then derive the equations of motion in causal diamond coordinates, and discuss the problem of lifting a causal diamond from the boundary to produce a proper string world-sheet in the bulk.
This procedure is the continuous version of a known discrete one we have already investigated in our recent paper\cite{Levay4}.
It is based on segmented strings.
For the convenience of the reader next we also include a clarification of the relationship between our sigma model and segmented strings. 
Section 5 is devoted to a detailed elaboration on the bulk gauge structure of our stringy dynamics that can also be reinterpreted in boundary terms.

Finally in Section 6 we present a case study of the $d=2$ case, namely AdS$_3$. In this case kinematic space is just of the following form\cite{Czech,Myers} ${\mathbb K}=dS_2\times dS_2$ i.e. two copies of de Sitter spaces. We show that in terms of $\mathcal P$ the two copies are governed by the self-dual and anti self-dual parts of this gauge invariant object. Moreover, these copies naturally live inside the Klein quadric the main actor of the usual twistor correspondence.
We then discuss cross-ratios and
twistors. In particular we show that two special lines in the twistor space ${\mathbb R}{\mathbb P}^3$ that represent planes of signature $(1,1)$ in the bulk intersect if and only if in the boundary the cross ratios satisfy a reality condition. The physical meaning of this condition is that the corresponding four boundary points associated with two intersecting causal diamonds are comprising events on the world lines of inertial observers or ones moving with a constant acceleration. 
The world lines in turn produce the flow lines for the modular flow of the embedding
diamond
which comprises the causal completion of the intersecting ones.

Next we turn to projective geometric considerations concerning the static slice of AdS$_3$. We discuss how the usual interpretation of kinematic space as the space of geodesic of the static slice\cite{Czech} shows up in our formalism.
Then we turn to an illustration of how the geometric regularization procedure for entanglement entropies via a regularization of geodesic lengths based on horocycles introduced in\cite{Levay1} emerges in our framework.
Finally a detailed elaboration on strings in AdS$_3$ is given.
We give explicit expressions for the lifting procedure of diamonds to strings comparing them to the segmented strings approach of Refs.\cite{DV1,DV2}.
In particular we produce an analysis of the special case when in the bulk strings coupled to B-fields  are present.
There is an extra parameter $\kappa$ describing the coupling to B-fields. For $\kappa\to 0$ we get back to free strings. The upshot is that the $\kappa\neq 0$  theory corresponds to a different choice of gauge.
This means that one can obtain different bulk string theories by pulling back the equations of motion representing the dynamics on kinematic space.

\section{Planes, geodesics, surfaces and null directions}

\subsection{${\rm AdS}_{d+1}$ and its Poincar\'e patch.}

In this paper we use the projective formalism as developed in the treatment of Refs.\cite{Barbot,Seppi} which is convenient for our purposes.
Let us consider the $d+2$ dimensional real vector space ${\mathbb R}^{d+2}$ equipped with a nondegenerate bilinear form $\cdot$ of index $2$. This means that for $X,Y\in ({\mathbb R}^{d+2},\cdot)$ we have
\beq
\cdot : {\mathbb R}^{d+2}\times {\mathbb R}^{d+2}\to {\mathbb R},\qquad (X,Y)\mapsto X\cdot Y\coloneqq {\eta}_{ab}X^{a}Y^{b}
\nonumber
\eeq
\beq
{\eta}_{ab}X^{a}Y^{b}
=-X^{-1}Y^{-1}-X^0Y^0+X^1Y^1+\dots + X^{d-1}Y^{d-1}+X^dY^d
\label{bili1}
\eeq
where $a,b=-1,0,1,\dots ,d-1,d$.
The line element is
\beq 
ds^2=\eta_{ab}dX^a dX^b.
\label{lineelement}
\eeq
One can also introduce light cone coordinates in the plane $(-1,d)$ as follows
\beq
X^{\pm}=X^{d}\pm X^{-1}
\label{lc0}
\eeq
then the $\cdot$ product takes the following form
\beq X\cdot Y=-X^0Y^0+X^1Y^1+\dots + X^{d-1}Y^{d-1}+\frac{1}{2}\left(X^+Y^-+X^-Y^+\right)
\label{newfrom}
\eeq
In this case instead of the standard one $(X^{-1},X^0,{\bf X},X^d)$ we are labelling the coordinates as $(X^+,X^0,{\bf X},X^-)$.

We also introduce in ${\mathbb R}^{d}$ a nondegenerate bilinear form $\bullet$ of index $1$ as follows
\beq
\bullet: {\mathbb R}^{d}\times {\mathbb R}^{d}\to {\mathbb R},\qquad (x,y)\mapsto x\bullet y\coloneqq {\eta}_{\mu\nu}x^{\mu}y^{\nu}
\nonumber
\eeq
\beq
{\eta}_{\mu\nu}x^{\mu}y^{\nu}
=-x^0y^0+x^1y^1+\dots + x^{d-1}y^{d-1}
\label{bili2}
\eeq
where $\mu,\nu=0,1,\dots , d-1$.
In the following we use the notation
\beq
(t,{\bf x})\equiv (x^0,x^i),\qquad i=1,2,\dots,d-1
\nonumber
\eeq
and we suppose that the vector space ${\mathbb R}^{d-1}$ is equipped with the usual norm.
Then we have
\beq
x^2\equiv x\bullet x=-t^2+\vert\vert{\bf x}\vert\vert^2
\label{minki}
\eeq

We define the $d+1$ dimensional anti deSitter space as the quotient
\beq
{\rm AdS}_{d+1}\coloneqq\{X\in {\mathbb R}^{d+2}\vert X\cdot X=-L^2\}/{\mathbb Z}_2
\label{AdS}
\eeq
where $L^2$ is the {\rm AdS} length and
${\mathbb Z}_2=\{\pm I\}$, i.e. under the action of this group we identify $X$ with $-X$.
In this way one can imagine $AdS_{d+1}$ as an object  living in the projective space ${\mathbb R}{\mathbb P}^{d+1}$.
An alternative definition in this spirit is
\beq
{\rm AdS}_{d+1}\coloneqq{\mathbb P}\{X\in {\mathbb R}^{d+2}\vert X\cdot X<0\}\subset {\mathbb R}{\mathbb P}^{d+1}
\label{AdSp}
\eeq
where ${\mathbb P}$ means projectivization.
We define the boundary of $AdS_{d+1}$ as the quadric
\beq{\partial}_{\infty} AdS_{d+1}\equiv{\mathcal Q}_d\coloneqq {\mathbb P}\{U\in{\mathbb R}^{d+2}\vert U\cdot U=0\}\subset {\mathbb R}{\mathbb P}^{d+1}.
\label{bdy}
\eeq
We will also use the double cover of $AdS_{d+1}$ denoted by
$\widetilde{{\rm AdS}}_{d+1}$ i.e.
\beq
\widetilde{{\rm AdS}}_{d+1}\coloneqq\{X\in {\mathbb R}^{d+2}\vert X\cdot X=-L^2\}
\label{doublecover}
\eeq
with the projection
\beq
\pi:\widetilde{{\rm AdS}}_{d+1}\to {\rm AdS}_{d+1}
\label{proji}
\eeq
Notice that in this projective geometric picture  $AdS_{d+1}$ and its boundary are both embedded into the projective space $\mathbb{RP}^{d+1}
$. The boundary ${\mathcal Q}_d$ is a quadric in $\mathbb{RP}^{d+1}$.

 Let us now define the half space model\cite{Seppi} of ${\rm AdS}_{d+1}$.
It is the upper half space
\beq
{\mathbb U}_{d+1}\coloneqq \{(t,{\bf x},z)\in {\mathbb R}\oplus {\mathbb R}^{d-1}\oplus{\mathbb R}^{+}\vert
ds^2=L^2\left[-dt^2+\vert\vert d{\bf x}\vert\vert^2+dz^2\right]/z^2\}
\}
\label{upper}
\eeq
equipped with the Poincar\'e metric coming from Eq.(\ref{lineelement}).
The $z=0$ boundary of this space in ${\mathbb R}^{d+1}$ will be denoted as $\partial {\mathbb U}_{d+1}$.

Now we define an embedding $\tilde{\mathcal E}:{\mathbb U}_{d+1}\to \widetilde{{\rm AdS}}_{d+1}$ as follows
\beq
\begin{pmatrix} t\\ {\bf x} \\ z\end{pmatrix}\mapsto \begin{pmatrix}X^{-1}\\ X^0 \\ {\bf{X}} \\X^d\end{pmatrix}
=\frac{1}{2z}\begin{pmatrix}-z^2-x^2-L^2\\ 2Lt \\ 2L{\bf x} \\ -z^2-x^2+L^2\end{pmatrix}
\label{embed}
\eeq
where we have used Eq. (\ref{minki}).
The inverse of $\tilde{\mathcal{E}}$ over its image is
\beq
\begin{pmatrix}t\\ {\bf x}\\ z\end{pmatrix}=\frac{L}{X^-}\begin{pmatrix} X^0\\ {\bf X}\\ L\end{pmatrix}
\label{inverz}
\eeq
hence  $\tilde{\mathcal{E}}$ is injective.
On the other hand our map is clearly not surjective since
$\tilde{\mathcal{E}}({\mathbb U}_{d+1})=\widetilde{\rm{AdS}}_{d+1}\cap\{X^->0\}$.
This means that the coordinates $(t,{\bf x},z)$ will not cover $\widetilde{\rm{AdS}}_{d+1}$ merely the so called "Poincar\'e patch" characterized by the constraint $X^{-}>0$.
One can now define the map
\beq
{\mathcal E}\coloneqq \pi \circ \tilde{{\mathcal E}}:{\mathbb U}_{d+1}\hookrightarrow {\rm AdS}_{d+1}
\label{mapembed}
\eeq
which is an isometric embedding, with its image being the complement of the intersection of the hyperplane ${\mathcal P}_{\infty}$ characterized by the constraint $X^{-}=0$ with 
${\rm{AdS}}_{d+1}\subset {\mathbb R}{\mathbb P}^{d+1}$.

\subsection{The boundary of ${\rm AdS}_{d+1}$ in the Poincar\'e patch}

Since ${\mathcal E}$ defined by Eqs.(\ref{embed}) and (\ref{mapembed}) 
is an embedding in the projective space ${\mathbb R}{\mathbb P}^{d+1}$
one can extend it to $\partial{\mathbb U}_{d+1}$ i.e. the boundary of the Poincar\'e patch by the formula
\beq
\left[\frac{-z^2-x^2-L^2}{z}:\frac{2Lt}{z}:\frac{2L\bf x}{z}:\frac{-z^2-x^2+L^2}{z}\right]=[-z^2-x^2-L^2:2Lt:2L{\bf x}:-z^2-x^2+L^2].
\nonumber
\eeq 
Indeed, after taking the $z\to 0$ limit we obtain a ray of the form $[U^{-1}:U^0:{\bf U}:U^d]$ with the property for its representatives $U\cdot U=0$.
After switching to $U^{\pm}$ coordinates a general representative for this ray can be choosen as 
\beq
\begin{pmatrix}U^+\\ U^0\\ {\bf U}\\ U^-\end{pmatrix}=\frac{1}{\Delta}\begin{pmatrix}

t^2-\vert\vert {\bf x}\vert\vert^2\\ Lt\\ L{\bf x}\\ L^2\end{pmatrix},\qquad\Delta\in {\mathbb R}\setminus\{0\}
\label{conf1}
\eeq
Notice that choosing a representative for a ray can be regarded as a gauge degree of freedom showing up as a {\it local rescaling} of the (\ref{conf1}) representatives. One can display this degree of freedom via a transformation of the form
\beq
\begin{pmatrix}U^+\\ U^0\\ {\bf U}\\ U^-\end{pmatrix}\mapsto \frac{1}{\Lambda _u}
\begin{pmatrix}U^+\\ U^0\\ {\bf U}\\ U^-\end{pmatrix},\qquad \Delta\mapsto \Lambda_u\Delta
\label{conf2}
\eeq
hence $\Lambda_u\in {\mathbb R}\setminus\{0\}$ acts as a local rescaling of $\Delta$, i.e. a one depending on the point $U\in \partial{\mathbb U}_{d+1}$ choosen.

The points of $\partial{\mathbb U}_{d+1}$ are comprising a copy of Minkowski space time ${\mathbb R}^{d-1,1}$ embedded in 
$\partial_{\infty}{\rm AdS}_{d+1}$.
Indeed ${\mathcal E}(\partial{\mathbb U}_{d+1})\subset \partial_{\infty}{\rm AdS}_{d+1}$. Moreover, one can check that ${\mathcal E}$ is injective when restricted to the boundary $\partial{\mathbb U}_{d+1}$ of the Poincar\'e patch. 
Hence the extension ${\mathcal E}:\partial{\mathbb U}_{d+1}\hookrightarrow \partial_{\infty}{\rm{AdS}}_{d+1}$ is an embedding as well.
Of course this embedding on the boundary is a non surjective one, hence in order to understand the structure of the full boundary $\partial_{\infty}{\mathbb U}_{d+1}$ of the half space model of ${\rm AdS}_{d+1}$ some extra work is needed.
Indeed, observe that
\beq
{\mathcal E}(\partial {\mathbb U}_{d+1})=\partial_{\infty}{{\rm AdS}_{d+1}}\cap\{U^-\neq 0\}\subset {\mathbb R}{\mathbb P}^{d+1}
\nonumber
\eeq
hence in order to construct the full boundary in the half space model the hyperplanes with the property $U^-=0$ should be characterized in terms of  ${\mathbb U}_{d+1}$ data.
Since we will restrict our attention to the Poincaré patch we will not go into the details of this construction, for the details see Refs.\cite{Barbot,Seppi}.

\subsection{Pl\"ucker coordinates and null directions}

Let us now consider two linearly independent vectors $A,B\in {\mathbb R}^{d,2}$. They are spanning a plane ${\mathcal P}$ 
. Define
\beq
P^{ab}\coloneqq A^{a}B^{b}-A^{b}B^{a}
\label{Plucker}
\eeq
called the Pl\"ucker coordinates of ${\mathcal P}$.
The $P^{ab}$ are independent of the representatives of ${\mathcal P}$.
Indeed, by choosing a new linear combination of the representatives of ${\mathcal P}$, say $A^{\prime}=\alpha A+\beta B$ and 
$B^{\prime}=\gamma A+\delta B$ the Pl\"ucker coordinates transform as 
\beq
(A,B)\mapsto (A^{\prime},B^{\prime})\implies 
P^{ab}\mapsto (\alpha\beta-\gamma\delta)P^{ab}
\label{transpluck}
\eeq
 Hence up to a nonzero constant for the characterization of ${\mathcal P}$ the same $P^{ab}$ will do.

One can also check that
\beq
{P^{a}}_{b}{P^{b}}_{c}A^{c}=DA^{a},\qquad 
{P^{a}}_{b}{P^{b}}_{c}B^{c}=DB^{a}
\label{eigen}
\eeq
where
\beq
D\coloneqq 
-\frac{1}{2}P\cdot P\equiv-\frac{1}{2}P^{ab}P_{ab}=
\left(A\cdot B\right)^2-\left(A\cdot A\right)\left(B\cdot B\right)
\label{Cayley}
\eeq
and it is understood that ${P^{a}}_{b}={\eta}_{bc}P^{ac}$.

Let us also define
\beq
\tilde{A}^{a}=-P^{ab}A_{b},\qquad
\tilde{B}^{a}=-P^{ab}B_{b}
\label{dual}
\eeq
These new vectors are also belonging to the same plane.
Moreover, we have
\beq
\tilde{A}\cdot\tilde{A}=-DA\cdot A,\qquad \tilde{B}\cdot\tilde{B}=-DB\cdot B
\label{spacetilde}
\eeq
and
\beq
\tilde{A}\cdot A=\tilde{B}\cdot B=0,\qquad \tilde{A}\cdot\tilde{B}=-DA\cdot B\qquad \tilde{A}\cdot B=-\tilde{B}\cdot A=-D.
\label{relations}
\eeq
Let us observe that for $A$ and $B$ timelike ($A\cdot A<0$ and $B\cdot B<0$)
and $D>0$, 
$\tilde{A}$ and $\tilde{B}$ are space like vectors.
In this case, in particular for $A,B\in\widetilde{{\rm AdS}}_{d+1}$,
\beq
A\cdot A=B\cdot B=-L^2,\qquad \frac{\tilde A}{\sqrt{D}}\cdot\frac{\tilde A}{\sqrt{D}}=
\frac{\tilde B}{\sqrt{D}}\cdot\frac{\tilde B}{\sqrt{D}}=L^2
\label{timelikeplane}
\eeq
Hence the plane spanned by the {\it orthogonal} vectors $A$ and $\tilde{A}/{\sqrt{D}}$ (or alternatively by $B$ and $\tilde{B}/{\sqrt{D}}$) is a time-like plane. 
As we will see later physically this important special case is realized for spacelike geodesics of  $\widetilde{\rm AdS}_{d+1}$ when the pairs $(A,\tilde{A})$ and $(B,\tilde{B})$ are spanning the same plane. Indeed, in this case the pair $(X(\lambda),\dot X(\lambda))$ is comprising the geodesic curve, and its tangent (velocity) for each value of the parameter $\lambda$. Clearly now
$X(\lambda)\cdot\dot{X(\lambda)}=0$.

Generally time-like planes spanned by the vectors $T$ and $S$ are characterized by the properties
\beq
S\cdot S>0,\qquad T\cdot T<0,\qquad T\cdot S=0\qquad D=-S^2T^2>0.
\label{genplane}
\eeq 
For such planes one can define {\it 
principal null directions}\cite{Krasnov} as follows
\beq
U=\frac{1}{\sqrt{2}}(T+S),\qquad V=\frac{1}{\sqrt{2}}(T-S)
\label{princi}
\eeq
satisfying
\beq
U\cdot U=V\cdot V=0,\qquad U\cdot V<0.
\label{propprinc}
\eeq
 In particular for $A$ and $B$ given one can define\footnote{Do not confuse these quantities with lower indices with the upper indexed ones of Eq.(\ref{lc0}).}
\beq
U_{\pm}=\frac{1}{\sqrt{2}}\left(A\pm\frac{\tilde{A}}{\sqrt{D}}\right), \qquad 
V_{\pm}=\frac{1}{\sqrt{2}}\left(B\pm\frac{\tilde{B}}{\sqrt{D}}\right)
\label{princnull}
\eeq
with the important property
\beq
U_{\pm}\cdot U_{\pm}=V_{\pm}\cdot V_{\pm}=0, \qquad U_+\cdot U_-=A\cdot A<0,\qquad V_+\cdot V_-=B\cdot B<0
\label{null}
\eeq
hence these vectors are also satisfying the (\ref{propprinc}) constraints. 
The rays associated to these null vectors determine two points on the boundary 
$\partial_{\infty}{\rm AdS}_{d+1}$ defined by Eq.(\ref{bdy}).

Geometrically this construction of principal null directions has the following geometric meaning.
The linearly independent vectors $A$ and $B$ in ${\mathbb R}^{d+2}$ define a plane. Projectively in ${\mathbb R}{\mathbb P}^{d+1}$ these vectors define a line. Our boundary $\partial_{\infty}{\rm AdS}_{d+1}\subset {\mathbb R}{\mathbb P}^{d+1}$ is a quadric ${\mathcal Q}\subset {\mathbb R}{\mathbb P}^{d+1}$. The two points $[A]$ and $[B]$ of ${\mathbb R}{\mathbb P}^{d+1}$ are not lying on ${\mathcal Q}$. However, it is easy to show that for $D>0$ the line connecting these two points intersects ${\mathcal Q}$ at precisely two points $[U_+]$ and $[U_-]$ or alternatively $[V_+]$ and $[V_-]$.
Indeed, let us consider the line defined by $L=\alpha A+\beta B$. Then the constraint $L\cdot L=0$ defines a quadratic equation for the ratio $\alpha/\beta$. This quadratic equation has {\it two real} solutions iff $D>0$. These solutions give rise to the principal null directions $[U_{\pm}]$. 
In the next section we will see that for $A,B\in {\rm AdS}_{d+1}$ this construction yields space-like geodesics connecting the boundary points $[U_+]$ and $[U_-]$.

Repeating this calculation for the ratio $\beta/\alpha$ instead yields $[V_{\pm}]$.
Of course we have the relation  
$U_{\pm} \simeq V_{\pm}$.
More precisely we have
\beq
U_{\pm}=\pm\frac{A\cdot A}{\sqrt{D}\pm A\cdot B}V_{\pm}=\mp \frac{\sqrt{D}\mp A\cdot B}{B\cdot B}V_{\pm}
\label{ratio}
\eeq
Hence for the representative of our line any of the pairs $(U_+,U_-)$, $(V_+,V_-)$, $(U_+,V_-)$ and $(V_+,U_-)$ can also be used.
Any one from such pairs of null vectors can be regarded as representatives of the {\it principal null directions} of our time-like plane. 
Later the pairs 
$(U_+,V_-)$ and $(V_+,U_-)$
will play an important role in our considerations. See Figure 5.
In this context we record for future use the relations
\beq
U_+\cdot V_- =A\cdot B-\sqrt{D},\qquad
U_-\cdot V_+ =A\cdot B+\sqrt{D}
\label{scalarnull}
\eeq
\beq
(U_+\cdot V_- )(U_-\cdot V_+)=(A\cdot A)(B\cdot B)
\label{fontos}
\eeq

\subsection{A geometric regularizaton for the length of spacelike geodesics}

Let us consider the Lagrangian ${\mathcal L}\equiv {\mathcal L}(X,\dot{X},C)$
\beq
{\mathcal L}=\dot{X}\cdot\dot{X}+C(X\cdot X+ L^2)
\label{Lag}
\eeq
where
$X^a(\lambda)$ is a curve in $\widetilde{\rm AdS}_{d+1}$,
$\dot{X}^a\coloneqq \frac{d}{d\lambda}X^a(\lambda)$ and $C(\lambda)$ is Lagrangian multiplier implementing the constraint
$X\cdot X=-L^2$.
From this constraint it follows that
\beq
X\cdot\dot{X}=0,\qquad \dot{X}\cdot\dot{X}+X\cdot \ddot{X}=0
\label{C}
\eeq
and the Euler-Lagrange equations read as
\beq
\ddot X^a=CX^a.
\label{EL}
\eeq
Multiplying Eq. (\ref{EL}) with $X$ we get $X\cdot\ddot{X}=-CL^2$. Using then the constraint equations (\ref{C})
we obtain
\beq
\dot{X}^2=C{L^2}.
\label{EOM2}
\eeq

One can then see that for trajectories $X^a(\lambda)$ satisfying the equation of motion the quantities
\beq
M^{ab}\equiv (X^a\dot{X}^b-X^b\dot{X}^a)/L^2
\label{conserved}
\eeq
are conserved ones i.e. $\dot{M}^{ab}=0$
reflecting the fact that the geodesic $X^a(\lambda)$ for all values of $\lambda$ is contained in a plane
through the origin spanned by $X$ and $\dot{X}$, characterized by the constant Pl\"ucker coordinates $M^{ab}$.
Moreover, one has the relation
\beq
C=-\frac{1}{2}M^{ab}M_{ab}.
\nonumber
\eeq
Hence $\dot{X}^2=C{L^2}={\rm const}$.
Clearly for space like, null and time like geodesics one has $C>0,0,<0$.

For space like geodesics, our main concern here, one has $C>0$. Then  one can define {\it two} principal null directions for our plane giving two linearly independent null vectors $U$ and $V$ spanning it. 
Using them the solution of the (\ref{EL}) equation of motion for space-like geodesics 
also satisfying the constraints $X^2=-L^2$ and $\dot{X}^2=C{L^2}={\rm const}$
is
\beq
X^a(\lambda)=\frac{L}{\sqrt{-2U\cdot V}}\left(Ue^{\sqrt{C}\lambda}+Ve^{-\sqrt{C}\lambda}\right)
\label{solutionspacelike}
\eeq
where
\beq
U\cdot V<0,\qquad U\cdot U=V\cdot V=0.
\label{relnorm}
\eeq

Let us now constrain our geodesic to go through the two fixed points $A$ and $B$.
Then in terms of the (\ref{princnull}) principal null directions associated with them
the length of a geodesic segment between $A$ and $B$ can be calculated as
\beq
X(\lambda_1)\equiv A=\frac{1}{\sqrt{2}}(U_-+U_+),\qquad
X({\lambda}_2)\equiv B=\frac{1}{\sqrt{2}}(V_-+V_+)
\nonumber
\eeq
 is
\beq
\ell(A,B)=\int_{\lambda_1}^{\lambda_2}\sqrt{\dot{X}^2}d\lambda =\sqrt{C}L(\lambda_2-\lambda_1)
\label{lengthab}
\eeq
On the other hand by virtue of (\ref{solutionspacelike}) one gets
\beq
X(\lambda_1)\cdot X({\lambda}_2)=A\cdot B=-L^2\cosh\left(\frac{\ell(A,B)}{L}\right).
\label{costav}
\eeq
Then using $A\cdot A=B\cdot B=-L^2$ and Eq. (\ref{Cayley}) from this
one obtains two solutions
\beq
e^{{\ell_{\pm}(A,B)}/L}=\frac{-A\cdot B\pm\sqrt{D}}{L^2}
=\frac{L^2}{-A\cdot B\mp\sqrt{D}}
\label{nearpenner}
\eeq
with the property
$e^{{\ell_{\pm}(A,B)}/L}=e^{\mp{\ell_{\mp}(A,B)}/L}$. Let us now defime ${\ell(A,B)}\equiv \ell_+(A,B)$.
Then by virtue of (\ref{scalarnull}) we get
\beq
e^{\pm\ell(A,B)/L}=\frac{\pm\sqrt{D}-A\cdot B}{L^2}=\frac{-U_{\pm}\cdot V_{\mp}}{L^2}.
\label{fontos1}
\eeq
From this one arrives at the following important formula
\beq
\ell(A,B)=\pm L\log\frac{\vert U_{\pm}\cdot V_{\mp}\vert}{L^2}.
\label{glength}
\eeq

In order to understand the geometric meaning of Eq.(\ref{glength}) we proceed as follows.
The length of an arbitrary spacelike geodesic is divergent.
This divergence can be regularized by cutting the ends of the geodesics in the (\ref{lengthab}) way 
where now $\lambda_1$ and $\lambda_2$ means some cutoff values for the geodesic parameters.
A geometrically well motivated choice for $\lambda_1$ and $\lambda_2$ is provided by the use of horospheres\cite{Seppi}. These objects are the generalizations of 
horocycles\cite{Penner1}. They were first used as geometric regulators in the holographic context in Ref.\cite{Levay1}. For a recent analysis see also Ref.\cite{Agrawal}.

 For two null vectors $U$ and $V$ satisfying Eq.(\ref{relnorm}) we define two horospheres
\footnote{ 
Note, that the use of the name "horosphere" is not precise, since in the literature these objects are usually defined as hyperplane sections of the upper sheet of the double sheeted hyperboloid by null vectors belonging to the future light cone. This hyperboloid is a hyperplane section of our  $\tilde{AdS}_{d+1}$. Especially in holography this amounts to taking the static $X^0=0$ slice. However, horospheres and their geometry can be generalized for our more general context. See Ref.\cite{Seppi}. Unlike in the static slice case in this general case they are topologically not spheres. }
as two sections  of $AdS_{d+1}$ by the hyperplanes defined by
\begin{equation}
    U\cdot X=-\frac{1}{\sqrt{2}}L^2,\qquad V\cdot X=-\frac{1}{\sqrt{2}}L^2
\label{pennerconv}
\end{equation}
where we used the normalization in a way that for the static slice of $AdS_3$ to be in accord with the book of Penner\cite{Penner2}.
Intersecting these horospheres with the spacelike geodesic defined by \eqref{solutionspacelike} one gets the values for $\lambda_1$ and $\lambda_2$
\begin{equation}
    \begin{aligned}
    &\qquad V\cdot X(\lambda_1)=-\frac{1}{\sqrt{2}}L^2\implies \lambda_1=-\frac{1}{\sqrt{C}}\log\frac{\sqrt{-U\cdot V}}{L}\\
    &\qquad U\cdot X(\lambda_2)=-\frac{1}{\sqrt{2}}L^2\implies \lambda_2=+\frac{1}{\sqrt{C}}\log\frac{\sqrt{-U\cdot V}}{L}
\end{aligned}
\end{equation}
Therefore the regularized length is
\begin{equation}
    \ell_{\rm reg}=L\log\frac{\vert U\cdot V\vert}{L^2}
\label{regike}
\end{equation}
Comparing this with Eq.(\ref{glength}) we see that  $\ell_{reg}=\ell(A,B)$,  
$A=X(\lambda_1)$, $B=X(\lambda_2)$ and $U=U_+$ and $V=V_-$. 
Hence the geometric meaning of the vectors ($A$ and $B$) and the null ones ($U_+$ and $V_-$), giving rise to the points
\beq
x^{\mu}_{u}\coloneqq L\frac{U^{\mu}}{U^-},\qquad
x^{\mu}_{v}\coloneqq L\frac{V^{\mu}}{V^-}
\label{kepike}
\eeq
on the boundary of the Poincaré patch where our spacelike geodesic is anchored, can be clarified as follows.

$U_+$ and $V_-$ are the principal null directions of the {\it plane} spanned by $A,B\in \widetilde{\rm AdS}_{d+1}\subset {\mathbb R}^{d+2}$. Projectively the light rays $[U]$ and $[V]$, with representatives $U_+$ and $V_-$,
 are represented by points on the quadric ${\mathcal Q}_d\subset\mathbb{RP}^{d+1}$.
Now in this projective picture our plane corresponds to a {\it line} in $\mathbb{RP}^{d+1}$ passing through the rays $[A]$ and $[B]$.
A spacelike geodesic is defined by the projective line connecting the points $[A]$ and $[B]$ in $\mathbb{RP}^{d+1}$ intersecting ${\mathcal Q}_d\subset {\rm AdS}_{d+1}\subset \mathbb{RP}^{d+1}$ in the points $[U]$ and $[V]$. 
According to Eq.(\ref{pennerconv}) $U_+$ and $V_-$ define two horospheres ${\mathcal S}_{u_{+}}$ and ${\mathcal S}_{v_{-}}$ orthogonal to our spacelike geodesic. A static slice of the $d=2$ situation
will be discussed in Section 5.

We have seen that each boundary point is represented by a light ray. 
Moreover, by choosing different representatives from such light rays amounts to choosing different cutoffs for spacelike geodesics.
According to Ref.\cite{Levay1} this ambiguity of different cutoffs can be regarded as a gauge degree of freedom.
One of the aim of the present paper is to explore the many important consequences of this gauge structure.
In order to do that we first reconsider the regularization problem for space-like geodesics also in the Poincaré patch.

\subsection{Space like geodesics in the Poincar\'e patch}

Let us also characterize space-like geodesics in the Poincar\'e patch.
We use the notation $e^{\pm}\equiv e^{\pm\sqrt{C}\lambda}$ in (\ref{solutionspacelike}) and write
\beq
x^{\mu}(\lambda)=L\frac{X^{\mu}(\lambda)}{X^-(\lambda)}=L\frac{e^+U^{\mu}+e^-V^{\mu}}{e^+U^++e^-V^{+}},
\qquad z(\lambda)=\frac{L^2}{X^-(\lambda)}
\label{kep1}
\eeq
Let us choose the representatives of the rays determined by $U$ and $V$ similar to (\ref{conf1})
\beq
\begin{pmatrix}
U^+\\U^{\mu}\\U^-\end{pmatrix}=\frac{1}{\Delta_u}\begin{pmatrix}-x_u^2\\Lx_u^{\mu}\\L^2\end{pmatrix},
\qquad
\begin{pmatrix}
V^+\\V^{\mu}\\V^-\end{pmatrix}=\frac{1}{\Delta_v}\begin{pmatrix}-x_v^2\\Lx_v^{\mu}\\L^2\end{pmatrix}
\label{kep2}
\eeq
where we have defined $x_u$ and $x_v$ by Eq.(\ref{kepike}).
Recall that the vectors $U$ and $V$ have to satisfy the constraints $U\cdot U=V\cdot V=0$ and $U\cdot V<0$.
Hence the rescalings serving as a local gauge degree of freedom showing up in Eq. (\ref{conf2}) 
\beq
U\mapsto \frac{1}{\Lambda_u}U,\qquad V\mapsto \frac{1}{\Lambda_v}V
\label{resc}
\eeq
have to satisfy the constraints
\beq
\Lambda_u\Lambda_v>0,\qquad \Lambda_u,\Lambda_v\in{\mathbb R}\setminus\{0\}
\label{gaugeconstr}
\eeq

Let us now introduce the quantities
\beq
x_0^{\mu}=\frac{1}{2}(x_v^{\mu}+x_u^{\mu}),\qquad \triangle x^{\mu}=x_v^{\mu}-x_u^{\mu}
\label{kep4}
\eeq
Since for an arbitrary vector $K\in{\mathbb R}^{d,2}$ we have
\beq
K\bullet K=K\cdot K-K^+K^-
\label{kep6}
\eeq
with the choice $K=UV^--VU^-$ we have $K^-=0$ hence one arrives at the formula
\beq
(UV^--VU^-)\bullet (UV^--VU^-)=-2U\cdot V U^-V^-.
\label{kep7}
\eeq
By virtue of Eqs.(\ref{kep2}) and (\ref{kep4}) this formula yields the important relation
\beq
\Delta x\bullet \Delta x=(x_u-x_v)^2=-2L^2\frac{U\cdot V}{U^-V^-}.
\label{relat}
\eeq
Using the parametrization of Eq.(\ref{kep2}) an alternative expression is 
\beq
\frac{(x_u-x_v)^2}{2\Delta_u\Delta_v}=-\frac{U\cdot V}{L^2}.
\label{fontoss}
\eeq
Since $U\cdot V<0$ this formula shows that either we have
\beq
U^-V^->0\quad (\Delta_u\Delta_v>0) \implies \triangle x\bullet \triangle x>0
\label{alt1}
\eeq
or
\beq
U^-V^-<0\quad (\Delta_u\Delta_v<0) \implies \triangle x\bullet \triangle x<0
\label{alt2}
\eeq
i.e when the signs of $U^-$ and $V^-$ are the same (not the same) we have space-like (time-like) separation between the corresponding boundary points.

Choosing in particular $U=U_+$ and $V=V_-$ one arrives at an alternative formula for the  (\ref{glength}) geodesic length of a space-like geodesic segment between $A$ and $B$
\beq
\ell(A,B)=L\log\frac{(x_u-x_v)^2}{2\Delta_u\Delta_v}.
\label{majdnemrt}
\eeq
Now under the (\ref{resc}) gauge transfromation
we have 
\beq
\Delta_u\mapsto \Lambda_u\Delta_u,
\qquad
\Delta_v\mapsto \Lambda_v\Delta_v,\qquad \Lambda_u\Lambda_v>0
\label{mertektrafo}
\eeq
Hence from Eq.(\ref{majdnemrt}) we learn that a gauge transformation amounts to choosing a different cutoff, serving as a regulator for the infinite length geodesic.

Let us finally define a new parameter $s\in{\mathbb R}$ instead of $\lambda\in[-\infty,\infty]$ as follows
\beq
e^{2\sqrt{C}\lambda}=\left(\frac{1-s}{1+s}\right)\frac{V^-}{U^-}=\left(\frac{1-s}{1+s}\right)\frac{\Delta_u}{\Delta_v}
\label{reparametrize}
\eeq
Then one can see that according to the alternatives (\ref{alt1})-(\ref{alt2}) one has the cases
\beq
U^-V^->0\quad  \implies s\in[-1,1]
\label{valt1}
\eeq
or
\beq
U^-V^-<0\quad  \implies s\in {\mathbb R}\setminus (-1,1)
\label{valt2}
\eeq

Let us now use (\ref{kep1}), (\ref{kep4}) and (\ref{relat}) to write
\beq
x^{\mu}(s)=\frac{(1-s)U^{\mu}V^-+(1+s)V^{\mu}U^-}{(1-s)U^-V^-+(1+s)V^-U^-}=\frac{1}{2}\Delta x^{\mu}s+x_0^{\mu}
\label{calculate}
\eeq
In case of (\ref{valt1}) $s\in [-1,1]$ , hence we have a line through $x_0$, starting from $x(-1)=x_u$ and arriving at $x(1)=x_v$.
In case of (\ref{valt1}) $s\in{\mathbb R}\setminus (-1,1)$, hence we have a line starting from $x(-1)=x_u$ going to the point at infinity and then coming back from the "other side" to $x(1)=x_v$.

Let us now calculate $z(s)=L^2/X^-(s)$!
Using Eq.(\ref{reparametrize}) and  (\ref{solutionspacelike}) with $a=-$ one obtains
\beq
e^{\sqrt{C}\lambda(s)}X^-(s)= \sqrt{\frac{1-s}{1+s}\frac{V^-}{U^-}}X^-(s) =\frac{L}{\sqrt{-2U\cdot V}}\frac{2}{1+s}V^-
\label{nana}
\eeq
From this one can express $X^-(s)$ in terms of the components of $U$ and $V$ needed for $z(s)$.
Indeed, using Eqs.(\ref{relat}) and (\ref{calculate}) one gets
\beq
z^2(s)=(\Delta x\bullet\Delta x) \frac{1}{4}(1-s^2)=(\Delta x\bullet\Delta x) \frac{1}{4}-(x-x_0)\bullet (x-x_0)
\label{kuka}
\eeq

	This gives us the final result for our spacelike geodesics in the Poincaré patch
\beq
(x(s)-x_0)^2+z(s)^2=\left(\frac{\Delta x}{2}\right)^2
\label{kep10}
\eeq
with keeping in mind the definitions of Eqs. (\ref{kep1})-(\ref{kep4}).
This result is showing that the space-like geodesics in the Poincar\'e patch are lying on a hypersurface in ${\mathbb U}_{d+1}$ given by an equation of the form
\beq
(x-a)^2+z^2=b,\qquad b\in{\mathbb R},\qquad a\in \partial \mathbb{U}_{d+1}
\label{kep11}
\eeq
According Ref.\cite{Seppi} the (\ref{kep11}) surfaces are comprising one type of totally geodesic hypersurfaces of $\mathbb{U}_{d+1}$.
One can combine this with Eq.(\ref{calculate}) which shows that we have a line in $\partial \mathbb{U}_{d+1}$, therefore a plane in $\mathbb{U}_{d+1}$. Hence the geodesics are intersections of hypersurfaces given by \eqref{kep10} and planes determined by the equation \eqref{calculate}.
Now the pair $(z,s)$ provides coordinates for this plane.
Then in this plane we have the equation
\beq
z^2+\left(\frac{\Delta x}{2}\right)^2s^2=\left(\frac{\Delta x}{2}\right)^2
\label{innen}
\eeq
From here one can see that in this plane for ${\Delta x}^2>0$ a space-like geodesic is a half-ellipse with foci on $\partial{\mathbb U}_{d+1}$.
On the other hand for ${\Delta x}^2<0$ we have half of a branch of a hyperbola with foci on $\partial{\mathbb U}_{d+1}$.
In this respect see also Appendix A of Ref.\cite{Danciger}.

It is important to emphasize again, that all these geodesics are space-like ones in the AdS sense.  However, the boundary lines determined by \eqref{calculate} can be both space-like and time-like in the boundary Minkowski sense. These alternatives are depending on the sign of the combination $U^-V^-$. Equivalently if $(\Delta x)^2>0$ the geodesic is a connected, space-like section of a one-sheeted hyperboloid, on the other hand for $(\Delta x)^2<0$  it is a disconnected, time-like section of a two-sheeted hyperboloid.

\subsection{Minimal surfaces of $AdS_{d+1}$}\label{sec:AdSMinimal}

In the following we are going to introduce codimension two minimal surfaces $X_{\mathcal R}\subset \widetilde{AdS}_{d+1}$ homologous to a boundary region ${\mathcal R}$ with $\partial {\mathcal R}\simeq S^{d-2}$. 
Such surfaces play an important role in the Ryu-Takayanagi formula\cite{RT1,RT2}.
The link to our previous considerations is provided by the fact\cite{Myers,Levay4} that a description of $X_{\mathcal R}$ can be provided  by our main actors, namely the two null vectors $U$ and $V$ satisfying 
$U\cdot V<0, U^-V^-<0$ subject to the constraint
\begin{equation}
X\cdot U=X\cdot V=0
\label{XR}
\end{equation}
Again by virtue of (\ref{kep2}) $U$ and $V$ define two boundary points $x_u$ and $x_v$.
Moreover, according to (\ref{relat}) $x_v$ and $x_u$ are timelike separated.
One can also see\cite{Myers,Levay4} that $x_u$ and $x_v$ are comprising the future and past tips of a boundary causal diamond.

In order to reconsider this result within our setup 
recall that $U\cdot X =U\bullet X+\frac{1}{2}(U^+X^-+U^-X^+)$
 then dividing by $U^-\neq 0$ and using the (\ref{embed}) coordinates of the Poincaré patch we get
\beq
2L\frac{U\bullet x}{U^-z}+\frac{1}{z}\left(-x\bullet x-z^2+\frac{U^+}{U^-}L^2\right)=0
\label{koztes}
\eeq
On the Poincaré patch $z\neq 0$ and $U\cdot U=U\bullet U+U^+U^-=0$
Taking this into account we can write (\ref{koztes}) as
\beq
2L\frac{U\bullet x}{U^-}-x\bullet x-z^2-\frac{U\bullet U}{U^-U^-}L^2=0
\label{koztes2}
\eeq
Hence using (\ref{kep2}) a more compact way of writing the constraints of Eq.(\ref{RT})  is

\begin{equation}\label{eq:minimal_eqs3}
\begin{aligned}
    z^2+(x-x_u)\bullet (x-x_u)&=0\\
    z^2+(x-x_v)\bullet (x-x_v)&=0
\end{aligned}
\end{equation}
These equations define two $d$ dimensional cones in the Poincaré patch. The codimension two minimal surface $X_{\mathcal R}$ is given by the intersection of these cones. For $z=0$ these cones boil down to boundary light cones. Hence in the boundary the intersection of the future light cone of $x_v$ and past light cone of $x_u$ gives rise to a causal diamond. For a pictorial representation of this situation in the $d=2$ case see Figure \ref{fig:diamond}. In this case the minimal surface $X_{\mathcal R}$ coincides with a spacelike geodesic familiar from the previous subsection and ${\mathcal R}$ is a line segment with its boundary comprising just two points. This is in accord with the fact that   
$\partial {\mathcal R}\simeq S^{0}={\mathbb Z}_2$.

\begin{figure}[!t]
    \centering
    \includegraphics[width=0.7\textwidth]{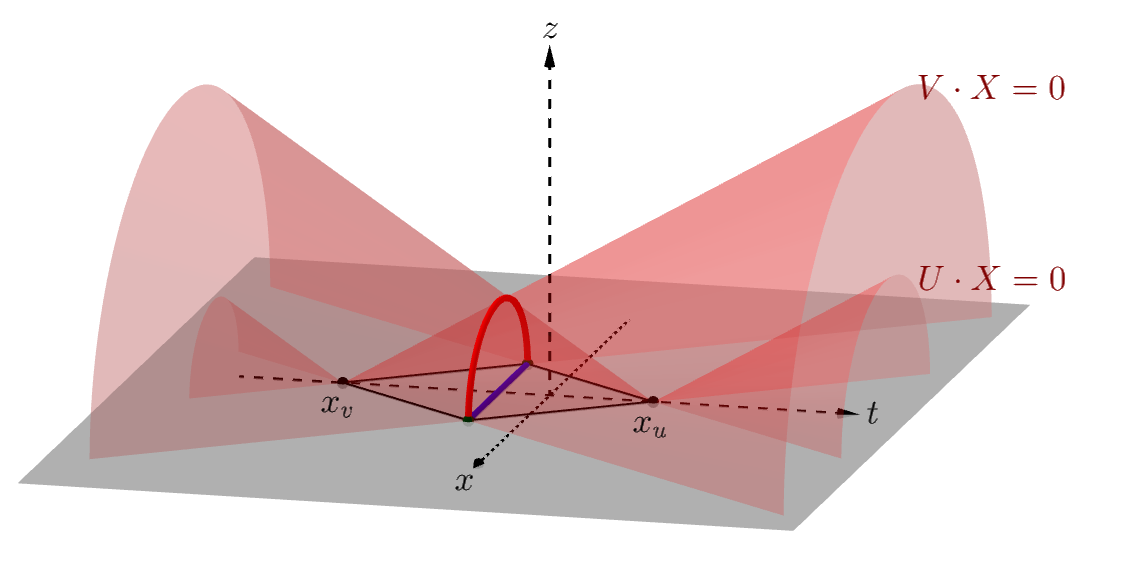}
    \caption{An $AdS_3$ causal diamond. A causal diamond is defined by the conditions $V\cdot X=U\cdot X=0$, where $U$ and $V$ are two distinct null vectors. These conditions give rise to a pair of cones in the bulk, whose intersection is a minimal surface, and the region enclosed by them and the boundary is the causal diamond itself. The tips of the diamond $x_u$ and $x_v$ lie on the boundary, and can be calculated from $U$ and $V$ by \eqref{kep2}. The figure was reproduced from \cite{Levay4}.
 }
 \label{fig:diamond}
\end{figure}

Let us subtract from each other the two equations  of (\ref{eq:minimal_eqs3}). Then we obtain
\beq
2x\bullet (x_v-x_u)+x_u\bullet x_u-x_v\bullet x_v=0
\label{kivonas}
\eeq

Now since $x_u\bullet x_u-x_v\bullet x_v=(x_u-x_v)\bullet (x_u+x_v)$ in terms of the quantities of (\ref{kep4}) one can
write this as
\beq
\triangle x\bullet (x-x_0)=0
\label{egyes}
\eeq
Moreover, one can write
\beq
x-x_u=x-x_0+\frac{1}{2}\triangle x,\qquad
x-x_v=x-x_0-\frac{1}{2}\triangle x
\label{kettes}
\eeq
Plugging this into either of Eqs.(\ref{eq:minimal_eqs3}) and using Eq.(\ref{egyes}) one gets
\beq
(x-x_0)\bullet(x-x_0)+z^2=-\frac{1}{4}\triangle x\bullet \triangle x
\label{harmas}
\eeq
Taken together eqs.(\ref{egyes}) and (\ref{harmas}) determine the structure of the minimal surfaces in the Poincar\'e patch
in terms of $x_0$ and $\triangle x$.
Eq.(\ref{egyes}) describes a hyperplane and Eq.(\ref{harmas}) a hyperboloid in the Poincar\'e patch.
Their intersection gives the minimal surface.

Now we see that each $U,V$ pair determines a plane in the embedding $\mathbb{R}^{2,d}$ or correspondingly a point in the two dimensional Grassmannian. These $U,V$ vectors also cut out $X_{\mathcal R}$ minimal surfaces from the AdS space. Therefore the Grasssmannian $Gr_2\left(\mathbb{R}^{2,d}\right)$ is related to the space of $AdS_{d+1}$ minimal surfaces in any dimensions. The $SO(1,d-1)\times SO(1,1)\times \mathbb{R}$ gauge invariance naturally arises from the invariance of the minimal surfaces. Indeed, notice that neither the center of the minimal hemisphere nor the radius of the surface depend on the scale of the null vectors $U$ and $V$. Therefore the minimal surfaces are invariant to the $U\mapsto aU$, $V\mapsto bU$ local transformations of the space of the Grassmannian. In the next section we will give an alternative derivation of the result that $SO(1,1)$ transformations of $U$ and $V$ and $SO(1,d-1)$ transformations in the perpendicular subspace leave invariant a given minimal surface\cite{Myers}.

In closing this section we elaborate on the well-known example of the minimal surface of the static case.
This is characterized by surfaces with $t={\rm constant}$.
Since Eq.(\ref{egyes}) takes the detailed form
\beq
(t_v-t_u)(t-t_0)=({\bf x}_v-{\bf x}_u)({\bf x}-{\bf x}_0)
\label{hyperdeteil}
\eeq
the left hand side can be a constant iff ${\bf x}_u={\bf x}_v={\bf x}_0$, and this constant is $t=t_0=\frac{1}{2}(t_v+t_u)$.
In this case Eq.(\ref{harmas}) gives the equations
\beq
t=t_0,\qquad\vert\vert{\bf x}-{\bf x}_0\vert\vert^2+z^2=R^2
\label{minsurfeq}
\eeq
where
\beq
R= \frac{1}{2}\vert \Delta t\vert=\frac{1}{2}\vert t_v-t_u\vert
\label{err}
\eeq
\beq
t_u=L\frac{U^0}{U^-},\qquad t_v=L\frac{V^0}{V^-}
\label{tekord}
\eeq
\beq
{\bf x}_0=L\frac{\bf U}{U^-}=L\frac{\bf V}{V^-}
\label{ikszkord}
\eeq
Hence we have a half sphere of radius $R$ localized at the $t=t_0$ hyperplane. The center of the half sphere is at ${\bf x}_0$.

\section{Twistor geometric derivation of the metric of kinematic space}

In our elaborations of correspondences between the geometric structures of the bulk and the boundary presented in Section 2. we have elucidated the pivotal role of two bulk light-like vectors $U$ and $V$ with the property
$U\cdot V<0$.
Such vectors have given rise to principal null directions for time-like planes. They were also compactly describing space-like geodesics.
Moreover, for $X\cdot X=-L^2$ the equations $U\cdot X=0$ and $V\cdot X$ have given rise to minimal surfaces $X_{\mathcal R}$ in $\widetilde{\rm AdS}_{d+1}$.  They have also defined spherical entangling regions on constant time slices of the boundary located inside
causal diamonds. 
Moreover, the equations $U\cdot X=V\cdot X=-L^2/{\sqrt{2}}$ have defined horospheres providing a natural geometrical method for regularizating diverging lengths of spacelike geodesics. In Section 6. we will see that in the ${\rm AdS}_3/{\rm CFT}_2$ context, via the Ryu-Takayanagi correspondence\cite{RT1,RT2}, this regularization also sheds some new light on the nature of boundary cutoffs. Indeed, in an earlier work one of us has already shown that horospherical cutoffs give rise to a geometrical way of regularizing boundary entanglement entropies for a ${\rm CFT}_2$ vacuum\cite{Levay1}.

An important comment is here in order.
A pair $(U,V)$ with $U\cdot V<0$, which gives rise to the minimal surface via the equations $X\cdot U=X\cdot V=0$, determines in the boundary the past ($x_u$) and future ($x_v$) tips of a causal diamond. 
Hence these tips should be {\it timelike separated}. Then by Eqs.(\ref{fontoss})-(\ref{alt2}) this implies that in this case we have to choose $U^-V^-<0$.
However, one can notice that the very same constraints on the pair $(U,V)$ also coincide with the (\ref{valt2}) one yielding for the spacelike geodesics the class featuring two disconnected branches of hyperbolae. Hence we conclude that in this case  {\it the same pair} encodes two different types of geometric objects in the Poincaré patch: {\it disconnected geodesics} and {\it minimal surfaces}. 

On the other hand for the regularization of spacelike geodesics belonging to the other type, i.e. the half ellipse ones of (\ref{valt1}), the a pair $(U,V)$ determining the horospheres used as regularizators are connected to two {\it spacelike separated points} $x_v$ and $x_u$. 
Then in this dual case we have to choose $U^-V^->0$. Now the pair $(U,V)$ encodes {\it connected geodesics}. 
Moreover, for this type of geodesic the pair $(U,V)$ also determines the regularizing ${\it horospheres}$.

In this section we would like to explore further the relevance of pairs of null vectors $(U,V)$ to interesting physics by presenting a new characterization of the metric structure on the moduli space of boundary causal diamonds. Such a space was first studied by de Boer et.al.\cite{Myers}. An alternative name for this mathematical object is also available in the literature as kinematic space\cite{Czech}.
We will see in the next section that if we upgrade the $(U,V)$ pair to fields defined on a $2D$ space then all of our geometric objects can naturally be related to strings propagating in $AdS_{d+1}$.
In order to achieve this goal we will regard the pair $(U,V)$ not merely as a pair of null vectors, but rather null vector fields parametrized by some set of coordinates. This will be the philosophy of the next section where the usual parametrization $U^a(\sigma, \tau)$ or $U^a(\sigma_+,\sigma_-)$ where $\sigma_{\pm}=\tau\pm\sigma$ shows up. 

Before embarking in such considerations the first step is a careful reconsideration of the metric on moduli space.
The novelty of our treatise of this metric structure is that it incorporates in an explicit manner the gauge structure, encapsulated in the local transformation rules of Eqs.(\ref{resc}). In our opinion this have not been elucidated properly yet. The hint that our twistor geometric setup of this moduli space of causal diamonds is naturally connected to string theory was motivated by our previous study\cite{Levay4} based on segmented strings \cite{Gubser,DV2, Dv3}.

First of all let us  introduce the notation $W^{a\alpha}$ with $a,b=-1,0,1,\dots d$ and $\alpha,\beta =1,2$. This quantity is a $(d+2)\times 2$ matrix containing the two $d+2$ component column vectors
$W^{a1}=U^a$ and $W^{a2}=V^a$. Then we have
$\sqrt{D}=\sqrt{(U\cdot V)^2}=\vert U\cdot V\vert=-U\cdot V$ hence
\beq
(W\cdot W)^{\alpha\beta}=\eta_{ab}W^{a\alpha}W^{b\beta}=(W^t\eta W)^{\alpha\beta}=
\begin{pmatrix}
    U\cdot U&U\cdot V\\
    U\cdot V&V\cdot V
    \end{pmatrix}=
-\sqrt{D}\sigma_1^{\alpha\beta}.
\label{prop1v}
\eeq
or 
omitting the $2\times 2$ matrix indices
$W\cdot W\sigma_1=-\sqrt{D}I$
Taking the trace of both sides of this equation and using the cyclic property of the trace one gets
\begin{equation}
    \sqrt{D}=-U\cdot V=-\frac{1}{2}\text{Tr}\left\{W\sigma_1 W^t\eta \right\}
\label{Cayleyvari}
\end{equation}

We have already seen in Eq.(\ref{resc}) that the $U\cdot V<0$ condition does not change under the local gauge transformation $U\mapsto U/{\Lambda_u}$, $V\mapsto V/{\Lambda_v}$, where now $\Lambda_u,\Lambda_v$ are real valued functions such that $\Lambda_u\Lambda_v>0$. This transformation acts on $W$ as
\begin{equation}
    W\mapsto W\tilde{\Sigma},\qquad \tilde{\Sigma}=\begin{pmatrix}
\frac{1}{\Lambda_u} & 0\\
0 & \frac{1}{\Lambda_v}
\end{pmatrix}
\end{equation}
One can rewrite $\tilde{\Sigma}$ as
\begin{equation}
    \tilde{\Sigma}=\pm \frac{1}{\sqrt{\Lambda_u\Lambda_v}}\begin{pmatrix}
\sqrt{\frac{\Lambda_v}{\Lambda_u}} & 0\\
0 &\sqrt{\frac{\Lambda_u}{\Lambda_v}}
\end{pmatrix}:=
\pm \frac{1}{\sqrt{\Lambda_u\Lambda_v}}\begin{pmatrix}
e^\xi & 0\\0&e^{-\xi}\end{pmatrix}=
\alpha\Sigma
\end{equation}
Here the matrix $\Sigma$ contains two dilatations.  Hence $\Sigma$ is acting on the new linear combinations $T=\frac{1}{\sqrt{2}}(U+V)$
and $S=\frac{1}{\sqrt{2}}(U-V)$ as a Lorentz boost with rapidity $\xi$, i.e. an element of $SO(1,1)$. These Lorentz transformations are belonging to the timelike plane with null directions $U$ and $V$.
This also shows that $\alpha$ and $\Sigma$ can be regarded as representatives of the local gauge group ${\mathbb R}^{\ast}\times SO(1,1)$ where ${\mathbb R}^{\ast}={\mathbb R}\setminus\{0\}$ i.e. the group of local Lorentz transformations and local rescalings of the null vectors $(U,V)$.
The case $\alpha=0$ is not included since in this case we have no nonzero null directions.

Now by working in the {\it bulk} we are able to build up a local  $\mathbb{R}^{\ast}\times SO(1,1)$ invariant metric on the moduli space of causal diamonds of the {\it boundary}. We call this space the {\it generalized kinematic space}. 
Let us first define the following field
\begin{equation}
    \Pi=\frac{-W\sigma_1 W^t\eta}{\sqrt{D}}=\frac{2}{\text{Tr}\left\{W\sigma_1 W^t\eta \right\}}W\sigma_1 W^t\eta
\label{projfield}
\end{equation}
It is easy to see that $\Pi W=W$ and $\Pi^2=\Pi$, therefore $\Pi$ is a projector-valued field. Moreover, due to $\Sigma\sigma_1\Sigma=\sigma_1$ it is invariant under $W\mapsto W\tilde{\Sigma}$, $\tilde{\Sigma}\in\mathbb{R}^{\ast}\times SO(1,1)$ transformations. One can also define $\Pi^{\perp}=I-\Pi$. Clearly $\Pi^{\perp}W=0$ and $\left(\Pi^{\perp}\right)^2=\Pi^{\perp}$.

Under the transformations $W\mapsto W\Sigma$ and $W\mapsto \alpha W$, $dW$ transforms as
\begin{equation}
    dW\mapsto dW\Sigma+Wd\Sigma,\qquad
    dW\mapsto \alpha dW+d\alpha W
\end{equation}
 Thanks to the property $\Pi^{\perp}W=0$ the following hold:
\begin{equation}
    \Pi^{\perp}dW\mapsto \Pi^{\perp}dW\Sigma
\end{equation}
Under $SO(1,1)$ transformations, and
\begin{equation}
    \Pi^{\perp}dW\mapsto \alpha dW
\end{equation}
under rescaling. Then it follows that the expression $\text{Tr}\left\{\left(\Pi^{\perp}dW\right)\cdot\left(\Pi^{\perp}dW\right)\sigma_1\right\}$ is $SO(1,1)$ invariant by the cyclic property of the trace. However, under rescaling this transforms as
\begin{equation}
    \text{Tr}\left\{\left(\Pi^{\perp}dW\right)\cdot\left(\Pi^{\perp}dW\right)\sigma_1\right\}\mapsto\alpha^2 \text{Tr}\left\{\left(\Pi^{\perp}dW\right)\cdot\left(\Pi^{\perp}dW\right)\sigma_1\right\}
\end{equation}
Hence by normalizing the expression with $-\sqrt{D}=U\cdot V=\frac{1}{2}\text{Tr}\left\{W^t\eta W\sigma_1\right\}$ and taking into account that $U\cdot V<0$ we get a candidate for an expression for the $\mathbb{R}^{\ast}\times SO(1,1)$ gauge invariant metric

\begin{equation}
    ds^2=\frac{\ell^2}{\sqrt{D}}\text{Tr}\left\{\left(\Pi^{\perp}dW\right)\cdot\left(\Pi^{\perp}dW\right)\sigma_1\right\}
\label{candidate}
\end{equation}
where we have introduced a length scale $\ell$ associated with kinematic space

In order to check that this formula indeed gives the correct expression for the metric on our generalized kinematic space we proceed as follows.
Let us first observe that
\beq
\Pi^t\eta=\eta\Pi,\qquad {\Pi^{\perp}}^t\eta=\eta\Pi^{\perp}
\label{perpeta}
\eeq
Then using the property $\left(\Pi^{\perp}\right)^2=\Pi^{\perp}$ 
we can write $\left(\Pi^{\perp}\right)^2=\Pi^{\perp}$
\begin{equation}
    ds^2=\frac{\ell^2}{\sqrt{D}}\text{Tr}\left\{\left(dW\cdot\Pi^{\perp}dW\right)\sigma_1\right\}
\label{can2}
\end{equation}
This formula can be rewritten as\footnote{Notice that this is a similar expression to the metric on the projective space of $\mathbb{R}^{2,d}$, namely if $X=[X^{-1},X^{0},X^{1},\dots X^{d}]$ are homogeneous coordinates on $\mathbb{RP}^{d+1}$, then the Fubini-Study\linebreak metric is $ds^2=L^2\frac{(X\cdot X)(dX\cdot dX)-(X\cdot dX)^2}{(X\cdot X)^2}$.}:

\begin{equation}
    ds^2=-2\ell^2\frac{\text{Tr}\left\{W\cdot W\sigma_1 \right\}\text{Tr}\left\{dW\cdot dW\sigma_1\right\}-2\text{Tr}\left\{\left(dW\cdot W\sigma_1\right)\left(W\cdot dW\sigma_1\right)\right\}}{\text{Tr}^2\left\{W\cdot W\sigma_1 \right\}}
\end{equation}
One can also express the line element in terms of the null vectors $U$ and $V$
\begin{equation}
    ds^2=2\ell^2\frac{(dU\cdot dV)(U\cdot V)-(dU\cdot V)(U\cdot dV)}{\left(U\cdot V\right)^2}.
\label{klassz}
\end{equation}

An even more compact formula for the line element can be obtained as follows.
Let us introduce the quantity
\beq
{\mathcal P}\equiv \frac {U\wedge V}{U\cdot V}
\label{celestial}
\eeq
\noindent
where
\beq
(U\wedge V)^{ab}=U^aV^b-U^bV^a
\label{celestialdetail}
\eeq
are the usual Pl\"ucker coordinates ${\mathcal P}^{ab}$ known from Eq.(\ref{Plucker}) of the plane spanned by $U$ and $V$.
Notice that thanks to the normalization factor $U\cdot V$ in the denominator our new quantity $\mathcal P$ is gauge invariant, i.e. it is invariant under local transformations of the form displayed in Eq. (\ref{resc}).
Now in terms of $\mathcal P$  one can express the (\ref{klassz}) line element in the nice and compact manifestly gauge invariant form
\beq
ds^2=-\frac{1}{2}\ell^2d{\mathcal P}\cdot d{\mathcal P}=-\frac{1}{2}\ell^2d{\mathcal P}^{ab}d{\mathcal P}_{ab}.
\label{endmetrik}
\eeq

One can then easily connect the (\ref{klassz}) version of (\ref{endmetrik}) to the canonical form
of the metric well-known in CFT and used in Ref.\cite{Myers}.
To this end we use formula (\ref{kepike}) and (\ref{relat}) to write
\beq
dx_u^{\mu}=L\frac{U^+dU^{\mu}-U^{\mu}dU^{+}}{(U^+)^2},\qquad
dx_v^{\mu}=L\frac{V^+dV^{\mu}-V^{\mu}dV^{+}}{(V^+)^2}
\label{reszlet}
\eeq
\beq
\frac{2L^2}{(x_v-x_u)^2}\eta_{\mu\nu}(x_u^{\mu}-x_v^{\mu})dx_u^{\nu}=\frac{V\cdot dU}{V\cdot U}-\frac{dU^+}{U^+}
\label{reszlet2}
\eeq
Then with the shorthand notation $x_u^{\mu}:=x^{\mu}$ and $x_v^{\mu}:=y^{\mu}$ one can prove that
\beq
ds^2=\omega_{\mu\nu}(x,y)dx^{\mu}dy^{\nu}:=\frac{4\ell^2}{(x-y)^2}\left[-\eta_{\mu\nu}+\frac{2(x-y)_{\mu}(x-y)_{\nu}}{(x-y)^2}\right]dx^{\mu}dy^{\nu}
\label{Myers}
\eeq
which is the usual form of the metric on the space of causal diamonds\cite{Myers}.

Having established that Eqs.(\ref{klassz}), (\ref{endmetrik}) and (\ref{Myers}) are different forms of the same metric structure let us see how our metric is related to cross ratios also discussed in Ref.\cite{Myers}.
Let us consider the four vector fields: $(U,U+dU,V,V+dV)$ with $U\cdot V<0$. They are arising from the original fields and their perturbations coming from a change of their parameters. Then we have
\beq
\frac{[U\cdot(V+dV)][(U+dU)\cdot V]}
{[U\cdot V][(U+dU)\cdot (V+dV)]}=1+\frac{(U\cdot dV)(dU\cdot V)}{(U\cdot V)^2}-\frac{dU\cdot dV}{U\cdot V}+\dots
\label{cross}
\eeq
Let us denote $\diamond$ the causal diamond related to $(U,V)$ and
$\diamond +\delta\diamond$ the causal diamond related to $(U+dU,V+dV)$ and ${\mathcal C}(\diamond_1,\diamond_2)$ the cross ratio that can be seen on the left hand side of Eq.(\ref{cross}).
Then we have
\beq
{\mathcal C}(\diamond,\diamond+\delta\diamond)=1-\frac{1}{2L^2}ds^2+\cdots =1+\frac{1}{4}d{\mathcal P}\cdot d{\mathcal P}\cdots.
\label{endcross}
\eeq

\section{Strings in ${\rm AdS}_{d+1}$}

Now we elevate our null vectors of the previous sections to fields implementing the Virasoro constraints of a bulk string theory in the conformal gauge.
Our aim is to establish an explicit connection between the dynamics of boundary causal diamonds and ${\rm AdS}_{d+1}$ bulk strings. In this way we will see that the motion of $AdS$ strings can be represented by dynamics on kinematic space.

The motion of strings embedded in ${\rm AdS}_{d+1}$ is described by the action
\begin{equation}\label{eq:polyakov}
    \mathcal{S}=-\frac{T}{2}\int d\tau d\sigma (\partial_{\sigma}X\cdot\partial_{\sigma}X-\partial_{\tau}X\cdot\partial_{\tau}X) 
\end{equation}
With the constraint $X\cdot X=-L^2$. The equation of motion of a string propagating in ${\rm AdS}_{d+1}$ is \cite{Maldacena}
\begin{equation}\label{eq:eom}
    \partial_+\partial_-X-\frac{1}{L^2}(\partial_-X\cdot\partial_+X)X=0
\end{equation}
Where $\sigma^-=\frac{\tau-\sigma}{2}$ and $\sigma^+=\frac{\tau+\sigma}{2}$ are null parameters of the worldsheet.
Moreover, the worldsheet points should satisfy the Virasoro conditions
\begin{equation}\label{eq:virasoro}
\partial_- X\cdot \partial_- X=\partial_+ X\cdot\partial_+ X=0
\end{equation}
From now on we will suppress the differentiation with respect to $\sigma_\pm$ before $X$ and simply denote $\partial_\pm X$ by $X_\pm$.

As we have previously seen, if we choose two $U,V\in\mathbb{R}^{2,d}$ null vectors (such that $U\cdot V<0$ and $U^-V^-<0$), the equations $U\cdot X=0$ and $V\cdot X=0$, $X\in {\rm AdS}_{d+1}$ determine two cones in $\mathbb{R}^{2,d}$ whose intersection is a minimal surface of the ${\rm AdS}$ space. Its boundary determine a causal diamond on $\partial AdS$. In Poincaré coordinates its past and future tips are given by the Minkowski vectors
\begin{equation}
    x^\mu=\frac{L}{U^-}(U^0,U^1,\dots,U^{d-1}),\qquad y^\mu=\frac{L}{V^-}(V^0,V^1,\dots,V^{d-1})
\end{equation}
According to the equation of motion and the Virasoro conditions, the null vectors $X_-$ and $X_+$ play an important role in describing string dynamics. Moreover, at each point of the string $X_-\cdot X=X_+\cdot X=0$, which are the same as the defining equations of a bulk minimal surface, or equivalently a boundary causal diamond. This fact gives a nice duality between strings and causal diamonds.

To formulate this duality, for each point of the string worldsheet let us define the Minkowski vectors in the Poincaré representation
\begin{equation}\label{eq:xy_def}
    x^\mu(\sigma^+,\sigma^-)=\frac{L}{X_-^-}(X_-^0,X_-^1,\dots,X_-^{d-1}),\qquad y^\mu(\sigma^+,\sigma^-)=\frac{L}{X_+^-}(X_+^0,X_+^1,\dots,X_+^{d-1})
\end{equation}
These quantities can be interpreted in the following way: at each point of the string worldsheet we define a pair of cones via the equations $X_-\cdot X=X_+\cdot X=0$. Their intersection is a spherical minimal surface which contains the worldsheet point $X$. These cones are projected to the boundary determining a causal diamond with tips $x^\mu$ and $y^\mu$. Therefore there is a unique correspondence between the points of the string worldsheet and boundary causal diamonds. Hence string motion is projected to a dynamics of causal diamonds.
For segmented strings a pictorial illustration of this connection see Figure \ref{fig:lattice}.

\begin{figure}[!t]
    \centering
    \includegraphics[width=0.7\textwidth]{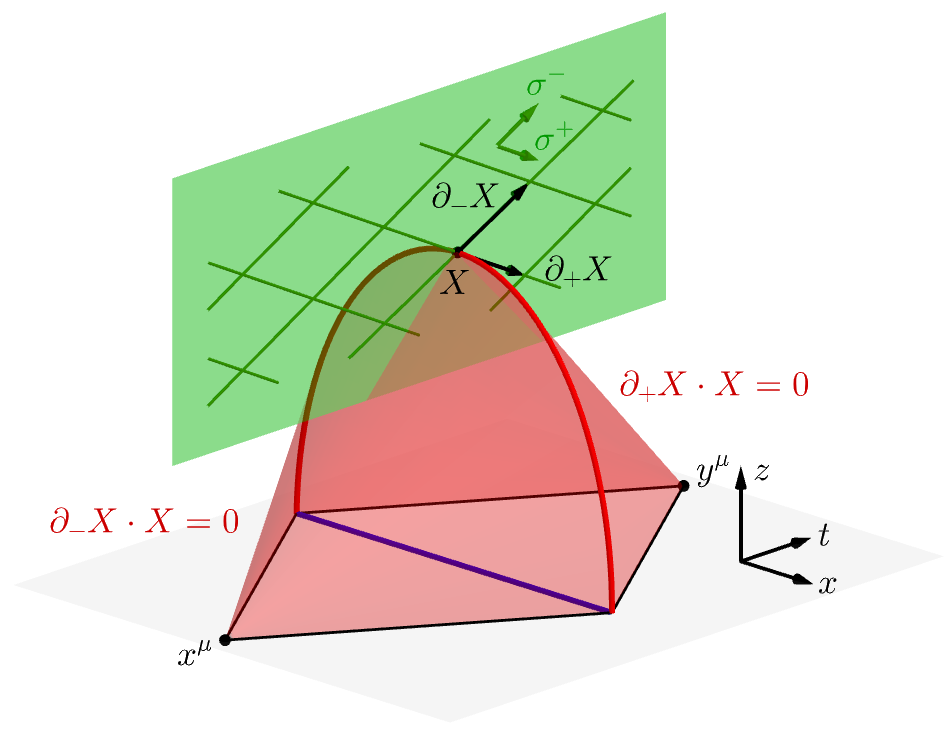}
    \caption{An illustration of the projection of a string worldsheet to causal diamonds in $AdS_3$. At each point of the worldsheet, the tangent vectors $\partial_\pm X$ define a causal diamond via equations $\partial_+ X\cdot X=\partial_-X\cdot X=0$. The past and future tips ($x^\mu$ and $y^\mu$ respectively), lying on the boundary, are given by \eqref{eq:xy_def}. The $X_{\mathcal R}$ Ryu-Takayanagi surface (red) and the corresponding boundary region (blue) is also shown. Reproduced from \cite{Levay4}. 
    }
    \label{fig:lattice}
\end{figure}

In the following we examine strings for which
\begin{equation}
    X_-\cdot X_+<0,\qquad X_-^-X_+^-<0
\end{equation}
From the bulk point of view these constraints define a timelike string. As we have seen before these also mean that the vectors $x^\mu$ and $y^\mu$ are timelike separated. We also parametrize the string in such a way that $x^0<y^0$, $(x-y)\bullet (x-y)<0$ and $\partial_+x\bullet\partial_+x,\partial_-y\bullet\partial_-y<0$. This means that $y^\mu$ is in the timelike future of $x^\mu$ and if we move on the worldsheet along the $\sigma^\pm$ parameters, then we get timelike separated causal diamonds on the boundary.

\subsection{Moduli space $\sigma$-model}

In this section we pull back the string action and equations of motion to boundary causal diamond coordinates. These equations display the dynamics in kinematic space dual to stringy dynamics in an explicit manner.

The string action \eqref{eq:polyakov} can be written in terms of the parameters $\sigma^\pm$ as
\begin{equation}\label{eq:action_vector}
    \mathcal{S}=T\int d\sigma^+ d\sigma^-X_-\cdot X_+
\end{equation}
For later purposes let us denote
\begin{equation}\label{eq:def_alpha}
    X_-\cdot X_+=-e^\alpha
\end{equation}
Using this definition the action becomes
\begin{equation}\label{eq:action_alpha}
    \mathcal{S}=-T\int d\sigma^+ d\sigma^- e^\alpha
\end{equation}

As it has been shown in \cite{
Maldacena} in the case of $AdS_3$, the string motion is governed by the field $\alpha$. There the authors have shown that by following the reduction method of \cite{ Pohlmeyer, Vega, Jevicki}. Later we will also demonstrate that the same holds for strings propagating in $AdS_{d+1}$. However, first we use the definition of $\alpha$ to verify that a natural connection between the motion of bulk strings and a dynamics of boundary causal diamonds emerges. As we have previously seen, the definition of the ${\rm AdS}$ space, the variation of the action with respect to the embedding coordinates, and the Virasoro constraints give the following contraints on the worldsheet
\begin{align}
    &X\cdot X=-L^2,\label{eq:emb_eq1}\\
    &X_-\cdot X_-=0,\label{eq:emb_eq2}\\
    &X_+\cdot X_+=0,\label{eq:emb_eq3}\\
    &X_{+-}-\frac{1}{L^2}(X_-\cdot X_+)X=0.\label{eq:emb_eq4}
\end{align}
By differentiating these equations with respect to $\sigma_\pm$  and using the definition \eqref{eq:def_alpha} of $\alpha$ it is easy to show that the following also hold
\begin{align}
    &X_+\cdot X_{--}=-\alpha_-e^{\alpha}\label{eq:deriv_eq1}\\
    &X_-\cdot X_{++}=-\alpha_+e^{\alpha}\label{eq:deriv_eq2}\\
    &X_{++}\cdot X_{--}=-\alpha_{+-}e^{\alpha}-\alpha_+\alpha_--\frac{1}{L^2}e^{2\alpha}\label{eq:deriv_eq3}
\end{align}
Now consider the following two expressions
\begin{equation}\label{eq:UVcomb}
\begin{aligned}
    &2L^2\frac{(\partial_+ U\cdot \partial_- V)(U\cdot V)-(\partial_+ U\cdot V)(\partial_- V\cdot U)}{\left(U\cdot V\right)^2},\\
    &2L^2\frac{(\partial_- U\cdot \partial_+ V)(U\cdot V)-(\partial_- U\cdot V)(\partial_+ V\cdot U)}{\left(U\cdot V\right)^2}
\end{aligned}
\end{equation}
with $U=X_-$ and $V=X_+$ and $a,b=\sigma^+,\sigma^-$. From these expressions one can build up the following formula reminiscent of the (\ref{klassz}) structure of the metric tensor of kinematic space.  By using \eqref{eq:deriv_eq1}-\eqref{eq:deriv_eq3} and introducing the world-sheet metric $h_{kl}:h_{+-}=h_{-+}=2,h_{++}=h_{--}=0$, after some algebraic manipulations it can be shown that
\begin{equation}
    2L^2\sqrt{-h}h^{kl}\frac{(\partial_{(k} X_{-}\cdot \partial_{l)} X_{+})(X_- \cdot X_+)-(\partial _{(k} X_{-}\cdot X_+)(\partial_{l)} X_{+}\cdot X_-)}{\left(X_-\cdot X_+\right)^2}=2L^2\alpha_{+-}+4e^\alpha
\end{equation}
where $(,)$ means symmetrization in the indices $k,l=\sigma_+,\sigma_-$\footnote{We use the convention $x_{(a}y_{b)}=x_{a}y_b+x_by_a$.} and $h$ denotes the determinant of $h_{ab}$\footnote{We will see later that the symmetrization is necessary to get the correct Virasoro constraints.}. If we rewrite the left-hand side in terms of the causal diamond coordinates $x^\mu,y^\mu$ or alternatively by the gauge invariant quantity of Eq.(\ref{celestial}) with $U=X_-$ and $V=X_+$ we get
\begin{equation}
-2L^2\sqrt{-h}h^{kl}\partial_{k}{\mathcal P}^{ab}\partial_{l}{\mathcal P}_{ab}=    \sqrt{-h}h^{kl}\omega_{\mu\nu}\partial_{(k} x^\mu\partial_{l)} y^\nu=4L^2\alpha_{+-}+8e^\alpha
\end{equation}
Where $\omega_{\mu\nu}$ is the (\ref{Myers}) metric on the space of causal diamonds. Notice that the right-hand side differs from the integrand of \eqref{eq:action_alpha} only by a total derivative. Therefore the on-shell string action is equivalent to
\begin{equation}
    \mathcal{S}=-\frac{T}{8}\int d\sigma^+ d\sigma^- \sqrt{-h}h^{kl}\omega_{\mu\nu}\partial_{(k} x^\mu\partial_{l)} y^\nu=\frac{T}{4}
\int d\sigma^+ d\sigma^- \sqrt{-h}h^{kl}
    \partial_{k}{\mathcal P}\cdot\partial_{l}{\mathcal P}
\label{twistact}
\end{equation}

What is the meaning of this form of the action?  Its derivation is based on three ingredients. The string equation of motion, the Virasoro and the $AdS$ constraints. The action featuring the metric $\omega_{\mu\nu}$ can be interpreted as a $\sigma$-model (in terms of the fields $x$ and $y$ or $P$) with target space the Grassmannian of planes through the origin of $\mathbb{R}^{2,d}$ (aka the kinematic space $\mathbb{K}$). There is a natural pull back of this model to $\mathbb{R}^{2,d}$ which in general does not necessarily encode an $AdS$ bulk string theory.  However, the Virasoro constraints reduce ${\mathbb K}$ to its subset
whose points are dual to timelike planes spanned by $X_\pm$ which are generated from the original coordinates $x$ and $y$. Then the $AdS$ constraint simply means that we take those points from this plane which are on the $AdS$ section of $\mathbb{R}^{2,d}$. Finally if we want the $\sigma$ model targeted on $\mathbb{K}$ to be dual to a $\sigma$ model targeted on $AdS$, then one should also take into account that the general string theory action \eqref{eq:action_vector} in the new coordinates $X_\pm$ should be extremal, which is equivalent to the equation of motion \eqref{eq:emb_eq4}. In Section \ref{sec:gauge}, we will see that the equations of motion in terms of the kinematic space variables $x^\mu$ and $y^\mu$ indeed contain the $SO(1,1)\times SO(1,d-1)$ gauge invariance of the kinematic space. 

From the twistorial point of view ${\mathbb K}$ can be regarded as the space of separable bivectors with coordinates given by ${\mathcal P}^{ab}$ also satisfying the  normalization constraint ${\mathcal P}\cdot{\mathcal P}=-2$. The space of separable bivectors is sitting inside the space of bivectors, i.e. the antisymmetric $(d+2)\times (d+2)$ matrices of dimension $N=\binom{d+2}{2}$. The space of these matrices forms the Lie algebra of the conformal group $SO(d,2)$. Then on ${\mathcal P}^{ab}$ we have the adjoint action of the conformal group. The projectivization of this space is ${\mathbb R}{\mathbb P}^{N-1}$ and the space of separable bivectors is described by  the Plücker relations
${\mathcal P}^{[ab}{\mathcal P}^{c]d}=0$ where $[abc]$ denotes antisymmetrization.
Hence ${\mathbb K}$ is the quadric ${\mathcal P}\cdot{\mathcal P}=-2$ inside the space of separable bivectors.  Kinematic space can also be regarded as a special orbit under the adjoint action of the conformal group. This is in accord with the result that ${\mathbb K}=SO(d,2)/SO(1,1)\times SO(d-1,1)$.
Notice that for $d=2$ the Plücker relations boil down to the (\ref{Plucker}) one which
 then defines the Klein quadric in ${\mathbb R}{\mathbb P}^5$. Then in this special case
 ${\mathbb K}$ is living inside the Klein quadric the space of the basic twistor correspondence.
 Within this context in Section 6. we recover the well-known result\cite{Czech,Myers} that for $d=2$ ${\mathbb K}\simeq dS_2\times dS_2$ i.e. the product of two de Sitter spaces.
 
Now if we look at the right hand side of Eq.(\ref{twistact}) we see that an alternative $\sigma$-model description of strings in $AdS_{d+1}$ can also be obtained by regarding the fields ${\mathcal P}^{ab}(\sigma^+,\sigma^-)$ as the basic entities of the theory . In this twistor theory motivated picture one can check that the constraints
\begin{equation}
{\partial}_-{\mathcal P}\cdot{\partial}_-{\mathcal P}=
{\partial}_+{\mathcal P}\cdot{\partial}_+{\mathcal P}_+=0
\end{equation}
also hold. These are just look like the Virasoro constraints in the conformal gauge. Then in this gauge instead of the Lagrangian showing up in (\ref{twistact}) one can use a Lagrangian of the form  ${\mathcal L}({\mathcal P},{\partial}_{\pm}{\mathcal P};\Lambda)\simeq{\partial}_-{\mathcal P}\cdot{\partial}_+{\mathcal P}
+\Lambda({\mathcal P}\cdot{\mathcal P}+2)$.
Now the equation of motion is 
of the form
$2{\partial}_+{\partial}_-{\mathcal P}=\Lambda{\mathcal P}$ where 
$\Lambda ={\partial}_-{\mathcal P}\cdot{\partial}_-{\mathcal P}$.

\subsection{Equations of motion in causal diamond coordinates}

The equations of motion in terms of the variables $x^\mu$ and $y^\mu$ can be derived by varying the action with respect to $x^\mu$ and $y^\mu$. From its variations one gets to the following equations
\begin{align}
    \frac{\delta \mathcal{S}}{\delta x^\mu}=0\qquad & \leftrightarrow\qquad \frac{\partial\omega_{\alpha\beta}}{\partial x^\mu}h^{kl}\partial_{(k} x^\alpha \partial_{l)} y^\beta=h^{ab}\partial_{(k}(\omega_{\mu\nu}\partial_{l)} y^\nu)\label{eq:boundary_eom_1}\\
    \frac{\delta \mathcal{S}}{\delta y^\nu}=0\qquad & \leftrightarrow\qquad \frac{\partial\omega_{\alpha\beta}}{\partial y^\nu}h^{kl}\partial_{(k} x^\alpha \partial_{l)} y^\beta=h^{ab}\partial_{(k}(\omega_{\mu\nu}\partial_{l)} y^\nu)\label{eq:boundary_eom_2}
\end{align}
One can also differentiate the action with respect to the worldsheet metric $h^{kl}$, which gives the stress-energy tensor $T_{kl}$ of the string. The explicit expression for $T_{kl}$ in terms of boundary data is
\begin{equation}
    T_{kl}=\frac{\delta \mathcal{S}}{\delta h^{kl}}=\frac{T}{4}\left(\omega_{\mu\nu}\partial_{(k} x^\mu\partial_{l)} y^\nu-\frac{1}{2}h_{kl}h^{mn}\omega_{\mu\nu}\partial_{(m} x^\mu\partial_{n)} y^\nu\right)
\end{equation}
If we require $T_{kl}$ to vanish, from $T_{++}$ and $T_{--}$ we get to the following conditions
\begin{align}
    &\omega_{\mu\nu}\partial_-x^\mu\partial_-y^\nu=0 \label{eq:vir1}\\
    &\omega_{\mu\nu}\partial_+x^\mu\partial_+y^\nu=0 \label{eq:vir2}
\end{align}
Due to the symmetrization in the variables $\sigma^\pm$ the other two conditions are trivially satisfied.
These are the Virasoro constraints in terms of boundary variables. We can also differentiate the definitions \eqref{eq:xy_def} of $x$ and $y$ and use equations \eqref{eq:emb_eq1}-\eqref{eq:deriv_eq3} to show that
\begin{equation}\label{eq:nullcond}
    \partial_-x\bullet \partial_+x=\partial_-y\bullet \partial_+y=0
\end{equation}
Equations \eqref{eq:boundary_eom_1}-\eqref{eq:nullcond} give a non-trivial dynamics of causal diamonds which are emerging as projections of a bulk string worldsheet. Later we will see in case of $AdS_3$ that these constraints indeed give the correct equations of motion and that they have a gauge invariance related to the moduli structure of the kinematic space.

So far we have derived the equations of motion for $x^\mu$ and $y^\mu$ from the bulk point of view. However, we have not given the explicit form of $X_\pm$ and $X$ in terms of $x^\mu$ and $y^\mu$ yet. This is the problem of how we can  "lift up" the causal diamonds to the bulk to get a proper string worldsheet. Since $X_+$ and $X_-$ are two null vectors, using the definitions in \eqref{conf1} we can write an ansatz for them in the form

\begin{align}
    X_-&=-\frac{e^\lambda}{\sqrt{2}L R}\left(\frac{-L^2-x\bullet x}{2},Lx^\mu,\frac{L^2-x\bullet x}{2}\right)\label{eq:xm}\\
    X_+&=\frac{e^\chi}{\sqrt{2}L R}\left(\frac{-L^2-y\bullet y}{2},Ly^\mu,\frac{L^2-y\bullet y}{2}\right)\label{eq:xp}
\end{align}
with
\begin{equation}
    R^2=-\frac{1}{4}(x-y)\bullet(x-y)>0
\end{equation}
Notice that in this ansatz
we used the gauge degree of freedom familiar from Eq.(\ref{conf1}) in an explicit manner by introducing the fields $\lambda$ and $\chi$.
Now from the equation $X_+\cdot X_-=-e^\alpha$ it can be easily shown that
\begin{equation}\label{eq:alpha_lambda_chi}
    \alpha=\lambda+\chi
\end{equation}
Now we see that the field $\alpha$ appearing in \cite{Maldacena} can be interpreted as the lift of boundary causal diamonds to bulk strings. If we use equation \eqref{eq:emb_eq4} and calculate the product $X_{+-}\cdot X_{+-}$ by differentiating $X_-$ with respect to $\sigma^+$ and $X_+$ with respect to $\sigma^-$ we get to the following identities
\begin{align}
    e^\alpha&=\pm \frac{e^\lambda L}{\sqrt{2} R}\sqrt{-\partial_+ x\bullet \partial_+ x}\\
    e^\alpha&=\pm \frac{e^\chi L}{\sqrt{2} R}\sqrt{-\partial_- y\bullet \partial_- y}
\end{align}
From now on we choose the branch with the upper sign. As we will see in Section 6.6, in three dimensions this choice ensures that we use the Poincaré patch in which $X^->0$. From the previous three equations we can explicitly calculate $\lambda$ and $\chi$. These are the following
\begin{equation}\label{eq:lambda-chi }
    e^\lambda=\frac{L}{\sqrt{2}R}\sqrt{-\partial_-y\bullet\partial_-y},\qquad e^\chi=\frac{L}{\sqrt{2}R}\sqrt{-\partial_+x\bullet\partial_+x}
\end{equation}
These expressions give the rules how to lift up the vectors built up from the causal diamond coordinates to the bulk to get the directional derivatives of the dual string worldsheet. Via the string equation of motion \eqref{eq:emb_eq4} these also determine the embedding coordinate $X$. However, notice that if we require the worldsheet to be smooth, namely $X_{+-}\equiv X_{-+}$, we should also take into account the following additional constraints
\begin{equation}
\begin{aligned}
    \partial_+\left(\frac{e^\lambda}{ R}\right)&=-\partial_-\left(\frac{e^\chi}{ R}\right),\\
    \partial_+\left(x^\mu\frac{e^\lambda}{ R}\right)&=-\partial_-\left(y^\mu\frac{e^\chi}{R}\right),\\
    \partial_+\left(x\bullet x\frac{e^\lambda}{R}\right)&=-\partial_-\left(y\bullet y\frac{e^\chi}{ R}\right)
\end{aligned}
\end{equation}
These conditions come from differentiating \eqref{eq:xm} and \eqref{eq:xp} with respect to $\sigma_+$ and $\sigma_-$ respectively, and requiring each of the components to be equal.

\subsection{Relation to segmented strings}\label{sec:segmented}

The relation between strings in $AdS_3$ and the so called "celestial" coordinates were first investigated in \cite{Dv3}. As we will see in Section \ref{sec:AdS3Strings} that the coordinates $b$ and $w$ in \cite{Dv3} are null coordinates for $x^\mu$ and $y^\mu$. If we can interpret $x^\mu$ and $y^\mu$ and $b,\bar{b}$ and $w,\bar{w}$ as boundary causal diamond coordinates, our duality naturally emerges from the formalism of \cite{Dv3}.

However the equations of motion there were derived from the segmented approximation of bulks strings \cite{Callebaut, DV1,Gubser,DV2, Levay4}. In \cite{Levay4} we investigated the $AdS_{d+1}$ generalization of two dimensional segmented strings and showed that there is a duality between string segments and boundary causal diamonds. In this section we summarize the basic concept of segmented strings in $AdS_{d+1}$  and following the idea of \cite{Dv3} we describe how one can take their continuous limit in arbitrary dimension to get the previous equations of motion for the fields $x^\mu$ and $y^\mu$.

A two dimensional segmented string is an approximate solution of the classical bulk string equations of motion. It is built up by elementary segments that are simply solutions of the equations of motion themselves. An elementary segment is defined through four $AdS_{d+1}$ vectors $V_1,V_2,V_3,V_4$ that stretch a quadrangle in the $AdS$ space. The edges of this quadrangle are given by the following null-vectors (see Figure \ref{fig:null_vectors}):
\begin{align}
p_1=V_2-V_1&&p_2=V_3-V_2 \label{eq:Vp_rel1}\\
p_3=V_3-V_4&&p_4=V_4-V_1 \label{eq:Vp_rel2}
\end{align}
It is easy to show that $p_i\cdot p_i=0$. With this data in hand, one can give an explicit solution for the points on the string:
\begin{equation}
X(\sigma^-,\sigma^+)=\frac{L^2+\sigma^-\sigma^+\frac{1}{2}p_1\cdot p_4}{L^2-\sigma^-\sigma^+\frac{1}{2}p_1\cdot p_4}V_1+L^2\frac{\sigma^-p_1+\sigma^+p_4}{L^2-\sigma^-\sigma^+\frac{1}{2}p_1\cdot p_4}
\end{equation}
It is straightforward to show that $X(\sigma^-,\sigma^+)$ satisfies \eqref{eq:eom} while the Virasoro constraints \eqref{eq:virasoro} also hold for $\partial_\pm X$. In $AdS_3$ the special property of this solution is that the normal vector of the worldsheet along the string segment is constant.

\begin{figure}[!t]
    \centering
    \includegraphics[width=0.3\textwidth]{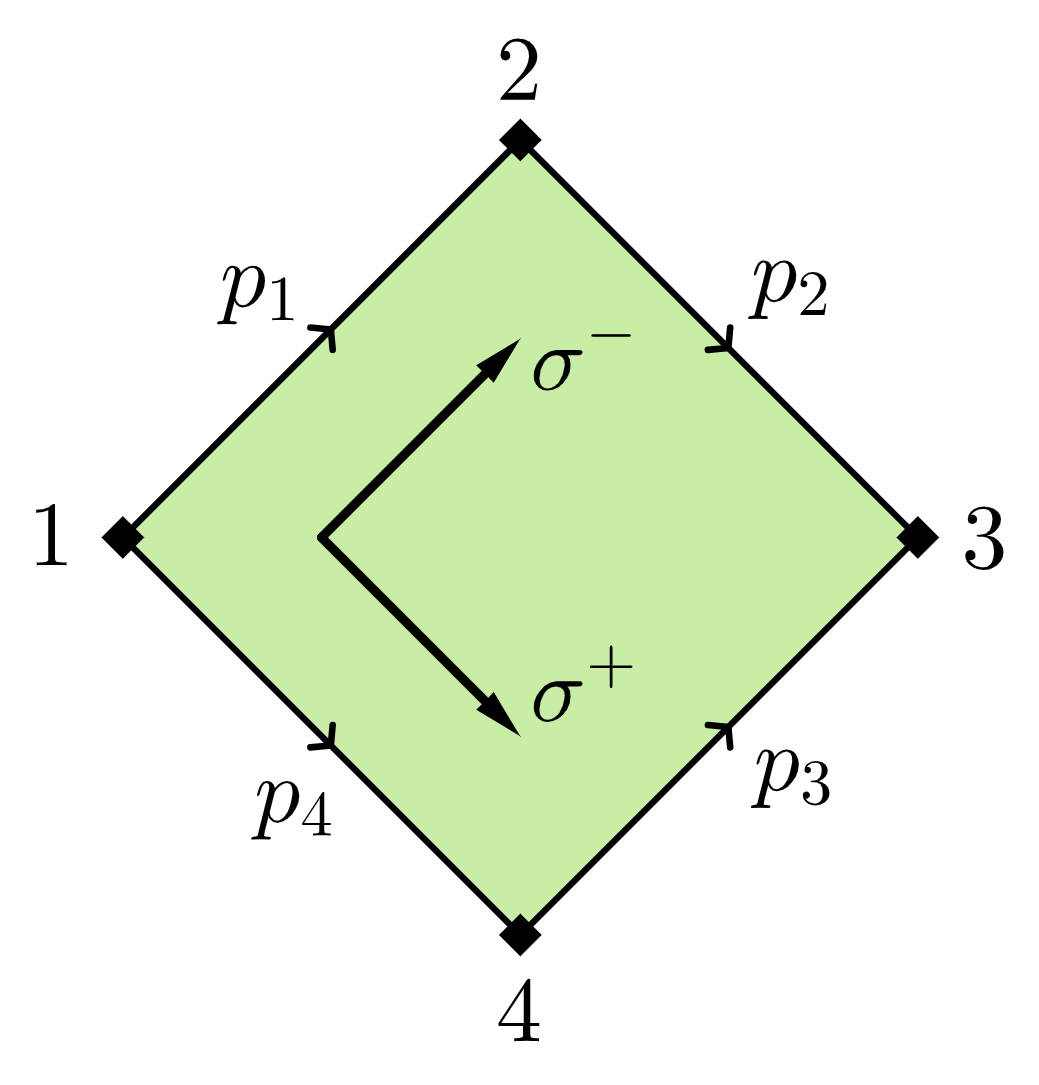}
    \caption{A string segment is defined by four $AdS$ vectors $V_i$ (correspond to the four vertices). They determine four null vectors $p_i$ (correspond to the four edges) via equations \eqref{eq:Vp_rel1} and \eqref{eq:Vp_rel2}. The figure was adapted from \cite{Levay4}}
    \label{fig:null_vectors}
\end{figure}

The null-vectors $p_1,p_2,p_3,p_4$ are the main ingredients to establish a duality between the string segments and boundary causal diamonds. As we saw in Section \eqref{sec:AdSMinimal}, if we take those $X$ points of the $AdS_{d+1}$ space for which $X\cdot p_1=0$, they determine a subspace of $AdS_{d+1}$, which is represented by a cone in the Poincaré patch. This cone then can be projected to the boundary determining also a lower dimensional causal cone in the boundary Minkowski space. Repeating the same with the vectors $p_2,p_3$ and $p_4$ we get  four different cones inside the bulk and on the boundary as well with their intersections determining a set of bulk minimal surfaces and boundary subsystems. The causality domains of these boundary subsystems stretched by the cones are the dual causal diamonds of the segmented string. For an  illustration of this connection between strings and causal diamonds in the ${\rm AdS}_3$ case see Figure \ref{fig:string_ent}. 

\begin{figure}[!t]
    \centering
    \includegraphics[width=0.6\textwidth]{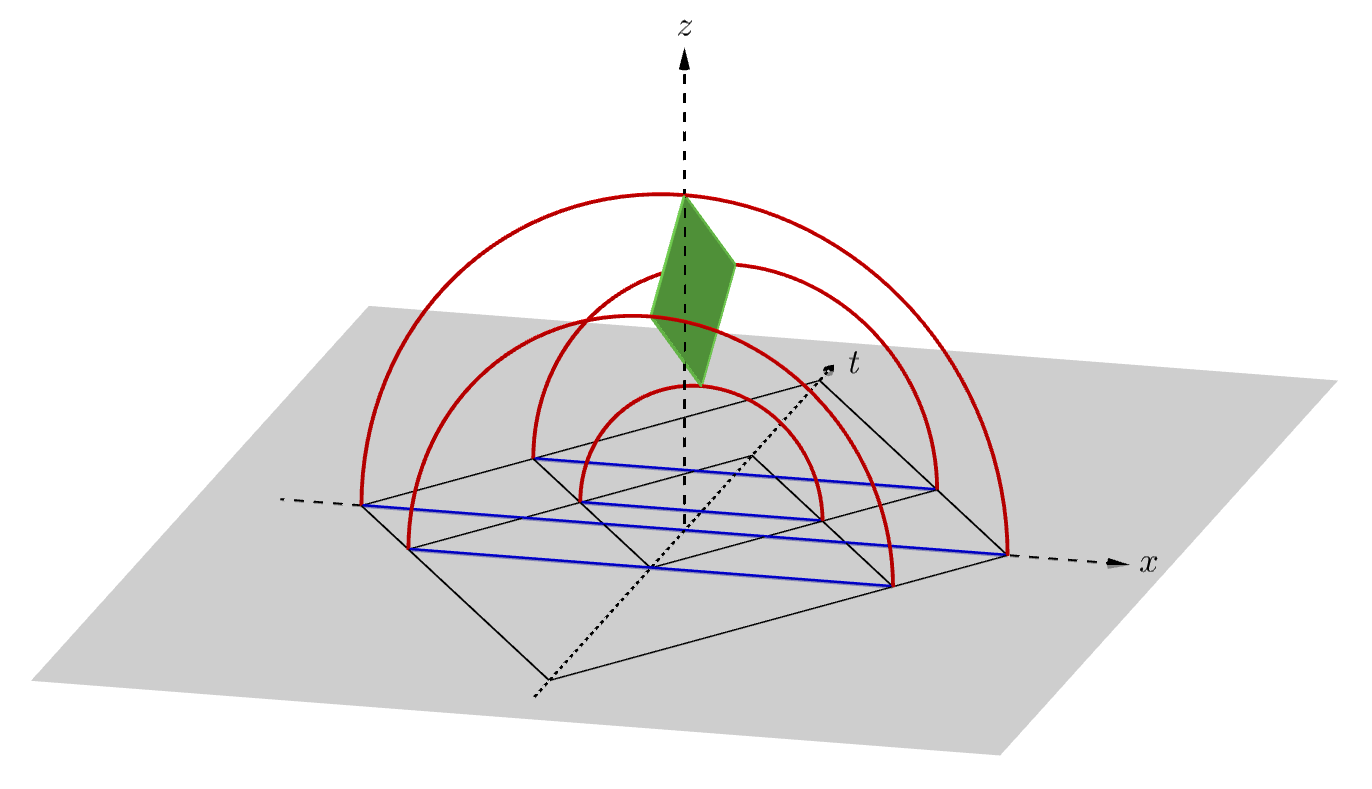}
    \caption{Duality between a string segment and boundary causal diamonds in $AdS_3$. A string segment is defined via four null vectors $p_1,p_2,p_3$ and $p_4$. The vertices of the string segment (green) lies on the top of four RT geodesics (red) related to bulk causal diamonds determined by pairs of the same null vectors via the conditions $p_i\cdot X=p_j\cdot X=0$. The RT geodesics are anchored to the endpoints of four (blue) boundary regions and the corresponding causality domains (black diamonds). Hence in the $z\to 0$ limit a connection between the string segment and these boundary causal diamonds is established. Reproduced from \cite{Levay4}.}
\label{fig:string_ent}
\end{figure}

As we discussed in \cite{Levay4}, there are certain conditions that the null-vectors should satisfy
\begin{enumerate}
    \item $V_i^->0$ for all $i=1,2,3,4$,
    \item $p_i\cdot p_j<0$ and $p_i^-p_j^-<0$ for neighbouring null vectors,
    \item $p_i\cdot p_j>0$ and $p_i^-p_j^->0$ for antipodal null vectors,
    \item And finally $p_1^0/p_1^-<p_3^0/p_3^-<p_4^0/p_4^-<p_2^0/p_2^-$.
\end{enumerate}
These are necessary conditions for the segment to be well-defined, timelike and representable in the Poincaré patch. It also gives an ordering of the boundary causal diamonds.

Now we show that the segmented string approach in the continuous limit is equivalent to what we discussed in the previous sections. As it was derived in \cite{DV1,Levay4} the surface of the segmented string is given by
\begin{equation}
A_{\diamond}=L^2\log\frac{(p_1-p_4)^2(p_3-p_2)^2}{(p_1+p_2)^2(p_3+p_4)^2}
\end{equation}
If we introduce the boundary coordinates for the null-vectors
\begin{equation}
    x_i^\mu:=\frac{L}{p_i^-}(p_i^0,p_i^1,\dots,p_i^{d-1}),\qquad i=1,2,3,4
\end{equation}
then the area can be explicitly expressed in terms of these variables as
\begin{equation}
    A_{\diamond}=L^2\log\frac{(x_1-x_4)^2(x_2-x_3)^2}{(x_3-x_4)^2(x_1-x_2)^2}
\end{equation}
where $(x_i-x_j)^2=(x_i-x_j)\bullet (x_i-x_j)$.

The question arises, how the equations of motion can be derived from the segmented approximation similarly to \cite{Dv3}. This can be done by gluing together elementary segments along their edges. If we label the null-vectors by $p_{i,j}$ following the convention of \cite{DV1} (see Figure \ref{fig:toda}), the total area of the string becomes
\begin{equation}
    A_{\text{tot}}=L^2\sum_{i, j} \log \frac{(x_{i,j}-x_{i,j+1})^2}{(x_{i,j}-x_{i+1,j})^2}=\frac{1}{2}\sum_{i, j}(S(i,j;i,j+1)-S(i,j;i+1,j))
\end{equation}
Where
\begin{equation}
    x_{i,j}^\mu:=\frac{L}{p_{i,j}^-}(p_{i,j}^0,p_{i,j}^1,\dots,p_{i,j}^{d-1})
\end{equation}
are boundary coordinates for the vectors $p_{i,j}$, and we defined
\begin{equation}
    S(i,j,k,l)=2L^2\log (x_{i,j}-x_{k,l})^2
\end{equation}
The sum can be split into elementary pieces in two different ways. One can calculate the sum by splitting it into combinations such as
\begin{equation}\label{eq:sum1}
    S(i-1,j-1;i-1,j)+S(i,j-1;i,j)-S(i-1,j;i,j)-S(i-1,j-1;i,j-1)
\end{equation}
which corresponds to summing up the areas of the string segments. Alternatively it is possible to sum up combinations like
\begin{equation}\label{eq:sum2}
S(i,j-1;i,j)+S(i+1,j-1;i+1,j)-S(i,j;i+1,j)-S(i,j-1;i+1,j-1)
\end{equation}
This corresponds to summing up contributions coming from edges not around string segments but around vertices. Let us examine these two combinations in the continuous limit.

\begin{figure}[!t]
    \centering\includegraphics[width=0.55\textwidth]{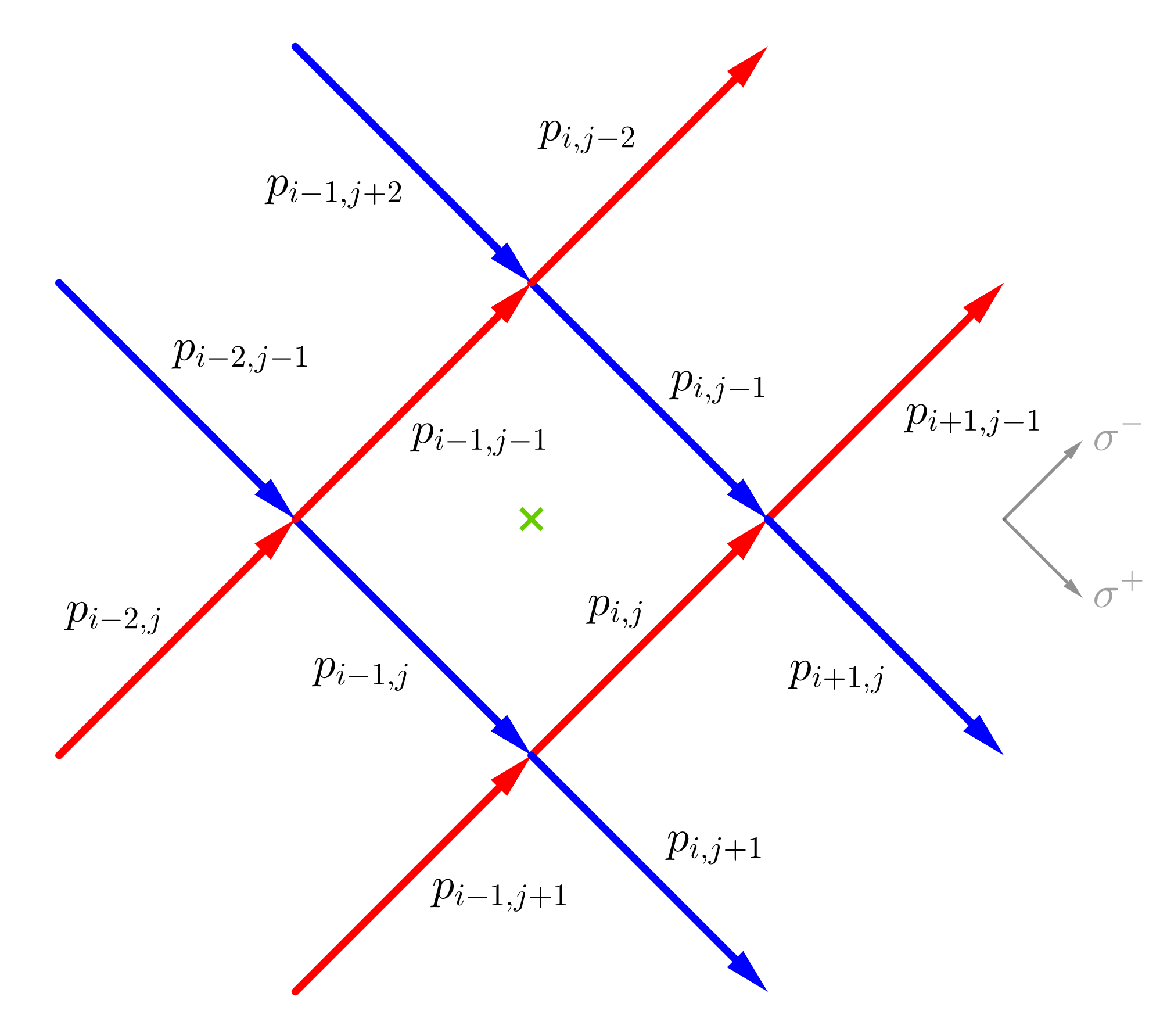}
    \caption{The lattice structure of string segments glued together along their defining null-vectors. Each null-vector $p_{i,j}$ is labeled by the horizontal $i$ and vertical $j$ indices. The red arrows denote edges that are represented by $x^\mu$ vectors, while the blue arrows are represented by $y^\mu$ vectors in the continuous limit. The parametrization of the world sheet in terms of $\sigma^\pm$ is also denoted in the figure. The green cross denotes the point where the vectors $x^\mu(\sigma^-,\sigma^+)$ and $y^\mu(\sigma^-,\sigma^+)$ (with no shift in their arguments) reside.}
    \label{fig:toda}
\end{figure}
The continuous limit is defined by taking each segment to be infinitesimally small. We can parametrize the string in such a way that in the infinitesimal limit the vectors $x^{\mu}_{ij}$ become \cite{Dv3}
\begin{equation}\label{eq:continuous_fields}
    \begin{aligned}
    x^\mu_{i-1,j-1}&=x^\mu(\sigma^-
    ,\sigma^+-\epsilon),&
    x^\mu_{i,j}&=x^\mu(\sigma^-,\sigma^++\epsilon),&
    x^\mu_{i+1,j-1}&=x^\mu(\sigma^-+2\epsilon,\sigma^++\epsilon),\\
    x^\mu_{i-1,j}&=y^\mu(\sigma^--\epsilon,\sigma^+),&
    x^\mu_{i,j-1}&=y^\mu(\sigma^-+\epsilon,\sigma^+),&
    x^\mu_{i+1,j}&=y^\mu(\sigma^-+\epsilon,\sigma^++2\epsilon),
\end{aligned}
\end{equation}
with $\epsilon$ small. The field $x^\mu$ corresponds to the "upmoving" edges, while $y^\mu$ for the "downmoving" ones (see Figure \ref{fig:toda}). Therefore in the continious limit one can write for example
\begin{equation}
\begin{aligned}
    S(i-1,j;i,j)&=S(x^\mu(\sigma^-  ,\sigma^+   +\epsilon),y^\mu(\sigma^-  -\epsilon,\sigma^+   ))=\\
    &=S(x^\mu(\sigma^-,\sigma^+   ),y^\mu(\sigma^-  ,\sigma^+))+\\
    &+\left[\partial_+x^\mu(\sigma^-  ,\sigma^+)\frac{\partial}{\partial x^\mu}S(x^\mu(\sigma^-,\sigma^+   ),y^\mu(\sigma^-  ,\sigma^+))-\right.\\
    &-\left.\partial_-y^\mu(\sigma^-  ,\sigma^+)\frac{\partial}{\partial y^\mu}S(x^\mu(\sigma^-,\sigma^+   ),y^\mu(\sigma^-  ,\sigma^+))\right]\epsilon+\mathcal{O}(\epsilon^2)
\end{aligned}
\end{equation}
with
\begin{equation}\label{eq:Sdef}
    S(x,y)=2L^2\log (x(\sigma^-,\sigma^+)-y(\sigma^-,\sigma^+))^2
\end{equation}
Repeating the same with the other contributions, one can rewrite \eqref{eq:sum1} and \eqref{eq:sum2} as
\begin{equation}
\begin{aligned}
    S(i,j-1;i,j)&+S(i-1,j-1;i-1,j)-S(i-1,j;i,j)-S(i-1,j-1;i,j-1)=\\
    &=\frac{\partial^2 S(x,y)}{\partial x^\mu \partial y^\nu}\partial_+ x^\mu \partial_- y^\nu (2\epsilon)^2+\mathcal{O}(\epsilon^3)\\
    S(i,j-1;i,j)&+S(i+1,j-1;i+1,j)-S(i,j;i+1,j)-S(i,j-1;i+1,j-1)=\\
    &=\frac{\partial^2 S(x,y)}{\partial x^\mu \partial y^\nu}\partial_- x^\mu \partial_+ y^\nu (2\epsilon)^2+\mathcal{O}(\epsilon^3)
\end{aligned}
\end{equation}
Hence summing over all segments the total area can be written in two equivalent ways
\begin{equation}\label{eq:cont_area1}
    A_{\text{tot}}=\frac{1}{2}\int \frac{\partial^2 S}{\partial x^\mu \partial y^\nu}\partial_+ x^\mu \partial_- y^\nu d\sigma^+ d\sigma^-=\frac{1}{2}\int \frac{\partial^2 S}{\partial x^\mu \partial y^\nu}\partial_- x^\mu \partial_+ y^\nu d\sigma^+ d\sigma^-
\end{equation}
Reinserting the wordlsheet metric $ds^2=h_{ij}d\sigma^i d\sigma^j=4d\sigma^+ d\sigma^-$, $i,j=\pm$ and taking the symmetric combination of the two different resulting expressions, we get for the total area that
\begin{equation}\label{eq:cont_action}
    A_{\text{tot}}=\frac{1}{8}\int \sqrt{-h}h^{ij}\omega_{\mu\nu}\partial_{(i}x^\mu \partial_{j)} y^\nu d\sigma^+ d\sigma^-
\end{equation}
With the action being $S=-TA_{\text{tot}}$ with string tension $T$. Then the action coincides with \eqref{twistact} if
\begin{equation}
    \omega_{\mu\nu}\equiv\frac{\partial^2 S}{\partial x^\mu \partial y^\nu}
\end{equation}
Where $\omega_{\mu\nu}$ is the kinematic space metric given by \eqref{Myers}. It is straightforward to show that by the definition \eqref{eq:Sdef} this relation indeed holds if we identify the kinematic space length scale $\ell$ with the $AdS$ scale $L$. 

With the action in terms of $x^\mu$ and $y^\mu$ and the causality conditions for the vectors $p_i$ in hand, all of the equations of motion and causality constraints for the vectors $x^\mu$ and $y^\nu$ (such as $x^0<y^0$ and $\partial_+x\bullet\partial_+x,\partial_-y\bullet\partial_-y<0$) can be derived giving the same relations as in the previous sections. As we will see in Section \ref{sec:AdS3Strings}, in case of $AdS_3$ the equations of motion coincide with those derived from the Toda like equation in \cite{DV1}.

One final remark is that investigating the relation between the segmented and continuous strings is not only useful to validate one or the other approach. As we have shown in \cite{Levay4}, there is also a duality between the geometry of segmented strings and boundary entanglement. Therefore the segmented string approach could give an idea about the physical content of our geometrical duality between bulk strings and boundary causal diamonds investigated in this paper.

\section{Gauge invariance}\label{sec:gauge}

In \cite{Maldacena} following \cite{ Pohlmeyer, Vega, Jevicki} it was shown that the dynamics of the propagating string in $AdS_3$ is governed by a single field $\alpha$ along with a holomorphic function $p(z)$. Furthermore, the equations of motion are given by the flatness conditions of two copies of $SL(2)$ connections. In this section, we show that a similar story holds for $AdS_{d+1}$ too. The string dynamics can be represented by $SO(1,d-1)\times SO(1,1)$ gauge invariant equations. In the previous sections we derived an explicit connection between bulk strings and boundary causal diamonds. The space of causal diamonds has a $SO(2,d)/SO(1,d-1)\times SO(1,1)$ factor space structure. Therefore the $SO(1,d-1)\times SO(1,1)$ invariant equations of motion for the bulk string is the manifestation of this duality.

In order to show this, it is convenient to introduce a set of $N_i$, $i=1,...,d-1$ normal vectors that are orthogonal to the string world-sheet in the embedding space. Let us define them as follows
\begin{align}\label{eq:Ndef}
    N^i\cdot N_j=L^2 \delta^{i}_j,&&
    N^i\cdot X=0,&&
    N^i\cdot X_\pm=0.
\end{align}
Where $\delta^{i}_{\,j}$ is the $d-1$ dimensional Kronecker delta and from now on we denote the differentiations with respect to $\sigma^{\pm}$ with subscripts (e.g.: $\partial_{+}X=X_+$). In the following we also use Einstein convention for the summations over $i,j$ indices and we denote $N^i=\delta^{ij}N_j$. Notice that the choice of $N_i$ is not unique if $d>2$, since we can choose any basis in the orthogonal subspace.

Following the notations of \cite{Maldacena}, let us introduce the dimensionless quantities
\begin{align}
    X_+ \cdot X_-&:=-e^{\alpha} \label{eq:alpha}\\
    N^i\cdot X_{--}&:=f^i\label{eq:defv}\\
    N^i\cdot X_{++}&:=g^i\label{eq:defu}
\end{align}
The fields $f^i$ and $g^i$ are generalizations of the holomorphic fields $p(z)$ and $\bar{p}(\bar{z})$ familiar from \cite{Maldacena}. The embedding space $\mathbb{R}^{2,d}$ is spanned by the pairwise orthogonal vectors $X_\pm,X$ and $N^i$, therefore their derivatives can be uniquely expressed in terms of these vectors giving a set of equations describing the motion of the string. Using $X\cdot X=-L^2$, the \eqref{eq:eom} equation of motion and the previous definitions it can be shown that these equations are the following
\begin{align}
    X_{+-}&=-\frac{1}{L^2}e^\alpha X \label{eq:eqmotion1}\\
    X_{++}&=\alpha_+ X_+ +\frac{1}{L^2} g_i N^i \label{eq:eqmotion2}\\
    X_{--}&=\alpha_- X_- +\frac{1}{L^2} f_i N^i \label{eq:eqmotion3}\\
    N_+^i&=g^i e^{-\alpha} X_- +(\omega^+)^{i}_{\,j} N^j\label{eq:eqmotion4}\\
    N_-^i&=f^i e^{-\alpha} X_+ +(\omega^-)^{i}_{\,j}N^j \label{eq:eqmotion5}
\end{align}
Where $(\omega^\pm)^i_{\,j}$ are $(d-1)\times (d-1)$ dimensional skew-symmetric matrices with indices $i,j=1,\dots,d-1$ and those depend on the choice of $N^a$.

The functions $\alpha,f,g$ and $\omega^\pm$ are also not independent from each other. From equations \eqref{eq:eqmotion2} and \eqref{eq:eqmotion3} it can be shown that
\begin{equation}
    X_{++}\cdot X_{--}=-\alpha_+\alpha_- e^{\alpha}+\frac{1}{L^2} f_i g^i
\end{equation}
However one can also rewrite $X_{++}\cdot X_{--}$ as:
\begin{equation}\label{eq:alpha_motion}
\begin{aligned} 
    X_{++}\cdot X_{--}=&\partial_-(X_{++}\cdot X_{-})-X_{++-}\cdot X_{-}=\\
    =&-\alpha_{+-}e^\alpha-\alpha_+\alpha_-e^\alpha-\frac{1}{L^2}e^{2\alpha}
\end{aligned}
\end{equation}
Where in the second line we used equation \eqref{eq:eqmotion1}. These two expressions lead to the following identity
\begin{equation}\label{eq:sinh}
    \alpha_{+-}+\frac{e^{-\alpha}}{L^2}f_i g^i+\frac{e^\alpha}{L^2}=0
\end{equation}
This is the generalization of the sinh-Gordon equation, familiar from \cite{Maldacena} and \cite{Dv3}, to arbitrary dimensions. From the definitions of $f^i$ and $g^i$, one can also show that the following equations hold
\begin{equation}\label{eq:uvmotion}
    f_+^i=(\omega^+)^{i}_{\,j}f^j,\qquad g_-^i=(\omega^-)^{i}_{\,j}g^j
\end{equation}
Which are also the generalizations of the holomorphicity condition $\partial_{\bar{z}}p(z)=0$ of \cite{Maldacena}.

Finally, one can also consider the mixed derivatives $N_{+-}^i$ and $N_{-+}^i$. Using the definitions and the equations of motions they can be written as
\begin{equation}
    \begin{aligned}
        N_{+-}^i=&(\omega^-)^{i}_{\,j}(\omega^+)^{j}_{\,k}N^k +\frac{1}{L^2}f^i e^{-\alpha}g_j N^j+(\omega_+^-)^{i}_{\,j}N^j\\
        &+e^{-\alpha}(\omega^-)^{i}_{\,j}g^j X_-+e^{-\alpha}(\omega^+)^{i}_{\,j}u^j X_+\\
        N_{-+}^i=&(\omega^+)^{i}_{\,j}(\omega^-)^{j}_{\,k}N^k +\frac{1}{L^2}g^i e^{-\alpha} f_j N^j+(\omega_-^+)^{i}_{\,j}N^j\\
        &+e^{-\alpha}(\omega^-)^{i}_{\,j}f^j X_-+e^{-\alpha}(\omega^+)^{i}_{\,j}u^j X_+
    \end{aligned}
\end{equation}
And as before, we can assume that $N^i(\sigma^+,\sigma^-)$ are smooth functions of the worldsheet parameters, hence we can require that $N^a_{+-}=N^a_{-+}$. Using the previously derived equations this condition translates to
\begin{equation}
    \left\{\frac{e^{-\alpha}}{L^2}f^i g_j-\frac{e^{-\alpha}}{L^2}g^i f_j+(\omega^-)^{i}_{\,k}(\omega^+)^{k}_{\,j}-(\omega^+)^{i}_{\,k}(\omega^-)^{k}_{\,j}+(\omega_+^-)^{i}_{\,j}-(\omega_-^+)^{i}_{\,j}\right\}N^j=0
\end{equation}
If we multiply both sides with $N_l$ with $l=1,\dots,d-1$, use \eqref{eq:Ndef} and relabel the indices we get that for arbitrary $i,j=1,\dots,d-1$:
\begin{equation}\label{eq:curvature}
    \frac{e^{-\alpha}}{L^2}f^i g_j-\frac{e^{-\alpha}}{L^2}g^i f_j+(\omega^-)^{i}_{\,k}(\omega^+)^{k}_{\,j}-(\omega^+)^{i}_{\,k}(\omega^-)^{k}_{\,j}+(\omega_+^-)^{i}_{\,j}-(\omega_-^+)^{i}_{\,j}=0
\end{equation}

Equations \eqref{eq:eqmotion1}-\eqref{eq:eqmotion5}, \eqref{eq:sinh},\eqref{eq:uvmotion} and \eqref{eq:curvature} completely describe the motion of a one dimensional string embedded into the $AdS_{d+1}$ space.

In \cite{Maldacena} the string motion in $AdS_3$ was described by an $SL(2)\times SL(2)$ invariant gauge theory. Notice that this group structure is similar to the local symmetry group $SO(1,1)\times SO(1,d-1)$ of the moduli space of causal diamonds. This similarity and the derived connection between bulk strings and causal diamonds gives the idea that in higher dimensions the string motion might also be described by an $SO(1,1)\times SO(1,d-1)$ gauge theory. The fields appearing in the equations should represent boundary causal diamonds.

To see this structure let us introduce two column vectors $t$ and $n$ with elements:
\begin{align}
    &t^0:=L e^{-\lambda} X_-,&t^1:&=L e^{-\chi} X_+,\\
    &n^{0}:=X,&n^{i}:&=N^{i},i=1,\dots, d-1
\end{align}
With dimensions $2$ and $d$ respectively. The two $t$ vectors describe the tangent subspace of the string worldsheet and locally defines a causal diamond on the boundary. This can be seen if we use expressions \eqref{eq:xm} and \eqref{eq:xp} to connect $t$ to the causal diamond coordinates:
\begin{align}
    t^0&=-\frac{1}{\sqrt{2} R}\left(\frac{-L^2-x\bullet x}{2},Lx^\mu,\frac{L^2-x\bullet x}{2}\right)\\
    t^1&=\frac{1}{\sqrt{2} R}\left(\frac{-L^2-y\bullet y}{2},Ly^\mu,\frac{L^2-y\bullet y}{2}\right)
\end{align}
The difference between the field $t$ and the vectors $X_-$ and $X_+$ is that the factors before the brackets in \eqref{eq:xm} and \eqref{eq:xp} are depending on the derivatives $\partial_\pm x^\mu$ and $\partial_\pm y^\mu$ via $\lambda$ and $\chi$. However, $t^0$ and $t^1$ includes only the causal diamond coordinates $x^\mu$ and $y^\mu$, hence $t$ encodes the causal diamond itself and its dynamics will only be determined by the corresponding equations of motion. Notice that $t$ is normalized in such a way that $t^0\cdot t^1=-L^2$. Finally $n$ represents the locally orthogonal subspace to the world-sheet spanned by $X_-$ and $X_+$.

Our equations of motion should be invariant under the local $SO(1,1)$ transformations in the subspace spanned by $t$ and $SO(1,d-1)$ transformations in the subspace spanned by $n$. It turns out that this is indeed the case. Using \eqref{eq:eqmotion1}-\eqref{eq:eqmotion5} and \eqref{eq:alpha_lambda_chi} the equations of motion for $n$ and $t$ become:
\begin{align}
    t_+^0&=-\lambda_+ t^0-\frac{e^\chi}{L}n^{0}&t_-^0&=\chi_- t^0+\frac{1}{L}e^{-\lambda} f_i n^{i}\\
    t_+^1&=\lambda_+ t^1+\frac{1}{L} e^{-\chi} g_i n^{i}&t_-^1&=-\chi_- t^1-\frac{1}{L}e^{\lambda}n^{0}\\
    n_+^{0}&=\frac{e^\chi}{L}t^1&n_-^{0}&=\frac{e^\lambda}{L}t^0\\
    n_+^{i}&=\frac{1}{L}e^{-\chi}g^i t^0+(\omega^+)_{\;j}^{i}n^{j}&n_-^{a}&=\frac{1}{L}e^{-\lambda}f^i t^1+(\omega^-)_{\;j}^{i} n^{j}
\end{align}
Where we have used the fact that $\omega^\pm$ are skew-symmetric, hence $(\omega^\pm)^{\;i}_j=-(\omega^\pm)^i_{\;j}$. The indices $i$ and $j$ go from $1$ to $d-1$. Now we can introduce the following matrices:
\begin{align}
    &{B^t_+}=
    \begin{pmatrix}
    \lambda_+&0\\
    0&-\lambda_+
    \end{pmatrix} &
    {B^t_-}&=
    \begin{pmatrix}
    -\chi_-&0\\
    0&\chi_-
    \end{pmatrix}\label{eq:Bt}\\
    &{B^n_+}=
    \left( \begin{array}{c|c}
    0&0\\
    \midrule
    0&-\omega^+
    \end{array}\right) &
    {B^n_-}&=
     \left( \begin{array}{c|c}
    0&0\\
    \midrule
    0&-\omega^-
    \end{array}\right)\label{eq:Bn}
\end{align}
Where $\omega^\pm$ are $(d-1)\times (d-1)$ matrices with elements $(\omega^\pm)^i_j$. $B_\pm^t$ have dimension $2\times 2$ while $B_\pm^n$ have dimension $d\times d$. Notice that since $\omega^\pm$ are skew-symmetric, the matrices $B^t_\pm$ and $B^n_\pm$ are $SO(1,1)$ and $SO(1,d-1)$ generators respectively. They represent the $SO(1,1)$ and $SO(1,d-1)$ invariance of the $\mathbb{R}^{2,d}$ subspaces spanned by $X_\pm$ and $X,N^i,\,i=1,d-1$ respectively. Furthermore if we introduce the following $2\times d$ matrices:
\begin{equation}\label{eq:Phi}
    \Phi_+=\left( \begin{array}{c|c}
    \frac{e^\chi}{L}&0\\
    \midrule
    0&\frac{e^{-\chi}}{L}g^T
    \end{array}\right),\qquad 
    \Phi_-=\left( \begin{array}{c|c}
    0&\frac{e^{-\lambda}}{L}f^T\\
    \midrule
    \frac{e^\lambda}{L}&0
    \end{array}\right)
\end{equation}
Where $f$ and $g$ are column vectors with elements $f^i$ and $g^i$ respectively. With these definitions in our hands, the equations of motions for $t$ and $n$ can be written as
\begin{align}
    \left(\partial_\pm +{B^t_\pm}\right)t&=\Phi_\pm\eta n\label{eq:gauge1}\\
    \left(\partial_\pm +{B^n_\pm}\right)n&=\Phi_\pm^T \sigma t\label{eq:gauge2}
\end{align}
Where $\eta$ and $\sigma$ are the $SO(1,d-1)$ and $SO(1,1)$ invariant metric tensors respectively. These two equations express the $SO(1,1)$ and $SO(1,d-1)$ gauge invariant equations of motions, and the matrices $B^t$ and $B^n$ are $SO(1,1)$ and $SO(1,d-1)$ connections. The first one corresponds to the dynamics of the subspace spanned by $X_\pm$ that is invariant under local $SO(1,1)$ transformations. Similarly the second one describes the dynamics of the perpendicular subspace. Introducing the covariant derivatives
\begin{equation}
    D_\mu^t=\partial_\mu+{B^t_\mu},\qquad
    D_\mu^n=\partial_\mu+{B^n_\mu}
\end{equation}
Therefore:
\begin{equation}
    D_\mu^t t=\Phi_\mu\eta n,\qquad D_\mu^n n=\Phi_\mu^T\sigma t
\end{equation}
Where $\mu=\sigma^{\pm}$.

The equations for $\alpha,f^i,g^i$ and $\omega^\pm$ derived previously in this section can be also written as gauge invariant equations. If we first consider equation \eqref{eq:sinh}, it can be shown that this is equivalent to
\begin{equation}\label{eq:flat1}
    \partial_+B^t_--\partial_-B^t_++[B^t_+,B^t_-]=(\Phi_-\eta)(\sigma\Phi_+)^T-(\Phi_+\eta)(\sigma\Phi_-)^T
\end{equation}
Similarly one can write \eqref{eq:curvature} as
\begin{equation}\label{eq:flat2}
    \partial_+B^n_--\partial_-B^n_++[B^n_+,B^n_-]=(\Phi_-\eta)^T(\sigma\Phi_+)-(\Phi_+\eta)^T(\sigma\Phi_-)
\end{equation}
Finally we have equation \eqref{eq:uvmotion}. It is easy to show that this gives the following conditions for $\Phi$
\begin{equation}\label{eq:flat3}
    (\partial_\pm+\overleftarrow{B}_\pm^t-(\overrightarrow{B}_\pm^n))\Phi_\mp=0
\end{equation}
Where the arrows denote the direction of the multiplication. 

Finally we would like to point out that fixing the gauge in the matrices $B^n_\pm$ and $B^t_\pm$ fix the functions $\lambda$ and $\chi$ (up to boundary conditions). These functions define the lifting from the boundary causal diamonds to the bulk string. Hence the choice of gauge in the gauge invariant equations give different dual solutions to the bulk worldsheet.

\section{The special case of $AdS_3$}

\subsection{Kinematic space for $d=2$}

Let us consider the special case when $d=2$ i.e. the case of ${\rm AdS}_3$. In this case we have the components of the four-vectors $a,b=-1,0,1,2$ and $\mu,\nu =0,1$, hence now we have $x^{\mu}=(x^0,x^1)=(t,x)$ used as the local coordinates for a $CFT_2$.
In this case it is known\cite{Myers} that the metric of Eq.(\ref{Myers}) will be the sum of two metrics on two copies of de Sitter space. Hence the moduli space of causal diamonds is $dS_2\times dS_2$.
In this section we would like to understand the twistor geometric meaning of this factorization of moduli space. Later subsections we consider the static slice of ${\rm AdS}_3$.

Explicitly the factorization of the moduli space means that if we define light cone coordinates for the causal diamond, namely
for the past tip
$u=t_u+x_u,\quad \bar{u}=t_u-x_u$, and for the future tip
$v=t_v+x_v,\quad \bar{v}=t_v-x_v$
then the metric (\ref{Myers}) takes the following form
\beq
ds^2=2L^2\frac{dv du}{(v-u)^2}+2L^2\frac{d\bar{v}d\bar{v}}{(\bar{v}-\bar{v})^2}
\label{fact}
\eeq
In order to understand this in terms of Pl\"ucker coordinates
let us introduce the Hodge dual $\ast{\mathcal  P}$ of ${\mathcal P}=\frac{U\wedge V}{U\cdot V}$ as follows
\beq
{\ast {\mathcal P}}^{ab}=\frac{1}{2}\epsilon^{abcd}{\mathcal P}_{cd}
\qquad {\mathcal P}_{ab}={\eta}_{ac}{\eta}_{bd}{\mathcal P}^{cd}
\label{Hodge}
\eeq
One can then check that $\ast^2=1$, hence the quantities
\beq
{\mathcal P}_{\pm}^{ab}=\frac{1}{2}({\mathcal P}\pm\ast {\mathcal P})^{ab}
\label{self}
\eeq
are self-dual and anti self-dual, i.e. $\ast {\mathcal P}_{\pm}=\pm {\mathcal P}_{\pm}$.

Observe that
\beq
{\mathcal P}\cdot{\mathcal P}=\frac{(U\wedge V)^{ab}(U\wedge V)_{ab}}{(U\cdot V)^2}=-2
\label{of1}
\eeq
Similarly one can show that $\ast{\mathcal P}\cdot\ast{\mathcal P}=-2$.
Notice also that
\beq
\ast{\mathcal P}\cdot{\mathcal P}=\frac{1}{2}\epsilon^{abcd}{\mathcal P}_{ab}{\mathcal P}_{cd}=0
\label{plucki}
\eeq
by virtue of the Plücker relation
\begin{equation}
    {\mathcal P}^{-10}{\mathcal P}^{12}-
    {\mathcal P}^{-11}{\mathcal P}^{02}+
    {\mathcal P}^{-12}{\mathcal P}^{01}
=0
\label{Pluckerrelation}
\end{equation}
valid for separable bivectors like $U\wedge V$.
Hence altogether we have the following set of relations
\beq
{\mathcal P}\cdot{\mathcal P}=\ast{\mathcal P}\cdot\ast {\mathcal P}=-2,\qquad
\ast{\mathcal P}\cdot{\mathcal P}=0
\label{kinpluck}
\eeq
From these equations for the self-dual and anti self-dual parts one obtains
\beq
{\mathcal P}_+\cdot{\mathcal P}_-=0,\qquad {\mathcal P}_{\pm}\cdot {\mathcal P}_{\pm}=-1
\label{kinpluck2}
\eeq
One also has
\beq
\ast{\mathcal P}^{-10}={\mathcal P}^{12},\qquad
\ast{\mathcal P}^{-11}={\mathcal P}^{-02},\qquad
\ast{\mathcal P}^{-12}=-{\mathcal P}^{01}
\label{hodgepart}
\eeq
hence
\beq
{\mathcal P}_{\pm}^{-10}=\frac{1}{2}\left({\mathcal P}^{-10}\pm {\mathcal P}^{12}\right)=\pm{\mathcal P}_{\pm}^{12}
\label{hp2}
\eeq
\beq
{\mathcal P}_{\pm}^{-11}=\frac{1}{2}\left({\mathcal P}^{-11}\pm {\mathcal P}^{02}\right)=\pm{\mathcal P}_{\pm}^{02}
\label{hp3}
\eeq
\beq
{\mathcal P}_{\pm}^{-12}=\frac{1}{2}\left({\mathcal P}^{-12}\mp {\mathcal P}^{01}\right)=\mp{\mathcal P}_{\pm}^{01}
\label{hp4}
\eeq
Then one can choose three independent coordinates for the self-dual and three ones for the anti self-dual part.
Let us choose the triples 
\beq
\begin{pmatrix}m^0\\m^1\\m^2\end{pmatrix}=
\begin{pmatrix}2{\mathcal P}_+^{-10}\\2{\mathcal P}_+^{-11}\\2{\mathcal P}_+^{-12}\end{pmatrix},\qquad
\begin{pmatrix}n^0\\n^1\\n^2\end{pmatrix}=
\begin{pmatrix}2{\mathcal P}_-^{-10}\\2{\mathcal P}_-^{-11}\\2{\mathcal P}_-^{-12}\end{pmatrix}
\label{emn}
\eeq 
Then 
\beq
-(m^0)^2+(m^1)^2+(m^2)^2=1,\qquad 
-(n^0)^2+(n^1)^2+(n^2)^2=1
\label{conifoldlike}
\eeq
Notice that the (\ref{Pluckerrelation}) Plücker relation can also be expressed in terms of the six independent  coordinates that can be found on the left hand side of (\ref{hp2})-(\ref{hp4}). This yields the quadratic constraint
\begin{equation}
    -(m^0)^2+(m^1)^2+(m^2)^2
+(n^0)^2-(n^1)^2-(n^2)^2=0
\label{Plucker22}
\end{equation}
which is further constrained by the extra (\ref{conifoldlike}) relations.

Now according to the Klein correspondence\cite{Ward}
to a line $\ell_p$ in ${\mathbb R}{\mathbb P}^3$ there corresponds a point $\mathcal P$ which is lying on a four dimensional quadric (the Klein quadric) living in ${\mathbb R}{\mathbb P}^5$.
The Klein quadric is given by the (\ref{Pluckerrelation}) Plücker relation. 
Now $\ell_p$ is special: it is going through the points  defined by the null vectors $(U,V)$ satisfying the constraint $U\cdot V<0$.
The point in the Klein quadric which corresponds to this line is just given by the formula
$U\wedge V/(U\cdot V)$.
In this special case the relevant part of the Klein quadric comprising such points is given by Eq.(\ref{Plucker22}). According to (\ref{conifoldlike}) this part also factorizes to the product of two quadrics.
Indeed, inside the Klein quadric we have two copies of a de Sitter space $dS_2$, which by virtue of Eq.(\ref{kinpluck2}) are independent.
Hence we have learned that the moduli space of causal diamonds, the kinematic space of $AdS_3$, is the space $dS_2\times dS_2$ an object which is living inside the Klein quadric.

In order to make this result even more explicit let us use the following parametrization for the vectors $-1,0,1,2$ components of the null vectors $U$ and $V$
\beq
U^a=\frac{1}{\Delta_u}\begin{pmatrix}u\bar{u}-L^2\\L(u+\bar{u})\\L(u-\bar{u})\\u\bar{u}+L^2\end{pmatrix},
\qquad
V^b=
\frac{1}{\Delta_v}\begin{pmatrix}v\bar{v}-L^2\\L(v+\bar{v})\\L(v-\bar{v})\\v\bar{v}+L^2\end{pmatrix}
\label{paralight1}
\eeq
A calculation then shows that
\beq
m^0=2{\mathcal P}_{+}^{-10}=\frac{L^2+\bar{u}\bar{v}}{L(\bar{u}-\bar{v})},\qquad
n^0=2{\mathcal P}_{-}^{-10}=\frac{uv+L^2}{L(u-v)}
\label{hodgeformulas1}
\eeq
\beq
m^1=2{\mathcal P}_{+}^{-11}=\frac{L^2-\bar{u}\bar{v}}{L(\bar{u}-\bar{v})},\qquad
n^1=2{\mathcal P}_{-}^{-11}=\frac{uv-L^2}{L(u-v)}
\label{hodgeformulas2}
\eeq
\beq
m^2=2{\mathcal P}_{+}^{-12}=\frac{\bar{u}+\bar{v}}{\bar{u}-\bar{v}},
\qquad
n^2=2{\mathcal P}_{-}^{-12}=\frac{u+v}{u-v}
\label{hodgeformulas3}
\eeq
i.e. the left and right moving coordinates factorize precisely into self duality and anti-self duality of the Plücker coordinates.

A calculation shows that
\beq
-(dm^0)^2+(dm^1)^2+(dm^2)^2=\frac{4d\bar{u}d\bar{v}}{(\bar{u}-\bar{v})^2}=-d{\mathcal P}_+\cdot d{\mathcal P}_+
\label{parsitt1}
\eeq
and similarly
\beq
-(dn^0)^2+(dn^1)^2+(dn^2)^2=\frac{4d{u}d{v}}{({u}-{v})^2}=-d{\mathcal P}_-\cdot d{\mathcal P}_-
\label{parsitt2}
\eeq
Now our (\ref{endmetrik}) formula gives $ds^2=-\frac{1}{2}L^2(d{\mathcal P}_+\cdot d{\mathcal P}_++d{\mathcal P}_-\cdot d{\mathcal P}_-)$ which is precisely
Eq.(\ref{fact}).
Hence we have shown that the factorization of the metric on the kinematic space of $AdS_3$ is represented by the factorization of Plücker coordinates into self-dual and anti self-dual parts.

\subsection{Cross ratios and twistors in $d=2$.}

Let us finally consider cross ratios in the twistor setting.
We consider two causal diamonds $\diamond_p$ and $\diamond_q$ in the boundary, corresponding to two planes $p$ and $q$ 
in the bulk with principal null directions $(U,V)$ and $(W,Z)$.
In the helicity formalism\cite{Levay4} we parametrize the four null vectors $U,V,W,Z$ according to the pattern
$U=(L,u,\overline{u},u\overline{u}/L)^T$ and similarly for the remaining vectors $V,W$ and $Z$ containing the light cone coordinates 
$u,v,w$ and $z$. Here for example $u=t_u+x_u$ and $\bar{u}=t_u-x_u$.
In the boundary we would like to have spacelike separated diamonds hence we demand that the diamonds are intersecting in the following way
\begin{equation}
u<w<v<z,\qquad
\bar{u}<\bar{w}<\bar{v}<\bar{z}
\label{causalkity}
\end{equation}
In the bulk we denote the gauge invariant Plücker bivectors of the corresponding planes as
\begin{equation}
{\mathcal P}=\frac{U\wedge V}{U\cdot V},\qquad
{\mathcal Q}=\frac{W\wedge Z}{W\cdot Z}.
\label{pluck}
\end{equation}

We wish to study the physically important special case when the planes $p$ and $q$ in ${\mathbb R}^{2,2}$ intersect.
In the projective context, known from twistor theory, we know that this means that the corresponding lines $\ell_p$ and $\ell_q$ in ${\mathbb R}{\mathbb P}^3$ intersect. 
Intersection of planes is translated to the fact that the four vectors $U,V,W,Z$ are not linearly independent. In other words we have the vanishing of the determinant: $\varepsilon_{\alpha\beta\gamma\delta}U^{\alpha}V^{\beta}W^{\gamma}Z^{\delta}=0$. Now a calculation shows that
\begin{equation}
\langle{\mathcal P},{\mathcal Q}\rangle :=\frac{1}{2}\ast{\mathcal P}\cdot{\mathcal Q}=\frac{1}{4}{\varepsilon}_{\alpha\beta\gamma\delta}{\mathcal P}^{\alpha\beta}{\mathcal Q}^{\gamma\delta}=
\frac{\varepsilon_{\alpha\beta\gamma\delta}U^{\alpha}V^{\beta}W^{\gamma}Z^{\delta}}{(U\cdot V)(W\cdot Z)}=\kappa-\bar{\kappa}
\label{imaginary}
\end{equation}
where
\begin{equation}
\kappa:=\frac{(z-u)(v-w)}{(v-u)(z-w)},\qquad \bar{\kappa}:=\frac{(\bar{z}-\bar{u})(\bar{v}-\bar{w})}{(\bar{v}-\bar{u})(\bar{z}-\bar{w})} 
\end{equation}
Moreover, one can use Eq.(\ref{causalkity}) and the relation $(u-v)(w-z)-(u-w)(v-z)+(u-z)(v-w)=0$ to show that
\begin{equation}
0<\kappa,\bar{\kappa}<1
\label{kisebb}
\end{equation}
Hence in the bulk ${\mathbb R}{\mathbb P}^3$  the lines $\ell_p$ and $\ell_q$ intersect iff in the boundary (which is a quadric in ${\mathbb R}{\mathbb P}^3$) the constraint $\kappa=\bar{\kappa}$ holds. We say that in this case the cross ratio $\kappa$ satisfies a {\it reality} condition.
In our previous paper\cite{Levay4} we have clarified the physical meaning of this condition. 
We have proved that a sufficient and necessary condition for the four boundary points $(u,\bar{u}),(v,\bar{v}),(w,\bar{w}),(z,\bar{z})$ comprising events on the world lines of inertial observables or ones moving with constant acceleration (executing hyperbolic motion) is precisely the reality condition for $\kappa$.
The world lines in turn produce the flow lines for the modular flow of the embedding
diamond
which comprises the causal completion of the intersecting ones\cite{Levay4}.
Now we have learned that this physical situation in the boundary is represented by the geometric condition of intersecting lines in the bulk.

Moreover, $\mathcal P$ and $\mathcal Q$ having six components, define two rays in ${\mathbb R}^6$, i.e. two points in ${\mathbb R}{\mathbb P}^5$. However, since these objects are separable bivectors at the same time they also satisfy the Plücker relations hence they define points in the Klein quadric which is a four real dimensional quadric living inside 
${\mathbb R}{\mathbb P}^5$.
This gives rise to the basic twistor correspondence which relates lines in 
${\mathbb R}{\mathbb P}^3$ to points in 
${\mathbb R}{\mathbb P}^5$. Moreover, it is also well-known that two lines in
${\mathbb R}{\mathbb P}^3$ intersect iff the corresponding points in 
${\mathbb R}{\mathbb P}^5$ are light like separated\cite{Penrose,Ward}.
It is easy to understand this finding in terms of our kinematic space as an object embedded into the Klein quadric.
First note that using the definition of Eq.(\ref{imaginary})  and Eqs.(\ref{hodgepart})-(\ref{emn}) one has
\begin{equation}
\langle {\mathcal P}-{\mathcal Q},{\mathcal P}-{\mathcal Q}\rangle ={\mathcal P}_-\cdot{\mathcal Q}_--{\mathcal P}_+\cdot{\mathcal Q}_+= n_p\circ n_q-m_p\circ m_q=(n_p-n_q)^2-(m_p-m_q)^2
\end{equation}
where $a\circ b:=-a^0b^0+a^1b^1+a^2b^2$ and $(a-b)^2=(a-b)\circ (a-b)$. Moreover, according to Eq.(\ref{emn}) the pairs $(m_p,n_p)$ and $(m_q,n_q)$ are the six coordinates for the points ${\mathcal P}$ and ${\mathcal Q}$ on the Klein quadric subject to the (\ref{conifoldlike}) constraints.
We notice that $(n_p-n_q)^2$ and $(m_p-m_q)^2$ give the Minkowski-separation  between the projections to the corresponding de Sitter spaces taken from the product $dS_2\times dS_2$. Then light-like separation of ${\mathcal P}$ and ${\mathcal Q}$ means that 
$\langle {\mathcal P}-{\mathcal Q},{\mathcal P}-{\mathcal Q}\rangle =0$
which boils down to the constraint that the separations of the corresponding projections to the respective de Sitter spaces are the same.
A naive illustration of this result is as follows. We have ${\mathcal P}={\mathcal P}_++{\mathcal P}_-$
and 
${\mathcal Q}={\mathcal Q}_++{\mathcal Q}_-$ and the two components of this decomposition are  projections to two "orthogonal" $2D$ de Sitter spaces. The fact that the distances of these "orthogonal" projections being the same is the hallmark of the two points residing on a light-cone of the Klein quadric. This reasoning also gives another interpretation of the reality condition $\kappa=\bar{\kappa}$.

As a reiteration of these results we would like to demonstrate that in the case when in the bulk $\ell_p$ and $\ell_q$ intersect and in the boundary the reality condition holds one can rewrite the geodesic distance
$d(\diamond_p,\diamond_q)$ in the form
\begin{equation}
\sin^2\frac{d(\diamond_p,\diamond_q)}{2\ell}=-\frac{1}{8}({\mathcal P}-{\mathcal Q})\cdot ({\mathcal P}-{\mathcal Q})
\label{geoddistpluck}
\end{equation} 
This formula expresses the distance of boundary causal diamonds in the bulk language in a gauge invariant form.

In order to show this first recall that from the helicity formalism we know\cite{Levay4} that for the calculation of quantities like $U\cdot V$ the metric 
\begin{equation}
g=
\varepsilon\otimes \varepsilon=\begin{pmatrix}0&1\\-1&0\end{pmatrix}
\otimes
\begin{pmatrix}0&1\\-1&0\end{pmatrix}.
\nonumber
\end{equation}
has to be used.
Then for example we have $U\cdot V=(u-v)(\bar{u}-\bar{v})$.
Now $\mathcal P\cdot\mathcal P=\mathcal Q\cdot\mathcal Q=-2$ and $\mathcal P\cdot\mathcal Q=2(U\cdot W)(V\cdot Z)-2(U\cdot Z)(V\cdot W)$.
Then one can use the relation $(u-v)(w-z)-(u-w)(v-z)+(u-z)(v-w)=0$ again to arrive at
\begin{equation}
-\frac{1}{8}({\mathcal P}-{\mathcal Q})\cdot ({\mathcal P}-{\mathcal Q})=1-\frac{1}{2}\left[
\frac{(z-u)(v-w)}{(v-u)(z-w)}+
\frac{(\bar{z}-\bar{u})(\bar{v}-\bar{w})}{(\bar{v}-\bar{u})(\bar{z}-\bar{w})}\right]:=1-\frac{1}{2}[\kappa +\bar{\kappa}]
\label{resszam}
\end{equation}
Moreover, using again this relation and the (\ref{causalkity}) one and then (\ref{kisebb}) we see that the left hand side of Eq.(\ref{resszam}) is positive.

Let us now recall formula
of Eq. (2.26) of Ref.\cite{Myers} 
\begin{equation}
d(\diamond_p,\diamond_q)=\ell\cos^{-1}\left( 2\sqrt{r(x_p,y_p;x_q,y_q)}-1\right),\qquad 0\leq r\leq 1
\label{mymyers}
\end{equation}
where the cross ratio is
\begin{equation}
r(x_p,y_p;x_q,y_q)=\frac{(y_q-x_p)^2(y_p-x_q)^2}{(y_p-x_p)^2(y_q-x_q)^2}
\end{equation}
where $(x_p,y_p)$ and $(x_q,y_q)$ are the past and future tips of the diamonds $\diamond_p$ and $\diamond_q$.
For $d=2$ this cross ratio boils down to
\begin{equation}
r(x_p,y_p;x_q,y_q)=\frac{(z-u)(v-w)(\bar{z}-\bar{u})(\bar{v}-\bar{w})}{(v-u)(z-w)(\bar{v}-\bar{u})(\bar{z}-\bar{w})}=\kappa\bar{\kappa}
\end{equation}
Now since in our physical situation the reality condition holds we have $r=\kappa^2$.
Using this in (\ref{mymyers}) we get
\begin{equation}
\kappa=\cos^2\left(\frac{d(\diamond_p,\diamond_q)}{2\ell}\right)
\label{kappacska}
\end{equation}
Using again the reality condition this time in the right hand side of (\ref{resszam}) yields $1-\kappa$. Combining this with (\ref{kappacska}) yields the result of (\ref{geoddistpluck}) as promised.

Notice finally that generally we have
\begin{equation}
\frac{1}{2}{\mathcal P}\cdot{\mathcal Q}=1-(\kappa +\bar{\kappa})
\label{realpart}
\end{equation}
One can compare this with Eq.(\ref{imaginary}). One can then see that (\ref{realpart}) is featuring the "real part" and
(\ref{imaginary}) the ǐmaginary part" of the cross ratio $\kappa$.

\subsection{The static slice}

Let us now define the {\it static slice} of $\widetilde{AdS}_3$ as the locus
\beq
{\mathbb H}\coloneqq\{X\in \widetilde{AdS}_3\vert X^0=0\}.
\label{static}
\eeq
We also define the light cone  ${\mathcal C}$
as 
\begin{equation}
    {\mathcal C}:=\{U\in{\mathbb R}^{2,1}\vert U\cdot U=0\}
\label{conci}
\end{equation}
Clearly in the Poincaré patch we have $X^{-1}\geq 0$ moreover, due to (\ref{AdS}) we have $X^{-1}\geq L$. Then the space ${\mathbb H}$ is the upper sheet of the two dimensional double sheeted hyperboloid embedded in
${\mathbb R}^{2,1}$.
Notice also that for arbitrary three component vectors arising from $A^{a}$ with $A^{0}=0$ the constraint $A\cdot A<0$ says that this three-vector is {\it time like} in ${\mathbb R}^{2,1}$.
Hence for two linearly independent vectors $X,Y\in {\mathbb H}$ we have $X\cdot X=Y\cdot Y=-L^2$ hence these are time-like and spanning a plane.
Moreover, since they have the same time orientation (both of them are lying on the upper sheet) then $X\cdot Y<0$.
Moreover, from the reversed Cauchy-Schwarz inequality\cite{Greg} we have
\beq
\vert X\cdot Y\vert\geq (X\cdot X)(Y\cdot Y)=L^4
\label{reversed}
\eeq
hence since $X$ and $Y$ are linearly independent we get from Eq.(\ref{Cayley})
\beq
D=(X\cdot Y)^2-L^4>0.
\label{geodcond}
\eeq
From Eq.(\ref{timelikeplane}) it then follows that
the vectors $\tilde{X}$ and $\tilde{Y}$ are spacelike and have the property $\tilde{X}\cdot \tilde{Y}>0$.
Notice moreover, that in this case
\beq
\frac{\tilde{X}}{\sqrt{D}},
\frac{\tilde{Y}}{\sqrt{D}}
\in dS_2
\label{kinematic}
\eeq
where
\beq
dS_2\coloneqq\{Y\in{\mathbb R}^{2,1}\vert Y\cdot Y=L^2\}
\label{desitter}
\eeq
is the two dimensional de Sitter space. We have already seen that a copy of $dS_2$ is the kinematic space in the static case\cite{Myers,Czech}.
Then the vectors $X$ and $\tilde{X}$ are spanning a timelike plane with its principal null directions  $(U,V)$.

\subsection{Projective representations of the static slice}

For $d=2$ the structure of the kinematic space ${\mathbb K}$ has been discussed in detail in Ref.\cite{Czech}. 
There the kinematic space was discussed as a space of geodesics in the static slice.
Here we would like to see how this approach fits into our projective geometric considerations. 

We have defined the static slice of $AdS_3$ space in \eqref{static}. We have seen that the spacelike geodesics of the static slice are given by the equation
\begin{equation}
    X(\lambda)=\frac{L}{\sqrt{-2U\cdot V}}\left(Ue^{\sqrt{C}\lambda}+Ve^{-\sqrt{C}\lambda}\right)
\end{equation}
Where $X(\lambda)\in \mathbb{H}$, and $U,V\in\mathcal{C}$ see Eq.(\ref{conci}) , $U\cdot V<0$ and $C>0$ is an arbitrary real positive parameter.

\begin{figure}[!t]
    \centering
    \includegraphics[width=0.4\textwidth]{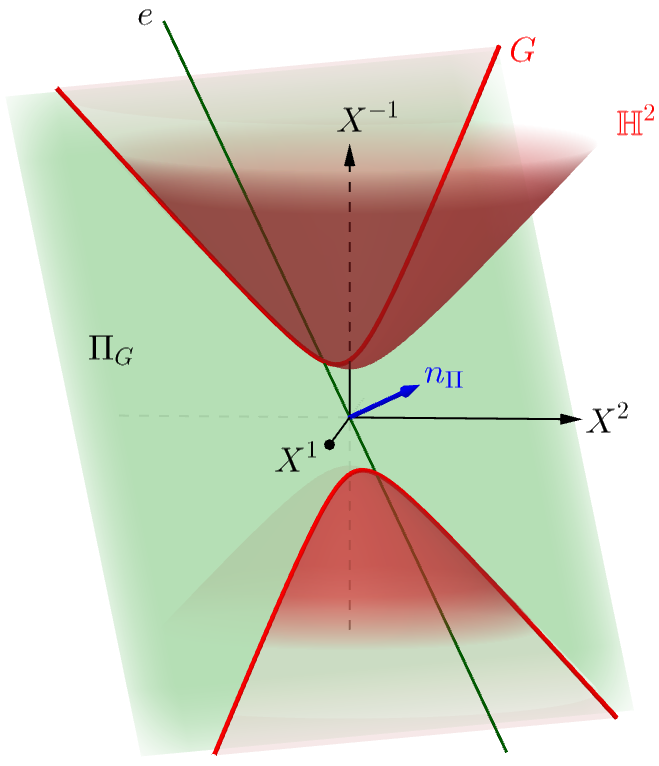}
    \caption{A geodesic $G$ of a static slice $\mathbb{H}^2$ of the $AdS_3$ space is given by the intersection of a plane $\Pi_G$ of the embedding space and $\mathbb{H}$ itself. The planes that define geodesics on $\mathbb{H}$ have spacelike normal vectors $n_\Pi$}.
    \label{fig:plane_geod}
\end{figure}

A particular geodesic can be realized as the intersection of $\mathbb{H}$ with a plane through the origin embedded in $\mathbb{R}^{2,1}$  spanned by the corresponding $U$ and $V$ null vectors. Consider a $G\subset \mathbb{H}$ geodesic and the plane $\Pi$ that cuts out $G$ from $\mathbb{H}$. Notice that for $X\in \mathbb{H}$ we have $X\cdot X<0$. So there are points in $X\in\Pi$ such that $X\cdot X<0$, which means the planes that define $\mathbb{H}$ geodesics are the ones with spacelike normal vectors $n_{\Pi}$ (see Figure 2). This means that there is a one-to-one correspondence between such planes in $\mathbb{R}^{2,1}$ and geodesics of $\mathbb{H}$.

It is also important to note that each timelike line $e$ (those who have timelike tangent vectors) that lies in $\Pi$ and goes through the origin intersects $\mathbb{H}$ exactly once. So the points of $G$ can be represented by timelike lines of $\mathbb{R}^{2,1}$. These properties give us an opportunity to have a projective geometric representation of $\mathbb{H}$ geodesics.

Now consider ordinary $\mathbb{R}^3$. The space of all $\mathbb{R}^{3}$ lines that go through the origin is the projective space $\mathbb{RP}^2$. We can represent any $e\subset\mathbb{R}^{3}$ line by any of its tangent vectors, so a suitable choice to visualize $\mathbb{RP}^2$ is to use the set of unit vectors  $\{n_e\}$. But the space of unit vectors in $\mathbb{R}^{3}$ is obviously the two dimensional unit sphere $S_2$. Hovewer, the antipodal points $n_e$ and $-n_e$ defines the same line so the same point in $\mathbb{RP}^2$. This leads us to the fact that $\mathbb{RP}^2\cong S^2/\mathbb{Z}_2$

Similarly the space of all $\mathbb{R}^{3}$ planes that go through the origin is the Grassmannian $Gr_{2,3}$. Each point in $Gr_{2,3}$ represents a plane in $\mathbb{R}^{3}$. 
In the projective picture one can alternatively consider the space ${\mathbb G}r_{1,2}$ which is the set of lines of ${\mathbb R}{\mathbb P}^2$.
By projective duality there is a correspondence between points and lines in
${\mathbb R}{\mathbb P}^2$ hence we expect that 
${\mathbb G}r_{1,2}$ gives rise to another copy of 
${\mathbb R}{\mathbb P}^2$.
Indeed, notice that in $\mathbb{R}^{3}$ any plane $\Pi$ can be uniquely determined by its unit (Euclidean) normal vector $n_\Pi$ (and by $-n_\Pi$ as well) through the origin. These normal vectors also uniquely define lines that are paralell to them and are going through the origin. This means that ${\mathbb G}r_{1,2}\cong\mathbb{RP}^2\cong S^2/\mathbb{Z}_2$.
This is the dual copy of the real projective plane.

The next step is to take into account that we are only interested in structures in $\mathbb{H}$. Moreover, we have to use the fact that there is an associated causal structure defined by our bilinear form of index one on ${\mathbb R}^{3}$ rendering it to ${\mathbb R}^{2,1}$. Then one can refine the picture above as follows.

 Take a point $X\in{\mathbb R}^{2,1}$. One can introduce homogeneous coordinates $[X]\in{\mathbb R}{\mathbb P}^2$  to represent a projective line going through $X$ and the origin.  By the causal structure of $\mathbb{R}^{3}$ we can split $\mathbb{RP}^2$ into three parts in the following manner (see Figure 3):
\begin{equation}\label{eq:causal}
    \begin{aligned}
        \mathbb{RP}^2_+&=\{[X]|X\in\mathbb{R}^{2,1} \land X\cdot X>0\}\\
        \mathbb{RP}^2_0&=\{[X]|X\in\mathbb{R}^{2,1} \land X\cdot X=0\}\\
        \mathbb{RP}^2_-&=\{[X]|X\in\mathbb{R}^{2,1} \land X\cdot X<0\}
    \end{aligned}
\end{equation}

\begin{figure}[!t]
    \centering
    \includegraphics[width=0.45\textwidth]{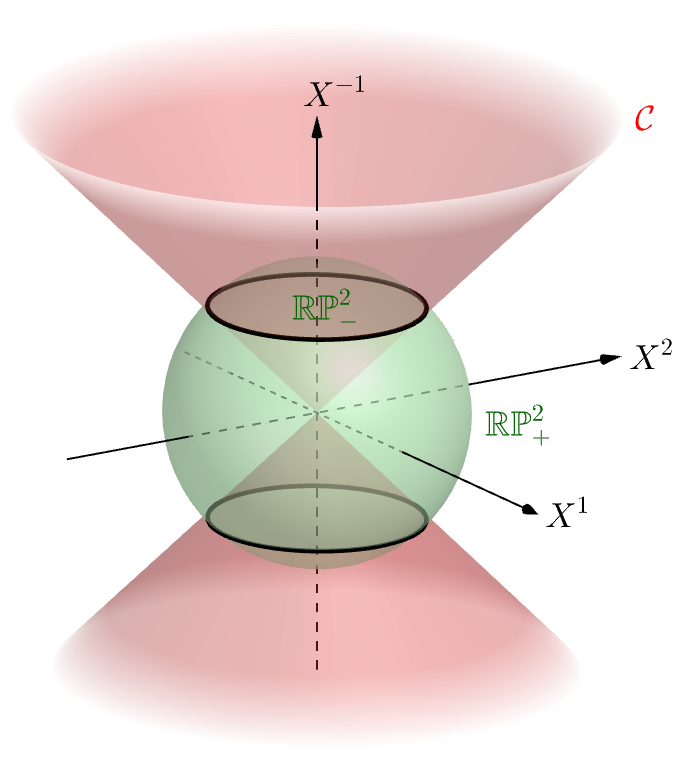}
    \caption{The lightcone $\mathcal{C}$ splits the space $\mathbb{R}^{2,1}$ into three different regions: a spacelike (outside the cone), a timelike (inside the cone) and a lightlike (the cone itself). By introducing homogeneous coordinates, one can also assign different parts of the projective space $\mathbb{RP}^2$ (represented by a green sphere) to these three regions.}
    \label{fig:proj_sphere}
\end{figure}
\noindent
As we have discussed the points of $\mathbb{H}$ are uniquely determined by timelike lines of $\mathbb{R}^{2,1}$. This means that the projective space of $\mathbb{H}$ is $\mathbb{PR}^2_-$. However, the geodesics of $\mathbb{H}$ are cut out by $\mathbb{R}^{2,1}$ planes with spacelike normal vectors. These normal vectors are comprising the space $\mathbb{RP}^2\cong Gr_{2,3}$ in different points, each one uniquely determining a plane in $\mathbb{R}^{2,1}$. But only spacelike vectors are considerable so we can say that the Grassmannian of $\mathbb{H}^2$ (more precisely in the static case the space of geodesics) is $\mathbb{RP}^2_+$. As we see planes define geodesics in $\mathbb{H}^2$ whose images are curves in $\mathbb{PR}^2_-$ or points in $\mathbb{PR}^2_+$ (see Figure 4).

\begin{figure*}[!t]
    \centering
    \begin{subfigure}[b]{0.405\textwidth}
        \centering
        \includegraphics[width=\textwidth]{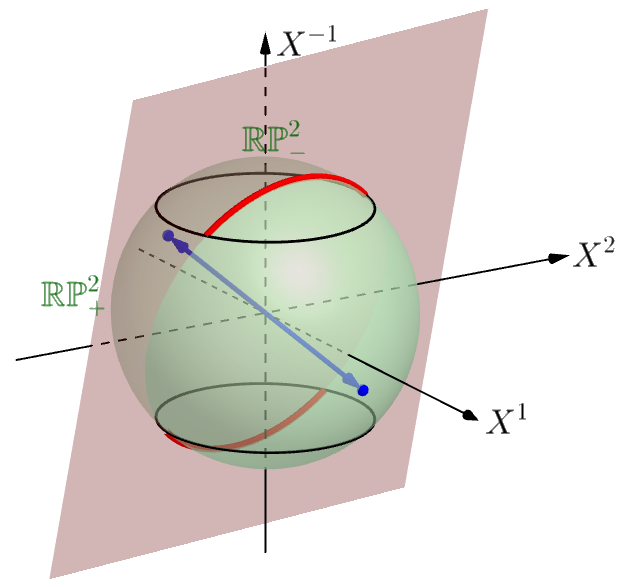}
    \end{subfigure}
    \hfill
    \begin{subfigure}[b]{0.425\textwidth}  
        \centering
        \includegraphics[width=\textwidth]{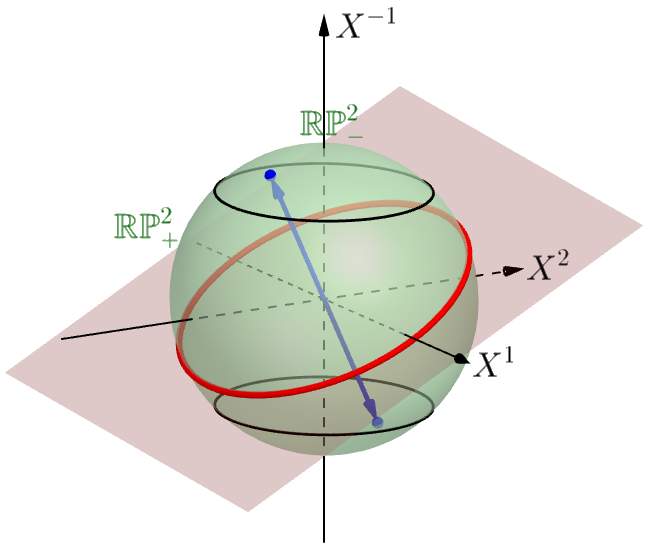}
    \end{subfigure}
    \vskip\baselineskip
    \begin{subfigure}[b]{0.405\textwidth}   
        \centering 
        \includegraphics[width=\textwidth]{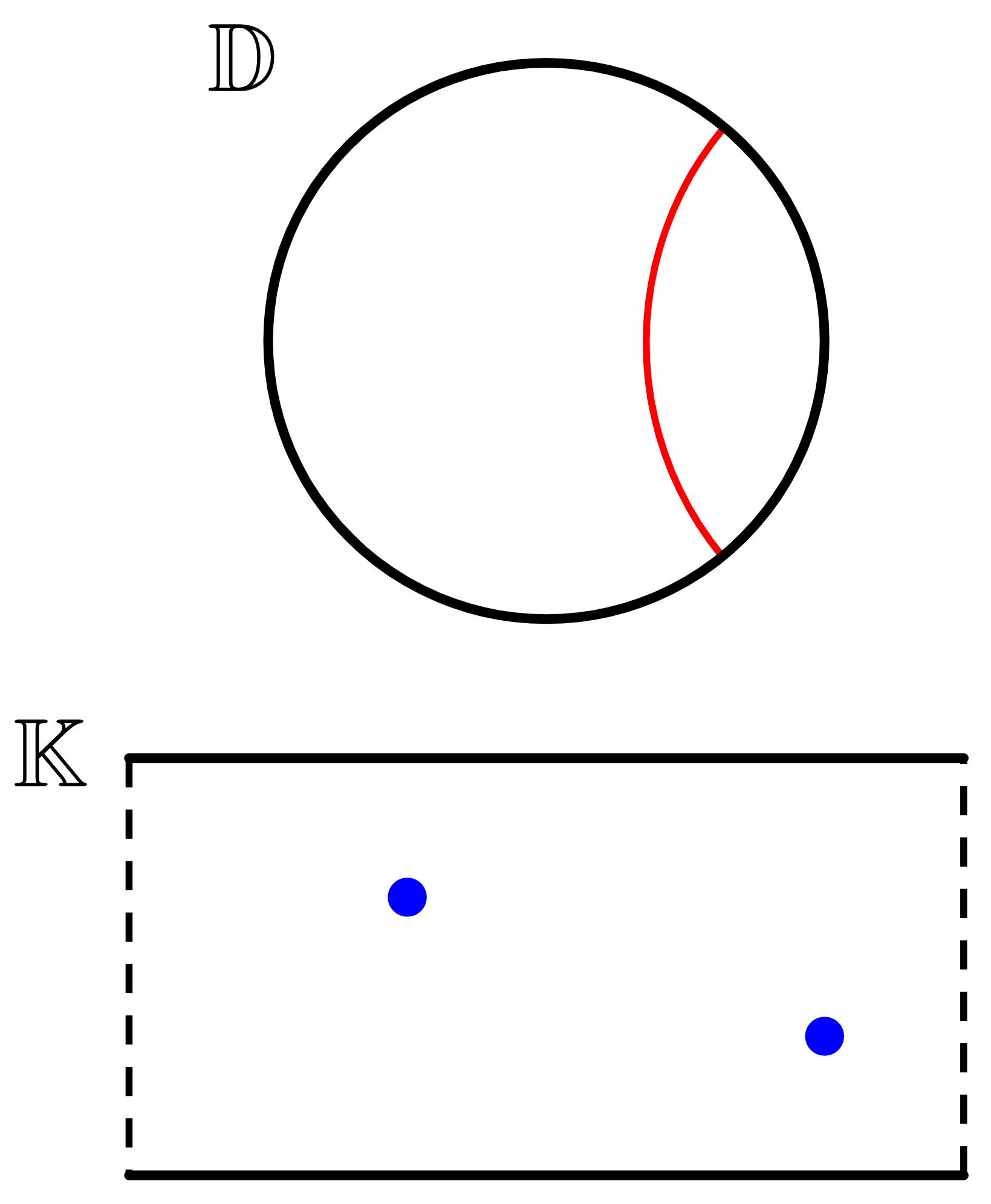}
    \end{subfigure}
    \hfill
    \begin{subfigure}[b]{0.405\textwidth}   
        \centering 
        \includegraphics[width=\textwidth]{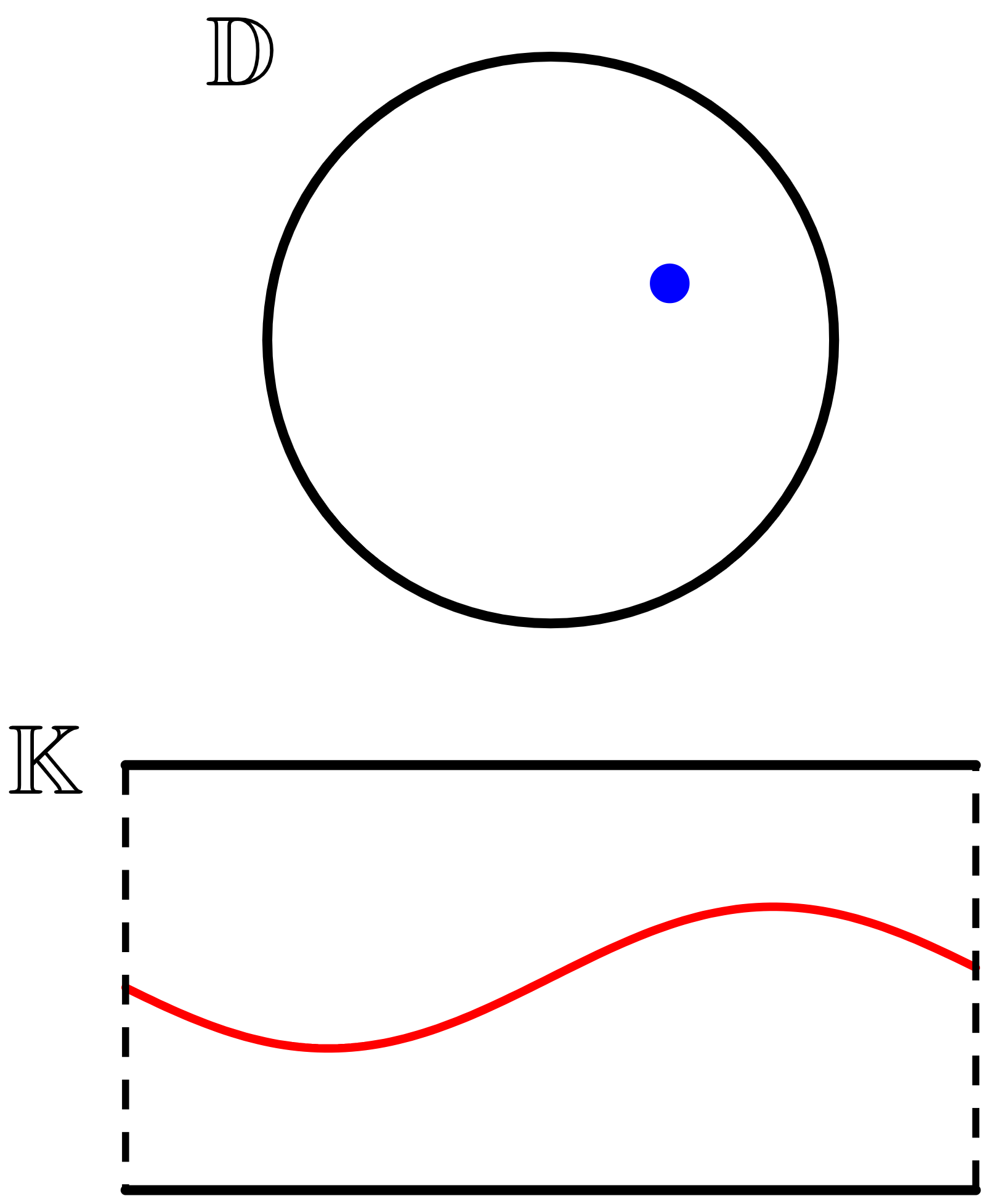}
    \end{subfigure}
    \caption{The static slice of the $AdS_3$ space $\mathbb{D}$ can be represented by $\mathbb{RP}^2_-$, while its kinematic space is $\mathbb{K}\cong \mathbb{RP}^2_+$. On the uppermost figures, the two spaces are illustrated at once on the projective sphere embedded into $\mathbb{R}^{2,1}$. On the left-hand side we show that a geodesic on $\mathbb{D}$ is dual to a point in $\mathbb{K}$ which can be easily seen by taking the normal vector of the defining plane of the geodesic. On the right-hand side we also illustrate the opposite. In the middle and the bottom of the figure this phenomena is shown from the points of view of $\mathbb{D}$ and $\mathbb{K}$.}
    \label{fig:kinematic_disk}
\end{figure*}

We can do the same argument in the other direction. Points of $\mathbb{PR}^2_-$ uniquely determine perpendicular planes and their internal lines define curves in $\mathbb{PR}^2_+$. What we got so far is that the projective space of $\mathbb{H}^2$ and the space of its geodesics are different parts of the same object, namely $\mathbb{RP}^2$. They are separated from each other by the causal structure of the embedding space $\mathbb{R}^{2,1}$. 

The Poincaré disk $\mathbb{D}$ by definition is the stereographic projection of $\mathbb{H}^2$ to the unit disk through the point $(-1,0,0)$. This is the same how we have constructed $\mathbb{RP}^2_-$. The only difference is that in the second case we projected $\mathbb{H}^2$ through the origin. Therefore we can say that $\mathbb H\cong\mathbb{D}\cong \mathbb{RP}^2_-$.
On the other hand the space of geodesics is just our kinematic space $\mathbb{K}$. By construction its points represent geodesics of the hyperboloid and its curves represent points on $\mathbb{H}$. This is again in complete accordance with the properties of $\mathbb{RP}^2_+$. Therefore $\mathbb{K}\cong \mathbb{RP}^2_+$.
    
The main conclusion is that the projective geometric viewpoint could be a useful tool to examine the objects of ${\rm AdS}_3$ space in a unified geometric manner. In the static slice we have found that the Poincaré disk and the kinematic space (both are important concepts in their own right) can both be considered as certain parts of a more general mathematical object, namely the projective plane providing a unified perspective.

\subsection{Entanglement and lambda length in the static case}

Let us consider now the static slice as modelled by ${\mathbb H}$ the upper sheet of the double sheeted hyperboloid. After stereographic projection it can be mapped to the Poincaré disc ${\mathbb D}$, and after a fractional linear transformation to the Poincaré upper half plane with coordinates $(z,x)$ where $z>0$. 
This is arising as the static slice i.e. the $t=0$ slice in the Poincaré patch.
The boundary is then obtained by taking the $z=0$ limit. 

Let us now consider a region ${\mathcal R}$ in the boundary which is an interval $[x_L,x_R]$ in the $x$-axis with left and right end points.
If we use the parametrization of causal diamonds of the previous two subsections one can then embed ${\mathcal R}$ inside a diamond as follows.
For $t>0$ the coordinates of the past and future tips of the diamond in the boundary ${\mathbb R}^{1,1}$ are:
$x^{\mu}_u=(t_u,x_u)=(-t,x)$ and $x^{\mu}_v=(t_v,x_v)=(t,x)$. Then the left and right end points of ${\mathcal R}$ are:  $x_L^{\mu}=(0,x-t)$ and 
$x_R^{\mu}=(0,x+t)$. Then $x_L=x-t$ and $x_R=x+t$. 

It is well-known that in a two dimensional conformal field theory $CFT_2$ with central charge $c$, the
entanglement entropy of the degrees of freedom lying in ${\mathcal R}$ with the ones that are lying in the complement of ${\mathcal R}$ is given by the formula\cite{calabrese}
\beq
S({\mathcal R})=\frac{c}{3}\log
\frac{\vert x_R-x_L\vert}{\delta}
\label{lambda2}
\eeq
The entanglement entropy is a divergent quantity hence we have to regularize it via introducing a cutoff $\delta$. 
Since $x_R-x_L=2t=t_v-t_u$ and according to Eq.(\ref{kepike}) we have $t_u=LU^0/U^-$ and $t_v=LV^0/V^-$
then one can write this as
\beq
S({\mathcal R})=\frac{c}{3}\log
\frac{ t_v-t_u}{\delta}
=\frac{c}{3}\log
\frac{L}{\delta}\left[\frac{U^0}{U^-}-\frac{V^0}{V^-}\right]
\label{lambda3}
\eeq

On the other hand we know from (\ref{majdnemrt}) that a formula for the  (\ref{glength}) horocycle regularized geodesic length $\ell_{\rm reg}(X_{\mathcal R})$ of a bulk geodesic $X_{\mathcal R}$ anchored to $\mathcal R$ can be defined.
By (\ref{regike}) this regularized length is
\beq
\ell_{\rm reg}(X_{\mathcal R})=L\log\frac{\Delta x\bullet\Delta x}{2\Delta_u\Delta_v}=L\log 
\left(\frac{(t_v-t_u)^2}{-2\Delta_v\Delta_u}\right)
=L\log\frac{\vert U\cdot V\vert}{L^2},\qquad \Delta x^{\mu}=x_v^{\mu}-x_u^{\mu}
\label{majdnemrt1}
\eeq
Note that in order to make sense for this expression the constraint of Eq.(\ref{alt2}) should hold.

Now according to the Ryu-Takayanagi (RT) prescription\cite{RT1,RT2,RT3}
this regularized spacelike geodesic can be regarded as a regularized extremal surface\footnote{In the static case this surface is a minimal one.} $X_{\mathcal R}$ in the bulk homologous to the region $\mathcal R$ in the boundary. Moreover, the length $\ell_{\rm reg}(X_{\mathcal R})$ is at the same time just the minimal area   ${\mathcal A}(X_{\mathcal R})$ of this "surface"\footnote{In the RT prescription one uses codimension two surfaces. For $d+1=3$ a codimension two object is one dimensional i.e. a curve.} hence ${\mathcal A}(X_{\mathcal R})=\ell_{\rm reg}(X_{\mathcal R})$ and 
\beq
S({\mathcal {\mathcal R})={\mathcal A}(X_{\mathcal R}})/4G\label{RT}
\eeq
Here when relating bulk and boundary constants we have to use the Brown-Henneaux\cite{BH} relation
$c=\frac{3L}{2G}$.
Comparing now (\ref{lambda3}) and (\ref{majdnemrt1}) we see that we are in line with (\ref{RT}) provided $\delta =\sqrt{-2\Delta_v\Delta_u}$. This formula then via RT gives a geometric meaning for the cutoff $\delta$.

Had we chosen a different parametrization for our causal diamonds we would have obtained a picture which is even more geometrical.
Indeed, if we happened to be relating  $U$ and $V$ to the left and right tips of our causal diamond by a formula similar to (\ref{kep2}) we would have run into alternative tips with a {\it spacelike} separation vector $\Delta x^{\mu}$ rather than a {\it timelike} one. In this case the constraint (\ref{alt1}) has to be used. Then one can introduce quantities $\Delta_L$ and $\Delta_R$ that can both be chosen positive.
Then one can see that the regularization process for the divergent entanglement entropy is connected to the regularized geodesic length via Penner's lambda length\cite{Penner1,Penner2} prescription.
Indeed, the lambda length of a geodesic in this case is given by the formula
\begin{equation}
    \lambda (X_{\mathcal R})=\frac{\vert x_R-x_L\vert}{\sqrt{\Delta_R\Delta_L}}
\end{equation}
where $\Delta_R$ and $\Delta_L$ are the Euclidean diameters of the regularizing horocycles.
The lambda length is related to the geodesic length\cite{Penner1,Penner2,Levay1}
as 
\begin{equation}
\ell(X_{\mathcal R})={\mathcal A}(X_{\mathcal R})=2\log\lambda(X_{\mathcal R})
\end{equation}
Comparing this with Eq.(\ref{lambda2}) gives for the cutoff
$\delta=\sqrt{\Delta_R\Delta_L}$.
Hence in this case the cutoff parameter $\delta$ is geometrized as the geometric mean of the diameters of the regularizing horocycles.

Finally we elaborate on the regularization process for the static case. Here the boundary of the horosphere is merely a point namely $x_u$. 
Generally however, the boundary of a horosphere ${\mathcal S}_u$ is a light cone in the boundary\cite{Seppi}, i.e. in the quadric ${\mathcal Q}_d$ of Eq.(\ref{bdy}). 
In order to see this let us consider the special case when $x_u=0$. In this case for the choice $U^{\mu}=0$  and $U^d=-U^{-1}=L^2/{\sqrt{2}}$ we have $U\cdot U=0$ and from $X\cdot U=-L^2/{\sqrt{2}}$ the result $X^{+}=X^d+X^{-1}=-1$ follows. Then from (\ref{embed}) we get $(x^2+z^2)/z=1$ i.e. our "horosphere" is $-t^2+\vert\vert{\bf x}\vert\vert^2+(z-1/2)^2=1/4$. Clearly this is indeed a sphere for $t=0$ (static slice). In the boundary ($z=0$) on the other hand one obtains a light cone $-t^2+\vert\vert{\bf x}\vert\vert^2=0$ as promised.

\begin{figure}[!t]
    \centering
    \includegraphics[width=0.65\textwidth]{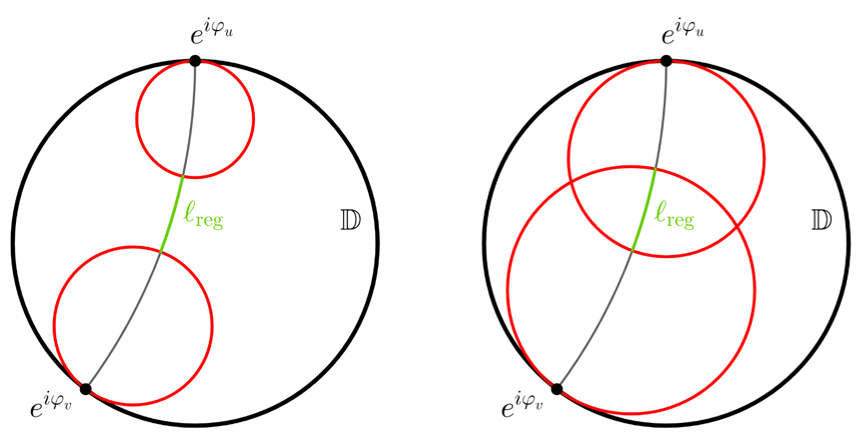}    
\caption{Horocycles regularizing the length of a geodesic in the Poincaré disc ${\mathbb D}$ arising as a static slice of $\widetilde{\rm AdS}_{3}$. The two different cases correspond to either choosing horocycles based on the pair $(U_+,V_-)$ (left), or $(V_+,U_-)$ (right). In accord with Eq.(\ref{glength}) in the first case the regularized length is positive, in the second it is negative.}
\label{horocycles}
\end{figure}

Let us gain some further insight visualized in Figure 5. Here the static slice is represented by the Ploincaré disc ${\mathbb D}$. Now a geodesic is anchored to the points $e^{\varphi_u}$ and $e^{i\varphi_v}$ lying on the circular boundary of the disc.  These boundary points represent the light rays $[U]$ and $[V]$ belonging to the $X^{0}=0$ slice of the quadric ${\mathcal Q}_2$ given by the projectivization of the equation $-(X^{-1})^2-(X^0)^2+(X^1)^2+(X^2)^2=0$.
Different representatives of $[U]$ and $[V]$ give rise to different horocycles. Two possible choices are given by the the pairs $(U_+,V_-)$ and $(V_+,U_-)$. Via Eq. (\ref{pennerconv}) in the Poincaré model they yield two pairs of circles tangent to the boundary at the points $e^{i\varphi_u}$ and $e^{i\varphi_v}$. These are our horocycles. The two pairs of cycles give rise to two different regularizators for the geodesic lengths. 
These cases showing up in the regularized length formula of Eq.(\ref{glength}) are displaying different sign combinations. They correspond to the situations where the horocycles are nonintersecting or intersecting. As discussed in detail in Refs.\cite{Penner2,Levay1} for these different cases the regularized geodesic length is positive or negative.

\subsection{Strings in $AdS_3$}\label{sec:AdS3Strings}
In the following we use our previous results to determine the equations of motion for boundary causal diamond coordinates in case of $AdS_3$. During the following two sections, in order to facilitate a direct comparison with the expressions obtained in \cite{Dv3} instead of our usual (\ref{kepike}) notation for the past and future tips of our diamond we revert to the notation of that paper. Namely, we will use the null coordinates $(b,\tilde{b})$ and $(w,\tilde{w})$ for the past and future tips $x^\mu$ and $y^\mu$ in the following way
\begin{align}\label{eq:continuous_null}
    b=x^0+x^1&&\tilde{b}=x^0-x^1\\
    w=y^0+y^1&&\tilde{w}=y^0-y^1
\end{align}
 Our aim is to check whether our general results give back the equations derived in \cite{Dv3}. The causality $\partial_+x\bullet\partial_+x,\partial_-y\bullet\partial_-y<0$ in terms of the null coordinates translates to $\partial_+ b\partial_+\tilde{b},\partial_-w\partial_-\tilde{w}>0$. Furthermore if we require that the $y$ tip of the causal diamond is in the timelike future of the $x$ tip, namely that $(x-y)\bullet(x-y)<0$ and $x^0<y^0$, then it implies that $w>b$ and $\tilde{w}>\tilde{b}$.

First we determine the action, and from its variation the equations of motion. If we use the coordinates $(b,\tilde{b})$ and $(w,\tilde{w})$, the action becomes
\begin{equation}
    \mathcal{S}\sim \int d\sigma^+ d\sigma^-\sqrt{-h}h^{ij}\left(\frac{\partial_{(i} b\partial_{j)} w}{(b-w)^2}+\frac{\partial_{(i} \tilde{b}\partial_{j)} \tilde{w}}{(\tilde{b}-\tilde{w})^2}\right)
\end{equation}
Where we have dropped the constant factors. From the variations of the action with respect to the fields $b,\tilde{b},w\tilde{w}$, we get the equation of motion
\begin{equation}
    \partial_{+-}b=2\frac{\partial_-b\partial_+b}{b-w},\qquad \partial_{+-}w=-2\frac{\partial_-w\partial_+w}{b-w}
\end{equation}
And the same with tilde. The Virasoro constraints \eqref{eq:vir1} and \eqref{eq:vir2} in terms of null fields become
\begin{equation}\label{eq:2eqmotion1}
    \frac{\partial_-b\partial_-w}{(b-w)^2}=-\frac{\partial_-\tilde{b}\partial_-\tilde{w}}{(\tilde{b}-\tilde{w})^2},\qquad \frac{\partial_+b\partial_+w}{(b-w)^2}=-\frac{\partial_+\tilde{b}\partial_+\tilde{w}}{(\tilde{b}-\tilde{w})^2}
\end{equation}
From equation \eqref{eq:nullcond} we get 
\begin{equation}\label{eq:2eqmotion2}
    \partial_-b\partial_+b=-\partial_-\tilde{b}\partial_+\tilde{b},\qquad \partial_-w\partial_+w=-\partial_-\tilde{w}\partial_+\tilde{w}
\end{equation}
From the previous two results it immediately follows that
\begin{equation}\label{eq:contbw}
    \frac{\partial_+b\partial_-w}{(b-w)^2}=\frac{\partial_+\tilde{b}\partial_-\tilde{w}}{(\tilde{b}-\tilde{w})^2},\qquad \frac{\partial_-b\partial_+w}{(b-w)^2}=\frac{\partial_-\tilde{b}\partial_+\tilde{w}}{(\tilde{b}-\tilde{w})^2}
\end{equation}
These are exatly the same equations that was derived in \cite{Dv3} from the segmented approximation of bulk strings.

In case of $d=2$, we can explicitly express all the quantites $\alpha,\lambda,\chi,u,v$ and the vectors $X_-,X_+,X,N$ in terms of the fields $b,\tilde{b},w,\tilde{w}$ as well. First we use the ansatze \eqref{eq:xm} and \eqref{eq:xp} for $X_+$ and $X_-$ in terms of the null fields
\begin{align}
    X_-&=-\frac{e^\lambda}{\sqrt{2}L R}\left(\frac{-L^2+b\tilde{b}}{2},\frac{L}{2}(b+\tilde{b}),\frac{L}{2}(b-\tilde{b}),\frac{L^2+b\tilde{b}}{2}\right)\\
    X_+&=\frac{e^\chi}{\sqrt{2}L R}\left(\frac{-L^2+w\tilde{w}}{2},\frac{L}{2}(w+\tilde{w}),\frac{L}{2}(w-\tilde{w}),\frac{L^2+w\tilde{w}}{2}\right)
\end{align}
Where $R^2=\frac{1}{4}(b-w)(\tilde{b}-\tilde{w})$. If we differentiate these expressions with respect to $\sigma^+$ and $\sigma^-$ respectively, and use \eqref{eq:eqmotion1}, after calculating the inner product $X_{+-}\cdot X_{+-}$ in both cases, it can be shown that
\begin{align}
    e^\alpha&=\pm \frac{L e^\lambda}{\sqrt{2}R}\sqrt{\partial_+ b\partial_+\tilde{b}}\\
    e^\alpha&=\pm \frac{L e^\chi}{\sqrt{2} R}\sqrt{\partial_- w\partial_-\tilde{w}}
\end{align}
However from the definition of $\alpha$, we also know that $e^\alpha=-X_+\cdot X_-$ (or equivalently $\alpha=\lambda+\chi$). Therefore it immediately follows that
\begin{equation}\label{eq:scalars}
\begin{aligned}
    e^\alpha&=2L^2\frac{\sqrt{\partial_+b\partial_+\tilde{b}\partial_-w\partial_-\tilde{w}}}{(b-w)(\tilde{b}-\tilde{w})},\\e^\lambda&=\pm\sqrt{2}L\sqrt{\frac{\partial_- w\partial_-\tilde{w}}{(b-w)(\tilde{b}-\tilde{w})}},\\ 
    e^\chi&=\pm\sqrt{2}L\sqrt{\frac{\partial_+ b\partial_+\tilde{b}}{(b-w)(\tilde{b}-\tilde{w})}}
\end{aligned}
\end{equation}
Using these expressions the derivative $X_{+-}$ can be easily determined
\begin{equation}\label{eq:xpm_bw}
    \begin{aligned}
        X_{+-}=&\frac{e^\lambda}{\sqrt{2} L R}\frac{\partial_+ b}{b-w}\left(\frac{-L^2+w\tilde{b}}{2},\frac{L}{2}(w+\tilde{b}),\frac{L}{2}(w-\tilde{b}),\frac{L^2+w\tilde{b}}{2}\right)+\\
        &\frac{e^\lambda}{\sqrt{2} L R}\frac{\partial_+ \tilde{b}}{\tilde{b}-\tilde{w}}\left(\frac{-L^2+b\tilde{w}}{2},\frac{L}{2}(b+\tilde{w}),\frac{L}{2}(b-\tilde{w}),\frac{L^2+b\tilde{w}}{2}\right)=\\
        =&\frac{e^\chi}{\sqrt{2} L R}\frac{\partial_- \tilde{w}}{\tilde{b}-\tilde{w}}\left(\frac{-L^2+w\tilde{b}}{2},\frac{L}{2}(w+\tilde{b}),\frac{L}{2}(w-\tilde{b}),\frac{L^2+w\tilde{b}}{2}\right)+\\
        &\frac{e^\chi}{\sqrt{2} L R}\frac{\partial_- w}{b-w}\left(\frac{-L^2+b\tilde{w}}{2},\frac{L}{2}(b+\tilde{w}),\frac{L}{2}(b-\tilde{w}),\frac{L^2+b\tilde{w}}{2}\right)
    \end{aligned}
\end{equation}
Where the first and second expressions are arising from differentiating $X_-$ and $X_+$ with respect to $\sigma^+$ and $\sigma^-$. Notice that equality holds between the two expressions if we impose the condition:
\begin{equation}
     \frac{\partial_+b\partial_-w}{(b-w)^2}=\frac{\partial_+\tilde{b}\partial_-\tilde{w}}{(\tilde{b}-\tilde{w})^2}
\end{equation}
Which is the same as the second equation of \eqref{eq:contbw}. Now we are able to determine the vector $X$. If we use for example the first expression for $X_{+-}$ from \eqref{eq:xpm_bw}, the equation of motion \eqref{eq:eqmotion1} and the explicit expressions in \eqref{eq:scalars} for $\alpha,\lambda$ and $\chi$, we obtain
\begin{equation}
\begin{aligned}
    X=&\mp\frac{1}{b-w}\sqrt{\frac{\partial_+b}{\partial_+\tilde{b}}}\left(\frac{-L^2+w\tilde{b}}{2},\frac{L}{2}(w+\tilde{b}),\frac{L}{2}(w-\tilde{b}),\frac{L^2+w\tilde{b}}{2}\right)\mp\\
    &\mp\frac{1}{\tilde{b}-\tilde{w}}\sqrt{\frac{\partial_+\tilde{b}}{\partial_+b}}\left(\frac{-L^2+b\tilde{w}}{2},\frac{L}{2}(b+\tilde{w}),\frac{L}{2}(b-\tilde{w}),\frac{L^2+b\tilde{w}}{2}\right)
\end{aligned}
\end{equation}
Now we are able to fix signs before all the previous quantites. We chose to use the Poincaré patch in which $X^->0$. This is consistent with the causality conditions $w>b$ and $\tilde{w}>\tilde{b}$ if we choose the upper signs. In higher dimensions we follow this convention to be consistent with the $d=2$ case. From now on we omit the other branch.

The only remaining quantities are the normal vector $N$ and the fields $f,g$ defined in \eqref{eq:defu} and \eqref{eq:defv}. Since in case of $AdS_3$ there is only one normal vector (up to an overall sign), it can be expressed explicitely as
\begin{equation}
    N_a=e^{-\alpha} \epsilon_{abcd} X^b X_-^c X_+^d
\end{equation}
Where the range of  indices $a,b,c,d$ is from $-1$ to $2$. Now we can use the expressions derived for $X$, $X_-$ and $X_+$ to express $N$ in terms of the fields $b$ and $w$. After some algebraic manipulations, it can be shown that the expression is similar to the one derived for $X$, namely:
\begin{equation}
\begin{aligned}
    N^a=&-\frac{1}{b-w}\sqrt{\frac{\partial_+b}{\partial_+\tilde{b}}}\left(\frac{-L^2+w\tilde{b}}{2},\frac{L}{2}(w+\tilde{b}),\frac{L}{2}(w-\tilde{b}),\frac{L^2+w\tilde{b}}{2}\right)-\\
    &+\frac{1}{\tilde{b}-\tilde{w}}\sqrt{\frac{\partial_+\tilde{b}}{\partial_+b}}\left(\frac{-L^2+b\tilde{w}}{2},\frac{L}{2}(b+\tilde{w}),\frac{L}{2}(b-\tilde{w}),\frac{L^2+b\tilde{w}}{2}\right)
\end{aligned}
\end{equation}
Where we have indicated that this expression is for the covariant form of $N$. Now without the detailed derivation, by differentiating $N$ with respect to $\sigma^+$ and $\sigma^-$ and we define:
\begin{equation}
    N\cdot X_{--}=f,\qquad N\cdot X_{++}=g
\end{equation}
we get for $f$ and $g$ the following:
\begin{equation}
    f=-L^2\frac{\partial_-b\partial_-w}{(b-w)^2}+L^2\frac{\partial_-\tilde{b}\partial_-\tilde{w}}{(\tilde{b}-\tilde{w})^2},\qquad g=L^2\frac{\partial_+b\partial_+w}{(b-w)^2}-L^2\frac{\partial_+\tilde{b}\partial_+\tilde{w}}{(\tilde{b}-\tilde{w})^2}
\end{equation}
It is also important to note that from the conditions $N_{+-}=N_{-+}$, $N_{+-}\cdot X=N_{-+}\cdot X=0$ all the other equations in \eqref{eq:2eqmotion1}, \eqref{eq:2eqmotion2} and \eqref{eq:contbw} can be derived.

In conclusion, we have derived the explicit connection between bulk string and boundary causal diamond data.

Finally we would like to point out that the connections in Section \ref{sec:gauge} for free strings propagating in $AdS_{3}$ are given by the following matrices:
\begin{equation}
\begin{aligned}
    B^t_+&=\left(\frac{\partial_+b+\partial_+ w}{b-w}+\frac{\partial_+\tilde{b}+\partial_+ \tilde{w}}{\tilde{b}-\tilde{w}}\right)\begin{pmatrix}
        -1&0\\
        0&1
    \end{pmatrix}\\
    B^t_-&=\left(\frac{\partial_-b+\partial_- w}{b-w}+\frac{\partial_-\tilde{b}+\partial_- \tilde{w}}{\tilde{b}-\tilde{w}}\right)\begin{pmatrix}
        -1&0\\
        0&1
    \end{pmatrix}
\end{aligned}
\end{equation}
While $B^n_\pm=0$. The relations are also satisfied giving rise to the conditions:
\begin{equation}
    \alpha_{+-}+\frac{e^{-\alpha}}{L^2}fg+\frac{e^\alpha}{L^2}=0
\end{equation}
With $\alpha=\lambda+\chi$, and:
\begin{equation}
    f_+=0,\qquad g_-=0
\end{equation}
This is the same as the  "analyticity" property that appeared in \cite{Maldacena} for the function $p(z)$.

In the following section we demonstrate that a different gauge choice solves the equations of motion for strings propagating in a background.

\subsection{Strings coupled to a $B$ field}

The last $AdS_3$ theory we are investigating in this section is the $SL(2,\mathbb{R})$ Wess-Zumino-Witten theory, in which the $AdS_3$ bulk string couples to the RR two-form field strength. The action for the classical string in this case is \cite{Gubser}:
\begin{equation}
    S\sim \int d\sigma^+ d\sigma^- \partial_k X^a \partial_l X^b\left(\sqrt{-h} h^{k l} \eta_{a b}+\kappa \epsilon^{k l} B_{ab}\right)
\end{equation} 
where $X^a$ are $AdS_3$ coordinates. The string is coupled the RR two-form $B_{ab}$ with coupling strength $0<\kappa<1$. Here $k,l=\sigma^+,\sigma^-$ and $a,b$ are $AdS_3$ coordinates. (We omitted the string tension, since it does not play a role in what follows.) The segmented string approximation of this model was considered in \cite{DV2} and \cite{Gubser2}. In this section we are investigating its continuous limit and show that similar gauge structures show up as previously. Moreover, the bulk geometrical objects (in particular the string itself) are naturally emerge again from boundary causal diamonds data.

In this case the equation of motion for the string can be written in a form
\begin{equation}
    \partial_{+} \partial_{-} X_a-\frac{1}{L^2}\left(\partial_{+} X\cdot \partial_{-} X\right) X_a+\frac{\kappa }{L^2}\epsilon_{abcd} X^b \partial_{-} X^c \partial_{+} X^d=0
\end{equation}
Provided with the Virasoro constraints
\begin{equation}
X_+\cdot X_+=X_-\cdot X_-=0
\end{equation}
Proceeding similarly as before, one can define the field $X_-\cdot X_+=-e^\alpha$ and the normal vector
\begin{equation}
    N_a=e^{-\alpha} \epsilon_{abcd} X^b X_-^c X_+^d
\end{equation}
Then the equation of motion becomes
\begin{equation}
    \partial_{+} \partial_{-} X+\frac{e^\alpha}{L^2}X+\frac{\kappa e^\alpha}{L^2}N=0
\end{equation}
Furthermore if we set
\begin{equation}
   \begin{aligned}   
        N\cdot X_{--}&=f\\
        N\cdot X_{++}&=g  
        \end{aligned}
\end{equation}
then the equation of the string along with the directional derivatives $\partial_\pm X$ and the normal vector $N$ is fully described by the set of equations
\begin{align}
    X_{+-}&=-\frac{1}{L^2}e^\alpha X-\frac{\kappa e^\alpha}{L^2}N \\
    X_{++}&=\alpha_+ X_+ +\frac{1}{L^2} v N \\
    X_{--}&=\alpha_- X_- +\frac{1}{L^2} u N \\
    N_+&=g e^{-\alpha} X_- -\kappa X_+\\
    N_-&=f e^{-\alpha} X_+ -\kappa X_-
\end{align}

To see the $SO(1,1)\times SO(1,1)$ gauge structure, we can introduce again the $t$ and $n$ vectors with elements
\begin{align}
    &t^0:=L e^{-\lambda} X_-,&t^1:&=L e^{-\chi} X_+,\\
    &n^{0}:=X,&n^{1}:&=N
\end{align}
Then it is straightforward to show that the equations of motion are equivalent to
\begin{align}
    \left(\partial_\pm +{B^t_\pm}\right)t&=\Phi_\pm\eta n\\
    \left(\partial_\pm +{B^n_\pm}\right)n&=\Phi_\pm^T \sigma t
\end{align}
with:
\begin{equation}
        {B^t_+}=
    \begin{pmatrix}
    \lambda_+&0\\
    0&-\lambda_+
    \end{pmatrix},\qquad
    {B^t_-}=
    \begin{pmatrix}
    -\chi_-&0\\
    0&\chi_-
    \end{pmatrix}
\end{equation}
\begin{equation}
    \Phi_+=\left( \begin{array}{c|c}
    \frac{e^\chi}{L}&-\kappa\frac{e^\chi}{L}\\
    \midrule
    0&\frac{e^{-\chi}}{L}g^T
    \end{array}\right),\qquad 
    \Phi_-=\left( \begin{array}{c|c}
    0&\frac{e^{-\lambda}}{L}f^T\\
    \midrule
    \frac{e^\lambda}{L}&-\kappa\frac{e^\lambda}{L}
    \end{array}\right)
\end{equation}
And $B_\pm^n=0$. As previously $B^t_\pm$ and $B^n_\pm$ are $SO(1,1)$ connections related to the tangent and "normal" space of the bulk string. It can be shown, that the relations \eqref{eq:flat1}-\eqref{eq:flat3} are also satisfied giving sligthly different conditions, eg.:
\begin{equation}
    \alpha_{+-}+\frac{1}{L^2}fge^{-\alpha}+\frac{1}{L^2}(1-\kappa^2)e^\alpha=0
\end{equation}
We see that the $SO(1,1)\times SO(1,1)$ invariance remains the same, however a different choice of $\Phi_\pm$ modified the bulk string theory.

Finally every previously investigated quantities can be expressed in terms of the celestial fields in case of a $B$ field as well, namely
\begin{equation}
\begin{aligned}  e^\alpha&=\frac{2L^2}{1-\kappa^2}\frac{\sqrt{\partial_+b\partial_+\tilde{b}\partial_-w\partial_-\tilde{w}}}{(b-w)(\tilde{b}-\tilde{w})},\\
e^\lambda&=\frac{\sqrt{2}L}{\sqrt{1-\kappa^2}}\sqrt{\frac{\partial_- w\partial_-\tilde{w}}{(b-w)(\tilde{b}-\tilde{w})}},\\
e^\chi&=\frac{\sqrt{2}L}{\sqrt{1-\kappa^2}}\sqrt{\frac{\partial_+ b\partial_+\tilde{b}}{(b-w)(\tilde{b}-\tilde{w})}}
\end{aligned}
\end{equation}
And
\begin{equation}
    f=-\frac{L^2}{1+\kappa}\frac{\partial_-b\partial_-w}{(b-w)^2}+\frac{L^2}{1-\kappa}\frac{\partial_-\tilde{b}\partial_-\tilde{w}}{(\tilde{b}-\tilde{w})^2},\qquad g=\frac{L^2}{1-\kappa}\frac{\partial_+b\partial_+w}{(b-w)^2}-\frac{L^2}{1+\kappa}\frac{\partial_+\tilde{b}\partial_+\tilde{w}}{(\tilde{b}-\tilde{w})^2}
\end{equation}
While the equations of motion and the continouity conditions are satisfied if the following relations hold for the celestial fields
\begin{equation}\label{eq:Bbw}
\begin{aligned}
       \partial_{+-}b&=2\frac{\partial_-b\partial_+b}{b-w},& \partial_{+-}w&=2\frac{\partial_-w\partial_+w}{w-b}\\
        \frac{\partial_+b\partial_-w}{(b-w)^2}&=\frac{\partial_+\tilde{b}\partial_-\tilde{w}}{(\tilde{b}-\tilde{w})^2},&
        \frac{\partial_-b\partial_+w}{(b-w)^2}&=\frac{\partial_-\tilde{b}\partial_+\tilde{w}}{(\tilde{b}-\tilde{w})^2}\\
        (1-\kappa)\frac{\partial_-b\partial_-w}{(b-w)^2}&=-(1+\kappa)\frac{\partial_-\tilde{b}\partial_-\tilde{w}}{(\tilde{b}-\tilde{w})^2},&(1+\kappa)\frac{\partial_+b\partial_+w}{(b-w)^2}&=-(1-\kappa)\frac{\partial_+\tilde{b}\partial_+\tilde{w}}{(\tilde{b}-\tilde{w})^2}\\
        (1-\kappa)\partial_-b\partial_+b&=-(1+\kappa)\partial_-\tilde{b}\partial_+\tilde{b},&
        (1-\kappa)\partial_-w\partial_+w&=-(1+\kappa)\partial_-\tilde{w}\partial_+\tilde{w}
\end{aligned}
\end{equation}
In the limit $\kappa\to 0$ we get back the equations derived for free strings. We can again interpret the $b$ and $w$ fields as boundary causal diamond coordinates. Therefore equation \eqref{eq:Bbw} encode a dynamics of boundary causal diamonds, which are dual to a $AdS_3$ bulk string theory coupled to $B$ field. We can also see that in the gauge invariant formalism what changed compared to the free string case is the form of $B_\pm^t$ in terms of the fields $b$ and $w$. This corresponds to a different choice of gauge. Therefore choosing a different gauge when pulling back the equations of motion from the boundary, or the kinematic space to the bulk one can get different bulk string theories.

One final remark is that all of these equations and expressions can be derived from a segmented string approach following the notations of Section \ref{sec:segmented} and the papers \cite{DV2,Gubser} and \cite{Dv3}.

\section{Conclusions}

In this paper we explored ideas of holography, strings and kinematic space in a unified framework borrowed from twistor theory. 
In our treatise of correspondences between the geometric structures of different spaces, such as the AdS$_{d+1}$ bulk, its boundary, the moduli space of boundary causal diamonds aka the kinematic space ${\mathbb K}$,
we reverted to the perspective offered by 
a projective geometry.
From this viewpoint it turned out that
certain planes spanned by two light-like vectors $U$ and $V$ in ${\mathbb R}^{d,2}$ are playing an important role.  In the projective picture they define special lines in ${\mathbb R}{\mathbb P}^{d+1}$.  In our geometric elaborations objects like Ryu-Takayanagi surfaces, spacelike geodesics, horospheres and the metric on kinematic space, familiar to physicists as the main actors on the holographic scene, all have found a natural place.

Then we have learned that the defining null vectors $U$ and $V$ of our projective lines can be upgraded to vector fields implementing the string theoretic Virasoro constraints in the conformal gauge.
This resulted in the realization that  our holographic objects can naturally be related to  strings that propagate in $AdS_{d+1}$. 
Then by relating the string action to an action incorporating the moduli space metric, we have shown that strings in the AdS bulk can alternatively be described by a Grassmannian $\sigma$ model targeted on ${\mathbb K}$.

We have shown that string motion in the bulk is projected to a dynamics of causal diamonds.
However, one can even succeed in the other way round, namely in the construction of the bulk world-sheet from the boundary data provided by the future and past tips of a causal diamond.
We have observed that the problem of how to lift up a diamond to get a proper string world sheet is reminiscent of the geometric one of lifting up certain objects from a base space to a bundle space which is a characteristic feature of gauge theories.
We have identified this  emerging gauge structure as the one that is modeled by a Stiefel bundle over the Grassmannian ${\mathbb K}=SO(2,d)/SO(1,d-1)\times SO(1,1)$. This structure is then responsible for the possibility of representing string dynamics in a form of a dynamics of causal diamonds.
We have also pointed out that a suitably discretized version of this theory contains the theory of segmented strings.

Finally we have worked out explicitly the special case of $d=2$.
Here we have found a direct link to a specialization of the basic twistor correspondence.
Namely, we have found that the special lines in ${\mathbb R}{\mathbb P}^{3}$ (i.e. in the projectivization of the bulk) correspond to special points comprising ${\mathbb K}=dS_2\times dS_2$ living inside the Klein quadric.
Intersecting bulk lines (also representing spacelike separated  intersecting boundary causal diamonds) correspond to points in light cones in kinematic space. Alternatively, light-like separation in ${\mathbb K}$ is represented by a reality condition for boundary cross ratios. 
For the static slice the projective geometric picture is even more suggestive. Here one can represent bulk boundary and kinematic space in a unified geometric manner.
For the $d=2$ case the uplifting of causal dynamics is a string theory in AdS$_3$. Different gauges correspond to different versions of this theory, namely either ones with or without B-fields.

Our treatise was entirely classical.
However, the question arises of what kind of quantum lessons one should draw from our investigations?
In this respect note that naively one expects that the full advantage of string theory as an agent of probing spacetime geometry  manifests itself merely in the quantum realm.
However, this is not the case. In  our recent paper\cite{Levay4} we have demonstrated that in the segmented string approximation {\it classical} segments of string world sheets are encoding information on {\it quantum} entanglement and complexity properties of the CFT$_d$ vacum.
More precisely, let us consider two time ordered intersecting causal diamonds: an initial and a final one ${\mathcal D}_i$ and ${\mathcal D}_f$.
Such diamonds are {\it spacelike} separated according to the metric of $\mathbb K$.
Let us also consider the associated reduced density operators $\varrho_i$ and $\varrho_f$ respectively. Then one expects that the classical causal dynamics discussed in this paper should somehow manifest itself in the quantum domain as a quantum computation taking $\varrho_i$ to $\varrho_f$. 
Then according to our findings in Ref.\cite{Levay4} one should have a correspondence between the complexity of this {\it quantum} computation and the area of the corresponding {\it classical} string segment.
In \cite{Levay4} we have indeed demonstrated this correspondence via using the so called quantum state space metric\cite{Provost}. However, this metric is not a genuine one on the space of density operators hence our argument was sketchy.
What one needs in this context is a careful study featuring the Bures-Uhlmann\cite{Bures} or other metrics that are inherently connected to Fisher information metrics.
In fact there is a full zoo of possibilities for such metrics\cite{Hijano}. As we have also shown infinitesimally close points on the string worldsheet are related to infinitesimally close spherical regions of the boundary via causal diamonds. In \cite{Faulkner} the authors investigated the relative entropy of the states of such infinitesimally close ball shaped regions. It is an interesting question if their results could give us a natural generalization of our segmented string-entanglement entropy duality in the continuous limit or whether there is a corresponding quantity on the string theory side that encodes the boundary relative entropy. Such quantum/classical aspects we are intending to explore in future works.

We also emphasize here that our approach was relevant merely in the context of pure AdS geometry. 
However, for interesting applications for more general scenarios one should pay attention to asymptotically AdS geometries. Indeed, originally these geometries are the main actors which play a role in the AdS/CFT correspondence.
It is well-known that pure AdS is dual to the CFT vacuum. 
Moreover, perturbations of the vacuum manifest themselves in certain aspects of asymptotically AdS geometries.
Even more general CFT states can dually describe topologically more complicated bulk objects like in the $AdS_3$ case multiboundary wormholes.
Can our approach account for them, or at least offer some new path to follow?

In this respect recall that
the results of our recent paper\cite{Levay4}
show that in the pure AdS context the area of the tile  swept out by a bulk string segment is proportional to
the distance between the corresponding overlapping time ordered causal diamonds of the boundary. Here the distance is understood with respect to the metric in kinematic space.
When the area of the bulk tile is small we even seem to have a relationship between the area of the tile and the complexity of a quantum computation in the boundary.
Now CFT excited states, at least locally, look like the vacuum.
Hence one expects that this corresponds to the fact that certain asymptotically AdS spacetimes locally look like pure AdS.
When we take string segments in this background also into account, one can conjecture that for small segments (i.e. locally) one should find a correspondence between strings propagating in an asymptotically AdS geometry and excited CFT states.
Hence our twistor picture in principle can be applied.
However, for more complicated CFT states corresponding to topologically nontrivial (e.g. multiboundary wormholes) spacetime configurations, the global aspects of strings (winding modes) have to be somehow taken into account.
In any case even in these more general cases intuitively the twistor theory motivated "space of strings of the bulk/ space of causal diamonds in the boundary" correspondence seems to be a useful guiding principle for further research.

Notice finally the amusing connection that was found here with twistor theory for the $d=2$ case.
Here AdS$_3$ geometry was regarded as a specialization of projective geometry
in the "projective twistor space" ${\mathbb R}{\mathbb P}^{3}$.   
On the other hand the kinematic space turned out to be the special manifold $dS_2\times dS_2$ living inside the Klein-quadric
residing in ${\mathbb R}{\mathbb P}^5$.
In the original twistor setting there is a correspondence between the {\it complex} projective twistor space ${\mathbb C}{\mathbb P}^3$ 
and the four dimensional Klein-quadric residing in ${\mathbb C}{\mathbb P}^5$. On the other hand the Klein quadric  was identified with the compactified and {\it complexified} Minkowski space-time.
Real spacetime manifolds of physical interest are coming out of this approach as ${\it real} sections$ of the Klein quadric\cite{Penrose,Hurd}.
It is thought provoking to realize in this respect that three dimensional anti de Sitter space seems to be an object which is rather living inside a twistor space, on the other hand kinematic space (comprising two copies of de Sitter spaces) is living inside the Klein quadric, a real section of a complexified space time.
Apart from recording this interesting coincidence, are there further lessons to draw (using e.g. the results of Ref.\cite{Krasnov}) for the $d>2$ cases?

\section{Acknowledgements}
This work was supported by the HUN-REN Hungarian Research Network
through the 
Supported Research Groups Programme,
HUN-REN-BME-BCE Quantum Technology Research Group (TKCS-2024/34) and
by the National Research Development and Innovation Office of Hungary within the Quantum Technology National Excellence Program (Project No. 2017-1.2.1-NKP-2017-0001).

\bibliography{references}
\bibliographystyle{JHEP}

\end{document}